\documentclass[12pt]{article}

\usepackage[margin=1in]{geometry}
\usepackage{amsmath,amssymb,mathtools,amsthm}
\usepackage{microtype}
\usepackage{comment}
\usepackage{placeins} %
\usepackage{ragged2e} %
\usepackage{array}
\usepackage{longtable}
\usepackage[authoryear,round]{natbib}
\usepackage{enumitem}
\usepackage{xcolor}

\usepackage{hyperref}
\hypersetup{
    hidelinks,
    pdftitle={Team Stability},
    pdfauthor={Anton Badev}
}

\usepackage{tikz}
\usepackage{graphicx}
\usetikzlibrary{positioning}

\definecolor{individualblue}{HTML}{3A6F8F}
\definecolor{projectamber}{HTML}{B78124}
\tikzset{
    individual node/.style={circle,draw=individualblue,fill=individualblue!8,thick,
        minimum size=15mm,inner sep=1pt,align=center,font=\small},
    project node/.style={rectangle,rounded corners=2pt,draw=projectamber,fill=projectamber!12,thick,
        minimum width=25mm,minimum height=15mm,inner sep=2pt,align=center,font=\small},
    diagram edge/.style={draw=black!75,thick}
}

\newtheorem{definition}{Definition}
\newtheorem{assumption}{Assumption}
\newtheorem{remark}{Remark}
\newtheorem{example}{Example}
\newtheorem{lemma}{Lemma}
\newtheorem{proposition}{Proposition}
\newtheorem{theorem}{Theorem}
\newtheorem{corollary}{Corollary}
\renewcommand{\qedsymbol}{\leavevmode\hbox to.77778em{\hfil\vrule width .9pt\vbox to.675em{\hrule height .9pt width.6em\vfil\hrule height .9pt}\vrule width .9pt\hfil}}

\newcommand{\cA}{\mathbb{A}}

\newcommand{\floatnotesfont}{\footnotesize}

\DeclareMathAlphabet{\mathmybb}{U}{bbold}{m}{n}
\newcommand{\1}{\mathmybb{1}}

\title{Team Stability:\\ Theory and Evidence}
\author{Anton Badev\thanks{For the latest version, visit \href{https://www.antonbadev.net/teams/}{https://www.antonbadev.net/teams/}.\endgraf
I am indebted to Steven Durlauf, James Heckman, Matthew Jackson, Michael K\"onig, George Mailath, Angelo Mele, Seth Richards-Shubik, and Petra Todd for discussions that have significantly shaped this project. I also thank audiences at Network Science and Economics conferences and meetings of the Econometric Society for helpful comments. Any remaining errors are my own. The analysis and conclusions are those of the author and do not indicate concurrence by other members of the research staff or the Board of Governors.}}
\date{November 14, 2026}

\begin{document}

\maketitle

\begin{abstract}
\justifying

Teams emerge organically when individuals decide on their own their activities and collaborators. I develop and test a theory in which teams form with consent, interactions reflect team composition, and payoffs feature externalities and several forms of congestion. Stability constraints concern group interactions and require each team member to be willing to remain rather than withdraw or pursue a project alone. Empirical analysis of research collaborations among economists at the Federal Reserve Board suggests that: (i.) these constraints help explain observed teams beyond the payoffs to collaboration; (ii.) estimated violations point to large teams of entrants that appear to be assigned;\linebreak (iii.) a compensatory mechanism mitigates the effects of variation in policy workload, which appears to affect team composition more than total research output; and \linebreak (iv.) shifting workload toward experienced economists increases entrants' research output and collaboration, at the cost of lower output among experienced economists who compensate for this shift in policy workload with increased interactions with entrants.
\end{abstract}

\clearpage

\section{Introduction}

Social groups and professional teams emerge organically when individuals decide on their own their activities and partners. The decisions to embark on a new project and join a team reflect individuals' interests, yet team collaborations typically involve cooperation and acceptance very much like friendships \citep{Jackson_Wolinsky_1996}. Choosing activities and partners shapes both the organization of social activity and the productivity of the teams that form. In this paper, my goal is to develop and test a theory of team formation and use this theory to study the determinants of aggregate activities, team sizes, and compositions.

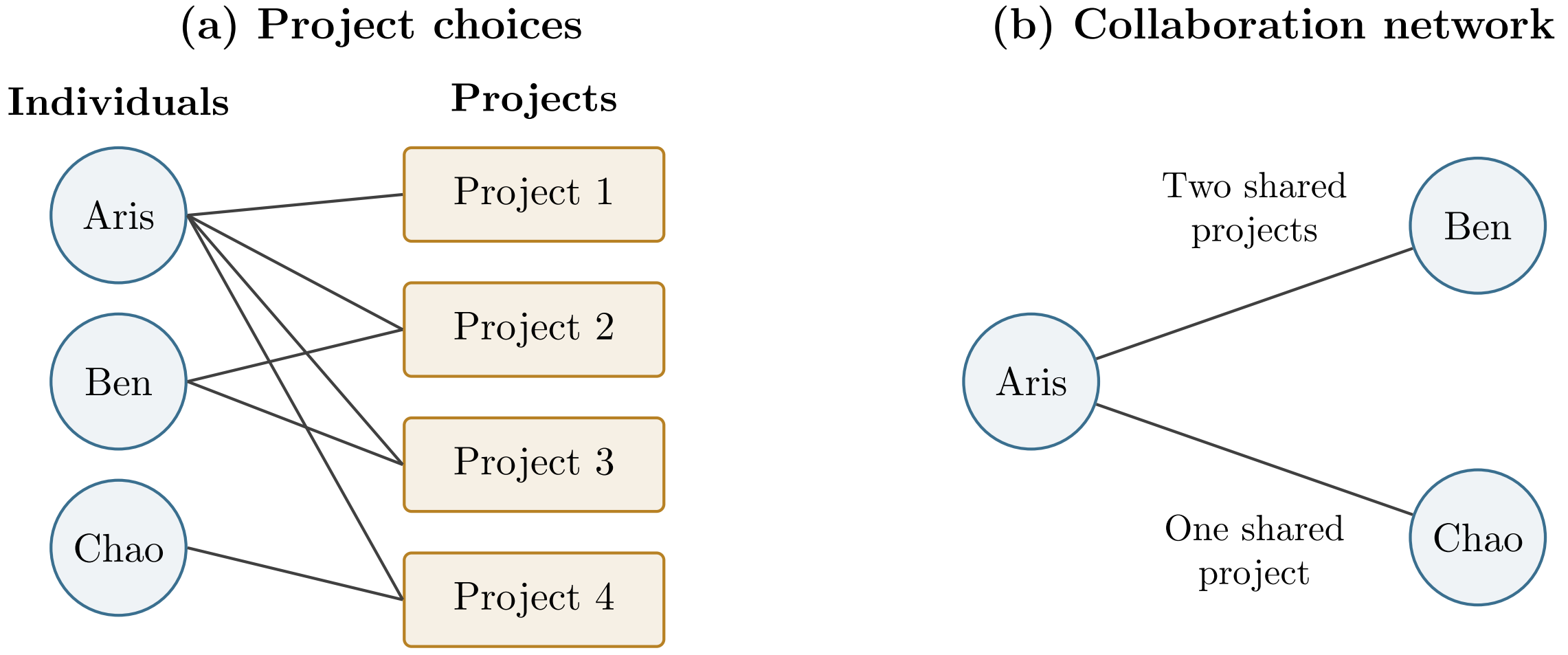
\begin{figure}[!ht]
\centering
\caption{Projects and Collaborations}
\label{fig:intro-setup}
\begin{minipage}[t]{0.48\textwidth}
\centering
\textbf{(a) Project choices}\\[6pt]
\begin{tikzpicture}[every node/.style={font=\small}]
\path[use as bounding box] (-0.9,-2.7) rectangle (6.0,2.95);
\node[font=\small\bfseries] at (0,2.7) {Individuals};
\node[font=\small\bfseries] at (4.0,2.7) {Projects};
\node[individual node,minimum size=13mm] (aris) at (0,1.6) {Aris};
\node[individual node,minimum size=13mm] (ben) at (0,0) {Ben};
\node[individual node,minimum size=13mm] (chao) at (0,-1.6) {Chao};
\node[project node,minimum height=9mm] (solo) at (4.0,1.8) {Project 1};
\node[project node,minimum height=9mm] (abone) at (4.0,0.5) {Project 2};
\node[project node,minimum height=9mm] (abtwo) at (4.0,-0.8) {Project 3};
\node[project node,minimum height=9mm] (ac) at (4.0,-2.1) {Project 4};
\draw[diagram edge] (aris.east) -- (solo.west);
\draw[diagram edge] (aris.east) -- (abone.west);
\draw[diagram edge] (aris.east) -- (abtwo.west);
\draw[diagram edge] (aris.east) -- (ac.west);
\draw[diagram edge] (ben.east) -- (abone.west);
\draw[diagram edge] (ben.east) -- (abtwo.west);
\draw[diagram edge] (chao.east) -- (ac.west);
\end{tikzpicture}
\end{minipage}\hfill
\begin{minipage}[t]{0.48\textwidth}
\centering
\textbf{(b) Collaboration network}\\[6pt]
\begin{tikzpicture}[every node/.style={font=\small}]
\path[use as bounding box] (-0.9,-2.7) rectangle (6.0,2.95);
\node[individual node,minimum size=13mm] (aris) at (0.2,0) {Aris};
\node[individual node,minimum size=13mm] (ben) at (4.5,1.5) {Ben};
\node[individual node,minimum size=13mm] (chao) at (4.5,-1.5) {Chao};
\draw[diagram edge] (aris) -- (ben);
\draw[diagram edge] (aris) -- (chao);
\node[align=center,font=\footnotesize] at (2.35,1.65) {Two shared\\projects};
\node[align=center,font=\footnotesize] at (2.35,-1.65) {One shared\\project};
\end{tikzpicture}
\end{minipage}
\par\smallskip
{\RaggedRight\floatnotesfont Notes. In panel (a), Aris works alone on Project 1, with Ben on Projects 2 and 3, and with Chao on Project 4. Panel (b) shows the collaborations implied by these choices; Ben and Chao share no project.\par}
\end{figure}

In the proposed model, individuals choose which projects to undertake and with whom to work (Figure~\ref{fig:intro-setup}). Their choices are subject to teams' stability constraints \citep{Badev2021}: an individual can join or remain on a team only if every member is willing to stay rather than withdraw or pursue a project alone. The model accommodates solo work, overlapping teams, and repeated collaboration, with both the number of projects and their membership determined endogenously. I estimate the model using data on research collaborations among economists at the Federal Reserve Board. The estimates support a distinct role for stability constraints: a model without them predicts too many team memberships that individuals would prefer to leave, even when its other parameters are re-estimated. Large violations of the estimated stability conditions point to large teams of entrants that appear to be assigned. Counterfactual experiments examine how the timing and allocation of policy workload affect research output and collaboration.

\FloatBarrier

When deciding whether to undertake a new project, an individual considers both her own interest in the activity and the benefits of working with others. Collaboration may be attractive because of positive experiences with past partners or because teammates bring complementary skills. Working alone remains an option. A team can therefore form and persist only if every member is willing to remain rather than withdraw or pursue a new project alone.

Team interactions involve collaboration and coordination among individuals with different skills, roles, and commitments. When a new member joins, incumbents may reallocate tasks and effort, redistributing the gains and costs of collaboration in ways the researcher does not observe. An incumbent's incremental payoff evaluated at a fixed allocation of effort may therefore differ from her payoff change after these internal adjustments. Collaboration may also involve altruistic motives, so an individual's direct gain need not fully capture her willingness to participate. These considerations motivate a stability requirement based on continued participation: entry need not increase every incumbent's direct payoff, provided that each member remains willing to stay rather than withdraw or pursue a project alone.

In a $k$-team stable equilibrium no individual can profitably revise her participation in a set of projects while satisfying the relevant stability constraints. The family of equilibria is indexed by the size of the consideration set, $k$, which governs the complexity of the considered deviations and ranks the $k$-team equilibria by set inclusion. The larger $k$ is, the richer is the set of alternatives individuals consider. The associated $k$-team consensual dynamics feature non-standard properties: the potential is non-monotone because individuals face stability constraints in adapting their project participations and a locally feasible revision can make a membership outside the consideration set unstable. Despite this non-monotonicity, the $k$-team dynamic converges almost surely to a $k$-team stable equilibrium (Proposition~\ref{prop:deterministic-path-bounds} and Theorem~\ref{thm:deterministic-team-convergence}). The key is that the instability is temporary and disappears in finite time because a dissatisfied teammate leaves or leaves and then starts a solo project, ending with a higher payoff than before the adjustment.

I reconcile the multiplicity of stable configurations of projects and teams---a primitive feature of models with strong complementarities---through a probabilistic ranking of project and team configurations following the tradition of logit models of adaptive dynamics.\footnote{See the contributions of \cite*{nakajima2007,mele2017structural,hsiehleeboucher2020specification,Badev2021,hsiehkonigliu2022coevolution}.} In particular, the $k$-team consensual dynamic is perturbed with logit shocks, to obtain a stationary distribution that serves as the likelihood.\footnote{As in random utility models, the stationary distribution assigns positive probability to every configuration in the model's state space, including those that are not stable equilibria. Suboptimality and violations of the stability constraints reduce a configuration's probability weight, all else equal. Evaluating this distribution at the observed project memberships yields the likelihood as a function of the model parameters.} I depart from this literature by introducing stochastic consent and an endogenous number of projects. At each period's update, a proposed configuration is screened across all team memberships; failed stability constraints reduce its probability of implementation through a channel separate from that of the payoff parameters (Proposition~\ref{prop:stationary-probabilistic-constraints}). The bounds on project counts and degrees hold in $k$-team stable states and along the unperturbed dynamics (Propositions~\ref{prop:number_projects}--\ref{prop:degree_bounds} and~\ref{prop:deterministic-path-bounds}), but not under logit perturbations: because the number of projects is endogenous, the stationary distribution assigns strictly positive probability to any non-negative project count. I obtain Gaussian-type tails for project counts and bounds on the likelihood distortion from capping the state space to a finite number in terms of the excluded probability mass (Theorem~\ref{thm:logit-project-count-tail}). Lastly, I analyze the convergence properties of adaptive dynamics, both because they provide an algorithmic template for simulating and estimating the model and because convergence can be slow in related network models (Theorems~\ref{thm:unconstrained-logit-speed} and~\ref{thm:unconstrained-logit-speed-comparison}).\footnote{On slow convergence in network models, see \citet{bhamidi2011mixing}. I substantially generalize the convergence comparison in \citet[Theorem~2]{Badev2021} to guide the design of the estimation algorithm. That result considers a zero-payoff benchmark with independent choices in stationarity. Theorem~\ref{thm:unconstrained-logit-speed-comparison} allows general potentials with heterogeneous returns, complementarities, and congestion, accommodating dependent memberships and multiple modes of the stationary distribution. Under nested, state-independent updates, larger meetings weakly increase the spectral gap, with convergence measured per meeting.} %

\subsection{Data, Identification, and Estimation}

The model is implemented on an unbalanced panel of economists at the Federal Reserve Board over 2007--2019. I combine bibliographic records from OpenAlex \citep{priem2022openalex} with archived staff rosters to identify authors, employment histories, and organizational affiliations. Manual checks correct author lists, merge duplicate versions, and distinguish research papers from institutional publications, which are excluded. The estimation sample contains 449 economists, 3{,}522 economist-years, and 1{,}918 papers. This sample contains economists who publish at least once during their observed tenure, retaining their years without papers. Entry into this research population is outside the model. I measure policy workload using the linguistic complexity of Board staff remarks in FOMC transcripts, providing a yearly proxy for demands on economists' research time.\footnote{\cite{Badev_Bardales_2026} present statistical evidence of collaboration cycles where periods of high policy workload are associated with a decrease in Board researchers' output.}

Pinning down model parameters takes a mix of variations. Repeated observations of economists' project counts reveal persistent productivity differences, while changes in the propensity to undertake another project as portfolios grow inform congestion. Conditional on these activity incentives, team size and section composition inform team values, and partner choices inform preferences over collaborators' productivity and tenure. Repeated projects within a year and the renewal of past collaborations distinguish the return to sharing several projects from the value of an established relationship. Variation in policy workload and job responsibilities informs how research incentives change with other demands on economists' time. The participation penalty is distinguished through changes in members' outside options: additional projects can move a membership across its empirical exit threshold without changing the team's composition. After accounting for the accompanying payoff changes, such contrasts separate the penalty from collaborator preferences. More generally, stability-constraint checks link the benefits of a particular team to commitments elsewhere in a member's portfolio, introducing variation that additive portfolio, team, and network payoffs cannot generally reproduce. Propositions~\ref{prop:sorting-consent-local-identification} and~\ref{prop:participation-nonabsorption} formalize the conditions for joint identification and separation of the participation penalty from additive payoffs.

I estimate the model from its stationary likelihood using Bayesian simulation methods, adapting the varying-$k$ double Metropolis--Hastings approach of \citet{Badev2021}. The estimation jointly accounts for persistent differences in individual productivity, team composition, repeated collaboration, workload, and the weight on stability violations. Direct evaluation of the likelihood is infeasible because its normalizing constants sum over all possible project assignments, which are in the order of $2^{P\times N}$. I instead simulate auxiliary configurations using Markov chains that preserve the model's stationary distribution. These simulations allow posterior sampling without evaluating the normalizing constants; their finite lengths are calibrated against much longer runs.

\subsection{Conclusions from the Empirical Analysis}

The estimates indicate that stability constraints help explain observed teams beyond the payoffs to collaboration. Each failed stability check lowers the probability of an outcome by about 29\% (90\% posterior interval 23--35\%). A model re-estimated with $\kappa=0$ reproduces overall participation and team composition but generates 26\% more memberships in violation than observed. The consent penalty closes this gap. Examining observed violations also draws attention to large teams of entrants that the model struggles to reproduce, suggesting that some projects reflect institutional assignment rather than decentralized team formation. For example, in 2017 the same two economists wrote a three-part FEDS Notes series. As three common projects of one pair within a year, these notes are among the deepest observed violations: in each part, both coauthors fall short of their solo option. Large teams provide other examples: three of the five members of a 2014 paper on the demand for short-term safe assets, and one member of the only six-member team in the sample, a 2018 paper on spending responses to the Earned Income Tax Credit, fall short of their solo option. See the appendix.

Counterfactual experiments examine how the timing and allocation of policy workload affect research output and collaboration. First, a compensatory response shifts research from solo projects into teams. Higher workload lowers own-project returns but raises team values, offsetting much of the direct decline in participation and relaxing the stability constraints. Raising workload across the observed range increases the team share of participations from 41\% to 52\%, while the sign of the change in total participation remains uncertain. Second, holding average workload fixed, bursts add at most 2.5\% to total papers because a quiet year raises output by more than a busy year lowers it. Their timing matters more for team composition: on the observed growing staff, placing the busy years last lowers the solo share by up to 2.8 percentage points and raises repeated coauthoring pairs by up to 38\%. More economists face the high workload, and the ties they form support repeated collaboration and larger teams. With a fixed staff, an early busy period builds more collaboration because its ties have longer to be reused. Third, holding total workload fixed, allocating more policy work to experienced economists raises total papers by 2.6\% but redistributes research much more substantially: entrants author 18\% more papers per economist-year, while economists with nine or more years at the Board author 7\% fewer. Entrants also gain collaborators, while the most experienced economists author fewer large-team papers and, less certainly, have fewer repeated collaborations. Their collaborations shift toward entrants and away from experienced peers.

\subsection{Related literature}

A long tradition of research, starting with \citet{beckermurphy1992division} and \citet{kremer1993oring}, develops theories of how individuals sort into productive teams and how team composition shapes output and individual returns \citep{carlierekeland2010teams,boerma2021sorting,kinateder2009team}. \citet*{jarosch2021learning} and \citet{herkenhoff2024production} extend this perspective to dynamic settings in which coworkers influence learning and future productivity. Related work studies how the gains from cooperation shape team formation and, conversely, how team durability and task assignments shape cooperation across activities \citep{chalkiadakis2010overlapping,boncinelli2022teams,brown2024team,bendor2026multiple}. My focus is on the microfoundations of team formation in an empirical framework that accommodates rich heterogeneity. Teams form with consent, interactions reflect team composition, and payoffs feature externalities and several forms of congestion. The empirical analysis has three goals: to test the theory, to match micro-level moments of team composition and activity, and to use the estimated model to predict equilibrium outcomes in counterfactual environments beyond those observed in the data.

Recent empirical studies of scientific teams provide several closer comparisons.\footnote{On coauthorship networks more broadly, \citet{goyal2006economics} document the emergence of a small world in economists' coauthorship networks, while \citet{ductor2014social} show that coauthor information modestly improves forecasts of individual research output over those based on past output alone. Bill cosponsorship in legislatures has the same bipartite structure: \citet{fowler2006connecting} maps cosponsorship networks in the U.S. Congress, \citet{battaglinileonepatacchini2020effectiveness} and \citet{canen2023social} relate legislators' effectiveness to their position in endogenous cosponsorship networks, and \citet{loolivellaimai2023bipartite} show that projecting legislator--bill data onto a legislator network induces aggregation bias.} \citet{andersonrichardsshubik2022collaborative} model endogenous project formation, research topics, and coauthor teams, using stability restrictions to partially identify collaboration preferences. \citet{allocca2023teams} jointly estimates entry into given projects and their probability of completion, while \citet{hsieh2025collaboration} model managerial assignment and individual effort across a given set of projects. \citet{bonhomme2026teams} uses repeated team assignments to recover individual contributions, sorting, and complementarities when only team output is observed. An extension of his model adds a stochastic blockmodel of team formation, allowing latent types to affect both output quantity and quality. In the proposed framework, the likelihood over project portfolios embeds a rich set of externalities and congestion mechanisms (including team size and section composition, collaborator sorting by productivity and tenure, and repeated and past collaboration), together with an estimable penalty for failed team stability constraints. Estimating this penalty allows me to assess whether stability constraints help explain observed teams beyond what payoffs alone can explain.

Individuals' project participation choices generate a collaboration network, linking the model to the strategic network formation literature surveyed by \citet{jackson2008networks}, \citet{graham2020network}, and \citet{depaula2020formation}.\footnote{For broader surveys of social interactions and peer effects, see \citet{brockdurlauf2001interactions}, \citet{blume2011identification}, \citet{jacksonrogerszenou2017consequences}, and \citet{bramoulle2020peer}.} In the proposed equilibrium, stability constraints adapt to multilateral interactions the notion of consent in \citet{Jackson_Wolinsky_1996}. Unobserved task and effort reallocations, along with altruistic motives, can make direct payoff changes an inaccurate measure of willingness to collaborate. A conservative stability criterion therefore permits entry to lower an incumbent's direct payoff, provided that every member remains willing to stay rather than withdraw or pursue a solo project. 

Stability constraints introduce non-monotonicity in the deterministic adaptive dynamics and additional terms in the model's likelihood, which belongs to the class of exponential random graph models (ERGMs). This literature derives ERGMs as stationary distributions of strategic network formation, first for links alone \citep{mele2017structural,mele2020desegregation}, then with endogenous social interactions \citep{hsiehleeboucher2020specification,Badev2021}, and for the joint evolution of R\&D collaboration networks and effort \citep{hsiehkonigliu2022coevolution,hsiehkonigliu2025spillovers}. My likelihood over project memberships combines the payoff potential with an estimable penalty for failed stability checks, imposed through global membership screening. The model without stability constraints is nested as a special case.

\section{A Model of Project and Team Selection}

\subsection{Setup and Notation}
\label{sec:preliminaries}

Individuals in a finite population $N=\{1,\ldots,n\}$ jointly choose projects and collaborations on a set of projects $P=\{1,\ldots,p\}$.  Let $a_{i,r}=1$ indicate $i$'s choice to participate in project $r$, otherwise $a_{i,r}=0$. The participation matrix $A=[a_{i,r}]_{n\times p}$ describes both the projects of each individual and the team of each project:
\begin{align*}
    P_i&=\{r\in P:a_{i,r}=1\},
    &p_i&=|P_i|=\sum_{r\in P}a_{i,r},\\
    N_r&=\{i\in N:a_{i,r}=1\},
    &n_r&=|N_r|=\sum_{i\in N}a_{i,r}.
\end{align*}
Thus, $P_i$ is the set of $i$'s projects and $p_i$ is $i$'s project count. Symmetrically, $N_r$ is the set of collaborators on project $r$ and $n_r$ is the size of team $r$. A project can be undertaken alone, $n_r=1$, or in a team, $n_r\geq2$. Individual $i$'s choices form row $\vec a_i=(a_{i,1},\ldots,a_{i,p})$ of $A$; $a_{-i}$ denotes the choices of all other individuals.

Participating in a joint project entails bilateral and team collaborations. For distinct $i$ and $j$, $g_{ij}=|P_i\cap P_j|=\sum_{r\in P}a_{i,r}a_{j,r}$ counts their shared projects, and $[g_{ij}>0]$ indicates that they share at least one. The weighted and unweighted collaboration networks are $G^W=(g_{ij})_{n\times n}$ and $G^U=([g_{ij}>0])_{n\times n}$, with $g_{ii}=0$. Individual $i$'s degree $d_i=\sum_{j\ne i}[g_{ij}>0]$ counts distinct collaborators, and her weighted degree counts collaborations across projects,
\begin{equation}
\label{eq:weighteddegree}
    d_i^{W}=\sum_{j\ne i}g_{ij}
    =\sum_{r\in P_i}(n_r-1)
    =\sum_{r\in P}a_{i,r}n_r-p_i.
\end{equation}
Appendix~\ref{app:setup-details} illustrates the setup with three individuals and four projects.

\subsection{Preferences and payoffs}
\label{sec:intro-payoffs}

Each individual $i$ receives utility from the projects in her portfolio $P_i$ and bears costs that depend on her project activity and collaborations. In a single period, her payoff is
\begin{equation}
\label{eq:payoffs-project}
\begin{aligned}
    U_i(A)={}&\sum_{r\in P_i}\biggl[
        \underbrace{\beta_i}_{\text{productivity}}
        +\underbrace{V(X_i,\{X_j\}_{j\in N_r})}_{\text{team value}}
        +\underbrace{\sum_{j\in N_r\setminus\{i\}}w(X_i,X_j)}_{\text{bilateral interactions}}
    \biggr]\\
    &-\underbrace{C_P(p_i)}_{\text{portfolio cost}}
    -\underbrace{\sum_{j\ne i}C_G(g_{ij})}_{\text{relationship costs}}
    -\underbrace{\psi_D\left[\frac{d_i(d_i+1)}{2}+\sum_{j\ne i}[g_{ij}>0]d_j\right]}_{\text{network congestion}}.
\end{aligned}
\end{equation}
Each project earns $i$ a baseline return $\beta_i$, her productivity, whether she works alone or with others. A team adds the team value $V(X_i,\{X_j\}_{j\in N_r})$, which depends on $i$'s characteristics $X_i$ and on those of all members, so members can value the same team differently and the value can depend on both size and composition; a solo project has no team value, $V(X_i,\{X_i\})=0$. Each teammate $j$ adds the pair value $w(X_i,X_j)$ on every shared project. The portfolio cost $C_P(p_i)$ is the burden of carrying $p_i$ projects. The relationship cost $C_G(g_{ij})$ is paid once per pair at its number of shared projects, so it can separate a first collaboration from repeated ones; in the application it also depends on the pair's collaboration history. The last term, scaled by $\psi_D$, is congestion in the unweighted network $G^U$: it rises with $i$'s number of distinct collaborators and with their degrees, and is unchanged by repeating an existing collaboration. Since each colleague $j$ appears in $g_{ij}$ of $i$'s projects, $U_i$ splits into own project activity $\beta_ip_i-C_P(p_i)$, team benefits, collaboration benefits net of costs $\sum_{j\ne i}[g_{ij}w(X_i,X_j)-C_G(g_{ij})]$, and network congestion.

For the marginal comparisons below, $\Delta C_P(p)=C_P(p)-C_P(p-1)$ is the cost of the $p$-th project, and $\Delta C_G(h)=C_G(h+1)-C_G(h)$ is the cost of one more common project for a pair that already shares $h$; thus $p$ includes the contemplated project, while $h$ counts common projects before its addition.

\begin{assumption}[Payoff terms]
\label{ass:payoffs}
\begin{enumerate}[label=(\roman*)]
\item \emph{Congestion.} $C_P(0)=0$, and the marginal cost $\Delta C_P(p)$ is nondecreasing in $p$ with $\Delta C_P(p)\to\infty$.
\item \emph{Team value.} There is a function $\Phi$ on finite sets of individuals with $\Phi(\varnothing)=\Phi(\{i\})=0$ for every $i$ such that the value of a project to each of its members is her marginal contribution to $\Phi$,
\[
V(X_i,\{X_j\}_{j\in S})=\Phi(S)-\Phi\bigl(S\setminus\{i\}\bigr),\qquad i\in S.
\]
$\Phi$ may depend on the characteristics of the members but not on the label of the project.
\item \emph{Pair value.} $w$ is symmetric, $w(X_i,X_j)=w(X_j,X_i)$.
\item \emph{Collaboration cost.} $C_G(0)=0$, and the cost of an additional common project is bounded below, $c_G=\inf_{h\geq0}\Delta C_G(h)>-\infty$.
\item \emph{Degrees.} $\psi_D\geq0$.
\end{enumerate}
\end{assumption}

Part (ii) is the structure of congestion payoffs \citep{rosenthal1973}, extended to values that depend on the identity of the members; it implies that a solo project has no team value, $V(X_i,\{X_i\})=\Phi(\{i\})=0$. The values of team size and of same-section membership used in Section~\ref{sec:structural} have this form (Appendix~\ref{app:setup-details}). The potential representation uses (ii) and (iii). Theorem~\ref{thm:existence} weakens the growth conditions in (i) and (iv) and places no sign restriction on $\psi_D$; the bounds on the number of projects use (i), (iv) and (v), and those on project size (iv) and (v), each with the solo normalization. In the application, $\beta_i$ is replaced by $b_{it}$ and the team value may depend on the year; the results hold year by year.

When deciding between working alone and joining team $r$, individual $i$ compares the return of a solo project net of the additional project cost, $\Delta_{a_{i,r}}U_i^{\mathrm{alone}}=\beta_i-\Delta C_P(p_i)$, with the gain from joining the team, which exceeds it by
\begin{align} \notag
\Delta_{a_{i,r}}U_i^{\mathrm{new}}
- \Delta_{a_{i,r}}U_i^{\mathrm{alone}}
={}&
V(X_i,\{X_j\}_{j\in N_r})
+\sum_{\substack{j\in N_r\\j\neq i}}
\Bigl[w(X_i,X_j) - \Delta C_G\bigl(g_{ij}\bigr)\Bigr]\\ \label{eq:new-vs-alone}
{}& \quad
-\psi_D
\left(d_i n_r^{(i)} -
\frac{\left(n_r^{(i)}\right)^2 - n_r^{(i)}}{2}
+ \sum_{j\in N_r^{(i)}}d_j
\right)
\end{align}
where the team $N_r$, its size $n_r$ and the degrees $d_i$ and $d_j$ are evaluated after entry, $g_{ij}$ counts the common projects before $i$ joins, and $n_r^{(i)}$ is the number of new collaborators $i$ gains by joining $r$,
\[
N_r^{(i)}(A^{-ir}):=\{j\in N_r(A^{-ir}):g_{ij}(A^{-ir})=0\},
\qquad n_r^{(i)}:=|N_r^{(i)}(A^{-ir})|
\]
with $A^{-ir}$ the profile with membership $(i,r)$ set to zero. These are also the relationships lost when the entry is reversed; if $i$ already belongs to $r$, the same set is $\{j\in N_r(A)\setminus\{i\}:g_{ij}(A)=1\}$. Her productivity and portfolio cost cancel, since both alternatives add one project. The team is preferable when its team value and net pair values outweigh the congestion created by new relationships, even if a particular teammate has a negative pair value. Appendix~\ref{app:setup-details} derives \eqref{eq:new-vs-alone} in steps, from a team of existing collaborators to one with new collaborators.

\subsection{Equilibrium}

\begin{definition}
\label{def:equilibrium}
A \textbf{\emph{$k$-team stable}} equilibrium is a pair $(P^*,A^*)$ of a set of projects $P^*$ and a participation profile $A^*=(\vec a_i^*)_{i\in N}$, $\vec a_i^*\in\{0,1\}^{P^*}$, such that, for every individual $i$ and every nonempty set $R\subseteq P^*\cup\{p^*+1\}$ of at most $k$ projects, $\vec a_i^*$ solves the problem below. Here $p^*+1$ denotes a new project, and $A^*$ is extended by an empty column for it. For each candidate choice, write $A=(\vec a_i,a_{-i}^*)$, holding $i$'s memberships outside $R$ fixed at their values in $A^*$:
\begin{align}
&\max_{\{a_{i,r}\}_{r\in R}}\ U_i(A)\notag\\
\label{stability_constraint}
&\text{s.t.}\quad a_{i,r}=1,\ n_r(A)>1\quad\Longrightarrow\quad
\sigma_{j,r}(A)\geq0\quad\forall j\in N_r(A),\ r\in R.
\end{align}
The participation slack is
\[
\sigma_{j,r}(A):=U_j(A)-\max\bigl\{U_j(A^{-jr}),\,U_j(A^{-jr,+js})\bigr\}.
\]
Here $A^{-jr}$ removes $j$ from $r$, and $A^{-jr,+js}$ additionally assigns her to a fresh solo project $s$, with all other memberships fixed. The project $s$ lies outside $P^*\cup\{p^*+1\}$ and is used only to evaluate the outside option; this comparison does not enlarge $R$.
\end{definition}

The team stability constraint \eqref{stability_constraint} lets $i$ join or keep a considered team only if every member weakly prefers remaining in it both to leaving and to replacing the membership with a solo project. The checks apply to both added and retained team memberships in $R$ and are evaluated at the complete profile resulting from all of $i$'s choices in $R$. Teams outside $R$, and teams that $i$ leaves, impose no additional participation checks. A team is unstable when a member prefers either outside option. For a fixed population, payoffs, project restrictions, and outside options, let $\mathcal E_k$ denote the set of $k$-team stable equilibria. A profile $(P,A)$ is $k$-team optimal in the absence of the constraint if each $\vec a_i$ solves the same problems without it.

\begin{lemma}
\label{lem:optimality_implies_stability}
Let $k\geq2$. Any $(P,A)$ that is unconstrained $k$-team optimal is a $k$-team stable equilibrium.
\end{lemma}

Leaving a team changes one membership, while replacing it with a solo project changes two. For $k\geq2$, unconstrained optimality rules out both improvements, so every member satisfies the constraint. For $k=1$, separate checks for leaving and adding a solo project need not rule out profitable solo replacement. The converse fails: the constraint can block an entry that benefits the entrant but leaves an incumbent preferring to exit, so a stable equilibrium need not be $k$-team optimal in the absence of the constraint, and can lie away from the maximum of the potential below. Example~\ref{ex:stable-not-unconstrained} in Appendix~\ref{app:setup-details} gives a case with three individuals.

\begin{theorem}
\label{thm:existence}
Suppose $N$ is finite, Assumption~\ref{ass:payoffs}(ii)--(iii) holds, $C_P(0)=C_G(0)=0$, and
\[
\lim_{p\to\infty}\frac{C_P(p)}{p}=\infty,
\qquad
C_G(g)\geq-ag-b\quad\text{for all integers }g\geq0,
\]
for some finite constants $a,b\geq0$. Then, for any $\psi_D\in\mathbb R$ and every integer $k\geq1$, there exists a $k$-team stable equilibrium $(P^\ast,A^\ast)$.
\end{theorem}

Existence rests on an exact potential $Q$: every unilateral change in an individual's project choices changes $Q$ by exactly her payoff change. Assumption~\ref{ass:payoffs}(ii) makes each member's team value her marginal contribution to the common project-level function $\Phi$, while (iii) makes pair values symmetric. These properties, together with the common relationship cost and the degree term, yield the potential representation verified in Proposition~\ref{prop:potential}. The growth conditions of Theorem~\ref{thm:existence} make $Q$ fall without bound as participation grows, so it attains a maximum over finitely many profiles up to project relabeling. At a maximizer, no individual gains from any unilateral change, including leaving a team or replacing a membership with a solo project, so the participation constraints hold. Applying Lemma~\ref{lem:optimality_implies_stability} at $\max\{k,2\}$ gives $\max\{k,2\}$-team stability and hence $k$-team stability for every $k\geq1$. Appendix~\ref{app:potential-existence} provides the formal argument. The potential collects own-project returns net of portfolio costs, team values, net bilateral returns, and network congestion:
\begin{equation}
\label{potential}
\begin{aligned}
Q(A)={}&\sum_{i\in N}\bigl[\beta_i p_i-C_P(p_i)\bigr]
+\sum_{r\in P}\Phi(N_r)\\
&+\sum_{i<j}\bigl[g_{ij}w(X_i,X_j)-C_G(g_{ij})\bigr]
-\frac{\psi_D}{2}\sum_{i\in N}\bigl(d_i+d_i^2\bigr).
\end{aligned}
\end{equation}

\subsection{Equilibrium Projects and Collaborations}
\label{sec:equilibrium_projects_collaborations}

Definition~\ref{def:equilibrium} combines an individual's incentive to change projects with the willingness of the affected team members to remain. A profitable change is feasible only if every participation constraint required by \eqref{stability_constraint} holds at the resulting profile. This section expresses those constraints as individual participation slacks, which restrict the number of projects an individual can sustain, the teams she can belong to, and her collaborations across those projects.

\subsubsection{Participation Slacks and Equilibrium}
\label{sec:participation-slacks}

Fix a membership $a_{j,r}=1$. Let $A^{-jr}$ remove $j$ from $r$, and let $B_{j,r}$ replace that membership with a new solo project, holding all other memberships fixed. Define $j$'s participation slack by
\begin{equation}
\label{eq:participation-slack}
\begin{aligned}
\sigma_{j,r}(A)
&:=U_j(A)-\max\bigl\{U_j(A^{-jr}),U_j(B_{j,r})\bigr\}\\
&=\Delta_{a_{j,r}}U_j(A)
-\max\bigl\{0,\beta_j-\Delta C_P(p_j)\bigr\}.
\end{aligned}
\end{equation}
Here $p_j$ includes membership in $r$, and the solo project in $B_{j,r}$ is evaluated after leaving $r$. Since a solo project has no team value and creates no relationships, its incremental payoff at $A^{-jr}$ is $\beta_j-\Delta C_P(p_j)$. Thus \eqref{eq:participation-slack} makes explicit the leave-or-replace comparison in Definition~\ref{def:equilibrium}: $\sigma_{j,r}(A)\geq0$ if and only if $j$ weakly prefers remaining in $r$ to either outside option.

The slack separates the return from undertaking a project from the return specific to its team. Writing $s_{j,r}(A):=U_j(A)-U_j(B_{j,r})$ for $j$'s team surplus gives
\begin{equation}
\label{eq:slack-team-surplus}
\sigma_{j,r}(A)
=s_{j,r}(A)+\min\bigl\{\beta_j-\Delta C_P(p_j),0\bigr\}.
\end{equation}
The team surplus $s_{j,r}$ has the team, collaboration, and degree terms of \eqref{eq:new-vs-alone}, evaluated for $j$'s membership in $r$. When a solo project has a nonnegative return, this team surplus is the entire slack. When the solo return is negative, the team must also compensate $j$ for the cost of undertaking the project at all.

\begin{remark}[Participation slacks and stability constraints]
\label{rem:slack-equilibrium}
For the payoffs in \eqref{eq:payoffs-project}, with $V(X_j,\{X_j\})=0$, member $j$ of a nonsingleton project $r$ satisfies the constraint in \eqref{stability_constraint} at profile $A$ if and only if $\sigma_{j,r}(A)\geq0$. Hence, in every $k$-team stable equilibrium, $k\geq1$, every member of each current nonsingleton project has a nonnegative slack, and a change in an individual's memberships is feasible if and only if the slacks of the constraints it requires are nonnegative at the resulting profile.
\end{remark}
The proof is in Appendix~\ref{app:slacks-entry}.

A constraint need not be binding before a deviation: an incumbent can bear a negative externality as long as her remaining slack is nonnegative. Appendix~\ref{app:slacks-entry} derives how entry into a team changes each incumbent's slack (Lemma~\ref{lem:participation-stability}). The entry test is individual: a favorable effect on one incumbent cannot compensate another for a negative slack, and an incumbent's negative slack can block privately profitable entry (Example~\ref{ex:stable-not-unconstrained}). Entry can also change the slacks of members of other projects; these spillovers are network externalities and add no constraints beyond those of Definition~\ref{def:equilibrium}. Nor do these results reduce the equilibrium test to single-membership changes when $k>1$: the constraints of a bundled change are checked at the profile that results from the whole change.

The bounds below are necessary conditions for the equilibria of Definition~\ref{def:equilibrium}. They use the nonnegativity of each current member's slack, together with solo-project deviations where needed, and impose no additional condition on the aggregate externality from entry.

Throughout, let $n=|N|<\infty$, with $n\geq1$, and normalize $V(X_i,\{X_i\})=0$. Let $c_G$ be the lower bound on marginal collaboration costs in Assumption~\ref{ass:payoffs}(iv). For each individual $i$ and team size $m\in\{1,\ldots,n\}$, define the upper envelope of team surplus
\begin{equation}
\label{eq:team-surplus-envelope}
\overline s_i(m):=
\max_{\substack{S\subseteq N,\ i\in S\\|S|=m}}
\left\{V(X_i,\{X_j\}_{j\in S})
+\sum_{j\in S\setminus\{i\}}\bigl[w(X_i,X_j)-c_G\bigr]\right\}.
\end{equation}
This evaluates team value and pair values on the same set of members. When $\psi_D\geq0$, it bounds the surplus $s_{i,r}$ of a membership in any team of size $m$: actual marginal collaboration costs are at least $c_G$, and the degree term is a cost. In particular, $\overline s_i(1)=0$, and
\[
W_i:=\max_{1\leq m\leq n}\overline s_i(m)\geq0
\]
is finite. These envelopes need not be attainable in equilibrium, because every member of a proposed team must satisfy her own participation condition.

\subsubsection{Number of Projects}

\begin{proposition}[Bounds on the equilibrium number of projects]
\label{prop:number_projects}
Let $k\geq1$, suppose Assumption~\ref{ass:payoffs}(i), (iv), and (v) holds, and retain the finite population and solo-team normalization above. In any $k$-team stable equilibrium,
\[
\underline p_i\leq p_i^*\leq\overline p_i,
\]
where
\begin{align*}
\underline p_i&:=\min\bigl\{p\in\mathbb N_0:\ \Delta C_P(p+1)\geq\beta_i\bigr\},\\
\overline p_i&:=\max\bigl\{p\in\mathbb N_0:\ p=0\ \text{or}\ \Delta C_P(p)\leq\beta_i+W_i\bigr\}<\infty.
\end{align*}
Consequently, the number of active projects satisfies
\begin{equation}
\label{eq:total-project-count-bound}
|P^*|\leq\sum_{i\in N}p_i^*\leq\sum_{i\in N}\overline p_i<\infty.
\end{equation}
\end{proposition}

The proofs of Propositions~\ref{prop:number_projects}--\ref{prop:degree_bounds} and Lemma~\ref{lem:team-stability-nesting} are in Appendix~\ref{app:bounds-nesting}. The lower bound concerns total project participation, rather than the number of solo projects. An individual may satisfy it entirely through team memberships. The upper bound follows because the greatest possible benefit from a team is finite for fixed $N$, while marginal project costs become arbitrarily large. Positive values of working with many potential collaborators therefore do not generate an unbounded number of projects. A team that gives $i$ a large surplus may still be infeasible for another member; the envelope is an individual bound, not a prediction that the maximizing team forms.

\subsubsection{Team Size and Individual Participation}

For $p\geq1$, write
\[
\lambda_i(p):=\bigl[\Delta C_P(p)-\beta_i\bigr]_+
\]
for the shortfall in the return to a solo project. Equation~\eqref{eq:slack-team-surplus} becomes $\sigma_{i,r}=s_{i,r}-\lambda_i(p_i)$: a team must cover this shortfall as well as its collaboration and degree costs. At a profile containing $i$ in $r$, let
\[
N_r^{(i)}(A^{-ir})=\{j\in N_r(A)\setminus\{i\}:g_{ij}(A)=1\},
\qquad n_r^{(i)}=|N_r^{(i)}(A^{-ir})|,
\]
be the collaborators whose relationship with $i$ would disappear if she left $r$, consistently with the new-collaborator set of Section~\ref{sec:intro-payoffs}. With counts and degrees evaluated at that profile, define
\begin{equation}
\label{eq:membership-degree-increment}
D^{\mathrm{degree}}_{i,r}:=
 d_i n_r^{(i)}-\frac{n_r^{(i)}(n_r^{(i)}-1)}{2}
 +\sum_{j\in N_r^{(i)}}d_j\geq0.
\end{equation}
This is the increment in $i$'s degree expression when she adds $r$ back after exit; $d_i\geq n_r^{(i)}$ makes it nonnegative.

\begin{proposition}[Individual participation and bounds on project size]
\label{prop:project_size}
Let $k\geq1$, suppose Assumption~\ref{ass:payoffs}(iv)--(v) holds, and retain the finite population and solo-team normalization above. Every nonsingleton equilibrium project satisfies, for each $i\in N_r^*$,
\begin{equation}
\label{eq:average-benefit-participation}
\frac{\displaystyle\sum_{j\in N_r^*\setminus\{i\}}
\bigl[w(X_i,X_j)-\Delta C_G(g_{ij}^*-1)\bigr]}{n_r^*-1}
\geq
\frac{\lambda_i(p_i^*)+\psi_DD^{\mathrm{degree}}_{i,r}(A^*)
-V(X_i,\{X_j\}_{j\in N_r^*})}{n_r^*-1}.
\end{equation}
This inequality is equivalent to $\sigma_{i,r}(A^*)\geq0$.

Define, for $p\geq1$,
\begin{align*}
\overline n_i(p)&:=\max\Bigl(\{1\}\cup
\{m\in\{2,\ldots,n\}:\overline s_i(m)\geq\lambda_i(p)\}\Bigr),\\
\overline n_i&:=\max\{m\in\{1,\ldots,n\}:\overline s_i(m)\geq0\}.
\end{align*}
Then every nonempty equilibrium project obeys
\begin{equation}
\label{eq:individual-project-size-bound}
1\leq n_r^*\leq\min_{i\in N_r^*}\overline n_i(p_i^*)
\leq\min_{i\in N_r^*}\overline n_i\leq n.
\end{equation}
\end{proposition}

The first inequality compares each member's average net pair value with the team benefit and the costs her membership must cover. A negative solo return raises the required team surplus through $\lambda_i(p_i^*)$; relationships unique to this project add degree costs through $D^{\mathrm{degree}}_{i,r}$. The size tests are necessary upper bounds, not sufficient conditions for a team or an equilibrium to exist. A size may pass $i$'s envelope test because of one composition and pass another individual's test because of a different composition. Each actual team must satisfy \eqref{eq:average-benefit-participation} for all its members, and equilibrium also requires the absence of profitable feasible deviations.

\subsubsection{Collaborations and Degrees}

\begin{proposition}[Bounds on equilibrium degrees]
\label{prop:degree_bounds}
Under the conditions of Proposition~\ref{prop:number_projects}, let $\overline p_i$ be its project-count bound and $\overline n_i$ the individual size bound of Proposition~\ref{prop:project_size}. In every $k$-team stable equilibrium,
\begin{equation}
\label{eq:weighted-degree-bound}
0\leq d_i^{W,*}
\leq p_i^*(\overline n_i-1)
\leq\overline p_i(\overline n_i-1),
\end{equation}
and
\begin{equation}
\label{eq:unweighted-degree-bound}
0\leq d_i^*\leq\min\{n-1,d_i^{W,*}\}
\leq\min\{n-1,\overline p_i(\overline n_i-1)\}.
\end{equation}
If $p_i^*>0$, the sharper size bound $\overline n_i(p_i^*)$ can replace $\overline n_i$ in the first upper bound of \eqref{eq:weighted-degree-bound}, and
\[
\left\lceil\frac{d_i^{W,*}}{p_i^*}\right\rceil\leq d_i^*.
\]
In particular, $\overline n_i=1$ implies $d_i^{W,*}=d_i^*=0$.
\end{proposition}

The degree bounds translate the participation restrictions into properties of the weighted and unweighted collaboration networks. They distinguish repeated work with the same collaborators, which raises $d_i^W$ without necessarily raising $d_i$, from expanding the set of collaborators. Their validity uses the individual size bounds and project counts, rather than an aggregate consent condition.

\subsection{Discussion}
\label{sec:team-stability-discussion}

The consideration bound $k$ determines how many project memberships an individual can reconsider together. It counts projects, rather than team members: even with two individuals, a profitable adjustment may require changing three memberships. Recall that $\mathcal E_k$ denotes the set of $k$-team stable equilibria of Definition~\ref{def:equilibrium}. We first show that $\mathcal E_2$ and $\mathcal E_3$ can differ, then establish the conditions under which these sets are ordered by inclusion, and finally relate team stability to network stability.

\subsubsection{Why the Consideration Bound Matters}
\label{sec:consideration-bound}

The bound can change the set of equilibria. In the following example, two individuals pay a fixed cost to establish their relationship, so shared projects are complements: one shared project does not pay, two do. A profile of solo projects is then $2$-team stable, but not $3$-team stable, because the profitable adjustment, in which one individual leaves her solo project and joins both of the other's projects, changes three memberships at once.

\begin{example}[Two-team stability without three-team stability]
\label{ex:two-versus-three-team-stability}
Consider two individuals with
\[
\beta_1=2,\quad \beta_2=4,\quad
C_P(p)=p^2,\quad w(X_1,X_2)=w(X_2,X_1)=3,
\quad C_G(g)=4[g>0],
\]
and set team value and degree congestion to zero. Repeated collaborations are allowed. Each shared project gives a collaboration benefit of $3$, while establishing the relationship costs $4$ once, independently of the number of shared projects. Thus
\begin{equation}
\label{eq:consideration-example-payoff}
U_i(A)=\beta_i p_i-p_i^2+3g_{12}-4[g_{12}>0],
\qquad i=1,2.
\end{equation}
Let $A^0$ consist of a solo project $r$ for individual 1 and two solo projects $s,t$ for individual 2. The payoffs are $(U_1(A^0),U_2(A^0))=(1,4)$. Then $A^0\in\mathcal E_2\setminus\mathcal E_3$.
\end{example}

\begin{figure}[!htbp]
\centering
\caption{Two-Team Stability and a Three-Membership Deviation}
\label{fig:two-versus-three-team-stability}
\begin{minipage}[t]{0.48\textwidth}
\centering
\textbf{(a) Stable for $k=2$: $A^0$}\\[6pt]
\begin{tikzpicture}[every node/.style={font=\small}]
\path[use as bounding box] (-0.9,-2.5) rectangle (4.7,2.5);
\node[individual node] (a1) at (0,1.3) {$i=1$\\$U_1=1$};
\node[individual node] (a2) at (0,-1.3) {$i=2$\\$U_2=4$};
\node[project node] (ar) at (3.2,1.7) {$r,\ n_r=1$\\$N_r=\{1\}$};
\node[project node] (as) at (3.2,0) {$s,\ n_s=1$\\$N_s=\{2\}$};
\node[project node] (at) at (3.2,-1.7) {$t,\ n_t=1$\\$N_t=\{2\}$};
\draw[diagram edge] (a1.east) -- (ar.west);
\draw[diagram edge] (a2.east) -- (as.west);
\draw[diagram edge] (a2.east) -- (at.west);
\end{tikzpicture}\\[-2pt]
$p_1=1,\quad p_2=2,\quad g_{12}=0$
\end{minipage}\hfill
\begin{minipage}[t]{0.48\textwidth}
\centering
\textbf{(b) Stable for every $k$: $A^1$}\\[6pt]
\begin{tikzpicture}[every node/.style={font=\small}]
\path[use as bounding box] (-0.9,-2.5) rectangle (4.7,2.5);
\node[individual node] (b1) at (0,1.3) {$i=1$\\$U_1=2$};
\node[individual node] (b2) at (0,-1.3) {$i=2$\\$U_2=6$};
\node[project node] (bs) at (3.2,0.9) {$s,\ n_s=2$\\$N_s=\{1,2\}$};
\node[project node] (bt) at (3.2,-0.9) {$t,\ n_t=2$\\$N_t=\{1,2\}$};
\node[text=black!65,font=\footnotesize] at (3.2,2.1) {$r$ removed};
\draw[diagram edge] (b1.east) -- (bs.west);
\draw[diagram edge] (b1.east) -- (bt.west);
\draw[diagram edge] (b2.east) -- (bs.west);
\draw[diagram edge] (b2.east) -- (bt.west);
\end{tikzpicture}\\[-2pt]
$p_1=p_2=2,\quad g_{12}=2$
\end{minipage}
\par\medskip
{\RaggedRight\floatnotesfont Notes. Blue circles represent individuals and amber boxes represent active projects. Solid edges are memberships. From panel (a) to panel (b), individual 1 changes $a_{1,r}:1\to0$, $a_{1,s}:0\to1$, and $a_{1,t}:0\to1$; individual 2's memberships stay fixed. Project $r$ is deleted when empty. Payoffs use the parameters of Example~\ref{ex:two-versus-three-team-stability}.\par}
\end{figure}
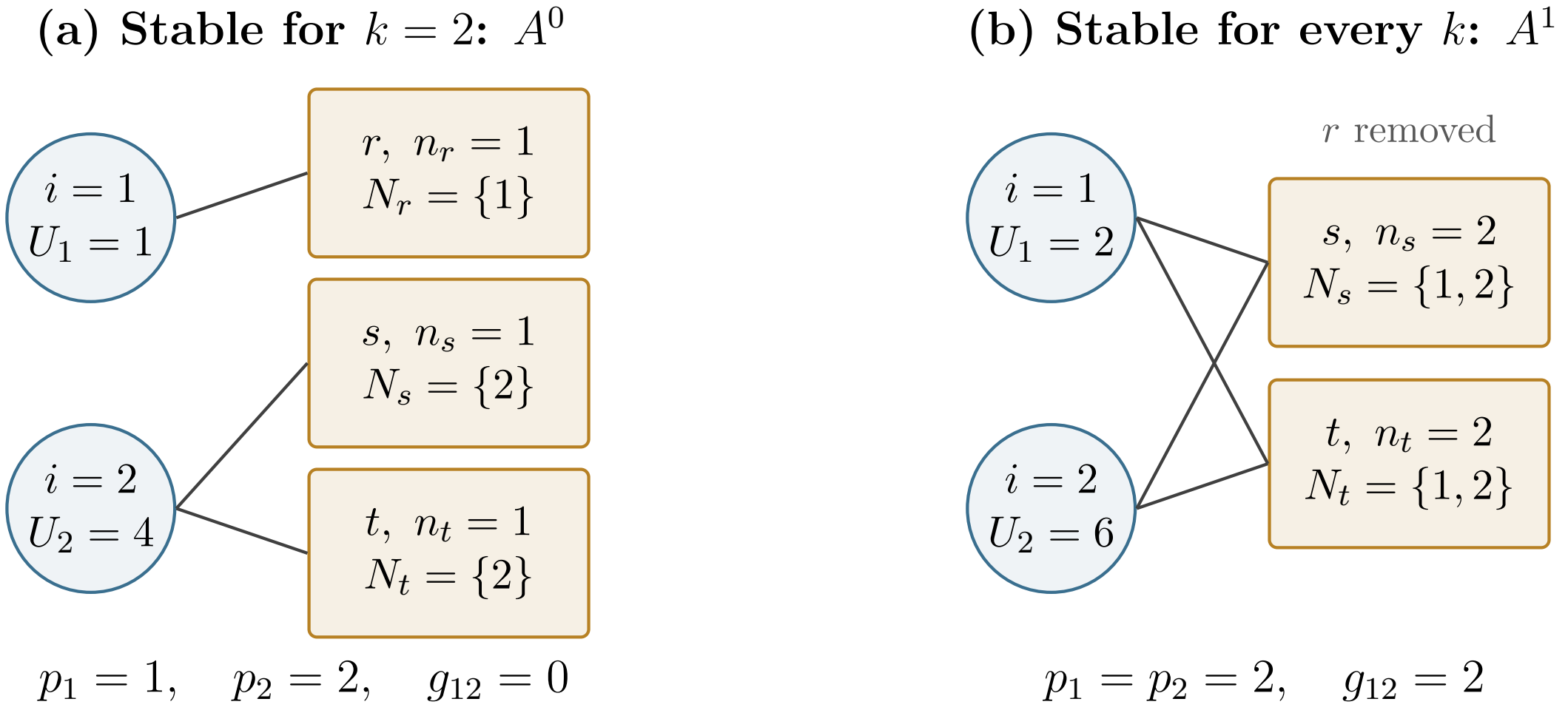

Figure~\ref{fig:two-versus-three-team-stability} illustrates the profitable adjustment. The fixed relationship cost makes shared projects complementary: one collaboration yields a net relationship payoff of $-1$, while two yield $2$. Retaining $r$ while adding both collaborations raises project congestion enough to eliminate the gain, so individual 1 must leave $r$ and join both $s$ and $t$. This package changes three memberships and raises her payoff from $1$ to $2$, while both individuals remain willing to participate. It is unavailable when $k=2$ but feasible when $k=3$, giving $A^0\in\mathcal E_2\setminus\mathcal E_3$, where $\mathcal E_k$ denotes the set of $k$-team stable equilibria. The resulting profile $A^1$ is stable for every $k$. Repeated collaboration supplies the complementarity here; Appendix~\ref{sec:network-correspondence} excludes repeated collaborations in its conditional network correspondence. Appendix~\ref{app:two-versus-three-team-stability} verifies the payoff comparisons and participation checks.

\subsubsection{Ordering the Equilibrium Family}

The inclusion argument concerns the feasible decision problems an individual faces. For a profile $A$ and a consideration set $R$, the candidate profiles that change only $i$'s memberships in $R$, and the individual slack constraints that Definition~\ref{def:equilibrium} requires for that problem, do not depend on $k$. Increasing $k$ therefore retains every smaller consideration problem, with its original participation checks.

\begin{lemma}[Nesting of team stable equilibria]
\label{lem:team-stability-nesting}
Hold the population, payoffs, project restrictions, and outside options fixed across consideration bounds $k$. Suppose that every consideration problem allowed at $k'$ is also allowed at $k>k'$, with the same candidate profiles and feasibility constraints. Then
\begin{equation}
\label{eq:team-stability-nesting}
\mathcal E_k\subseteq\mathcal E_{k'}
\qquad\text{for all integers }1\leq k'<k.
\end{equation}
In particular, these conditions hold under Definition~\ref{def:equilibrium}, and
\[
\mathcal E_1\supseteq\mathcal E_2\supseteq\mathcal E_3\supseteq\cdots.
\]
No additional restrictions on the payoff functions are needed for this inclusion.
\end{lemma}

Example~\ref{ex:two-versus-three-team-stability} makes the inclusion strict: $\mathcal E_3\subsetneq\mathcal E_2$. Increasing the consideration bound can remove equilibria because it admits additional bundles of entries and exits. The participation rule itself remains unchanged. Existence is supplied separately by Theorem~\ref{thm:existence}: under its assumptions, a global potential maximizer belongs to every $\mathcal E_k$. The equilibrium family can therefore be nested and nonempty while still changing with $k$.

\subsubsection{Team Stability and Network Stability}
\label{sec:discussion-network-stability}

Team stability is a property of the bipartite participation graph $A$, which links individuals to projects. Intuitively, the collaboration network $G^W$, in which $g_{ij}=\sum_{r\in P}a_{i,r}a_{j,r}$ counts the projects that $i$ and $j$ share, inherits a stability property. Appendix~\ref{sec:network-correspondence} formalizes this intuition in a characterization. When every project has at most two members and each pair shares at most one project, so that $G^W=G^U$, a profile in which everyone has a solo project is $k$-team stable, for any $k\geq2$, if and only if individual project counts are optimal and the induced collaboration network is $n$-stable in the sense of \citet{Badev2021}: each individual can reconsider all of her links, subject to the consent of her partners. Details are in Proposition~\ref{prop:team_implies_network_stability}. Because two project changes already let an individual replace a collaborator, $2$-team stability rules out profitable replacements, which pairwise stability does not test. Larger teams and repeated collaborations require the richer project representation.

\section{$k$-Team Consensual Dynamics ($k$CD)}
\label{sec:team-dynamics}

This section develops adaptive processes for project participation. At each date, one individual reconsiders her membership in at most $k$ projects, including the possibility of starting one new project. The bound applies to the whole consideration set, including memberships retained during a revision. I first study payoff-maximizing revisions subject to the participation conditions of Definition~\ref{def:equilibrium}. Under the stated assumptions, this process converges almost surely to a $k$-team stable equilibrium despite temporary instability outside the consideration set and occasional decreases in the potential.

I then introduce logit proposals and stochastic screening of all team memberships in the candidate state. The resulting stationary distribution combines the payoff potential with an estimable penalty for failed participation checks and provides the likelihood used in estimation. Its invariance to $k$ and state-independent meeting probabilities allows the sampler's meetings to be chosen for computational efficiency.

\subsection{Deterministic Updates and Convergence}
\label{sec:deterministic-team-dynamics}

Consider a finite, nonempty population $N$, with $n=|N|$, a consideration bound $k\geq1$, and an initial participation profile $A_0$ with finitely many active projects. At date $t$, the active projects are $P_t$. To allow a new project to start, append one empty column $r_{\mathrm{new}}$ to the current profile $A_t$ and write $\widehat P_t=P_t\cup\{r_{\mathrm{new}}\}$.

A random meeting $(i,R)$ gives individual $i$ the opportunity to reconsider her participation in a nonempty set $R\subseteq\widehat P_t$ of at most $k$ projects. She can join, leave, or retain projects in $R$, while her memberships elsewhere and every other individual's choices remain fixed. Meetings are drawn afresh at each date, with positive probability for every allowed $(i,R)$, including singleton consideration sets. For example, let $m_t=\min\{k,|P_t|+1\}$, choose the individual and the consideration-set size uniformly, and then choose uniformly among sets of that size. The resulting meeting probabilities are
\begin{equation}
\label{eq:deterministic-team-meetings}
\rho_k(i,R\mid A_t)
=\frac{1}{n\,m_t\binom{|P_t|+1}{|R|}},
\qquad 1\leq |R|\leq m_t.
\end{equation}

The individual's choice is subject to the participation conditions of Definition~\ref{def:equilibrium}. Every considered team in which she participates after the revision must leave all its members willing to remain. These conditions apply to retained memberships as well as new entries and are evaluated at the complete candidate profile, including the resulting project loads and collaboration network. Let $\mathcal F_i(A_t;R)$ collect the profiles $B$ that differ from $A_t$ only in $i$'s memberships in $R$ and satisfy the stability constraints, i.e.:
\begin{equation}
\label{eq:deterministic-team-feasibility}
b_{i,r}=1,\quad n_r(B)>1
\quad\Longrightarrow\quad
\sigma_{j,r}(B)\geq0
\quad\text{for every }j\in N_r(B),\qquad r\in R.
\end{equation}
Teams outside $R$, and teams in $R$ that $i$ leaves, impose no additional participation check on her revision. Leaving all considered projects is always feasible, so $\mathcal F_i(A_t;R)$ is nonempty.

Given these feasible alternatives, individual $i$ chooses the participation pattern that gives her the highest payoff, keeping the current profile whenever it is feasible and attains the maximum. Denoting her choice by $B_t$, this gives
\begin{equation}
\label{eq:deterministic-team-update}
B_t\in\operatorname*{arg\,max}_{B\in\mathcal F_i(A_t;R)}U_i(B).
\end{equation}
A fixed deterministic rule resolves any remaining ties. Projects left without members are then deleted, and the remaining columns are canonically relabeled to obtain $A_{t+1}$. Project labels have no payoff consequences, so profiles that differ only by relabeling are treated as the same state. Conditional on the meeting, the update is deterministic; randomness enters through the sequence of meetings.

The current profile need not be a feasible alternative. A previous revision may have changed project loads or collaborations enough to make a membership outside its consideration set unstable. When that team is subsequently considered, the mover must satisfy its participation conditions, even if doing so lowers her payoff. This possibility explains why the potential can fall along a realized path, as discussed below.

\subsubsection{Bounds Along the Path}

The load-independent upper bounds in Section~\ref{sec:equilibrium_projects_collaborations} are invariant when the initial profile satisfies them. More generally, arbitrary finite initial conditions yield the following bounds, without an externally imposed limit on the number of project slots.

\begin{proposition}[Bounds on deterministic paths]
\label{prop:deterministic-path-bounds}
Suppose Assumption~\ref{ass:payoffs}(i), (iv), and (v) holds, with finite $N$ and $V(X_i,\{X_i\})=0$. Using $\overline p_i$ and $\overline n_i$ from Propositions~\ref{prop:number_projects} and~\ref{prop:project_size}, define
\begin{align*}
M_i&:=\max\{p_i(A_0),\overline p_i\},\\
L_i&:=\max\Bigl(\{\overline n_i\}\cup
\{n_r(A_0):r\in P_i(A_0)\}\Bigr),
\qquad M:=\sum_{i\in N}M_i.
\end{align*}
Every path of \eqref{eq:deterministic-team-update} satisfies, at every date,
\begin{equation}
\label{eq:deterministic-path-bounds}
p_i(A_t)\leq M_i,\qquad
n_r(A_t)\leq\min_{j\in N_r(A_t)}L_j,\qquad
|P_t|\leq M,
\end{equation}
and
\begin{equation}
\label{eq:deterministic-degree-bounds}
d_i^W(A_t)\leq M_i(L_i-1),\qquad
d_i(A_t)\leq\min\{n-1,d_i^W(A_t)\}.
\end{equation}
Consequently, only finitely many states are reachable up to project relabeling. If $p_i(A_0)\leq\overline p_i$ and $n_r(A_0)\leq\overline n_i$ for every initial membership, then $M_i=\overline p_i$ and $L_i=\overline n_i$.
\end{proposition}

The bounds also hold at forced payoff-reducing updates: a new team membership must pass its participation test, a solo membership beyond $\overline p_i$ is dominated by omitting it, and exit reduces team size. Thus a per-person cap of at least $M_i$ leaves every deterministic path unchanged. Appendix~\ref{app:deterministic-team-dynamics} gives the proof.

The other restrictions in Section~\ref{sec:equilibrium_projects_collaborations} have different roles along a path. The lower bound $\underline p_i$ is an equilibrium restriction: a transient individual can have too few projects and later add a profitable solo project. The sharper bounds $\overline n_i(p_i(A_t))$ follow for memberships whose current individual slacks are nonnegative; local updates do not guarantee that all outside memberships retain this property. At a limiting equilibrium, all the bounds in that section apply.

\subsubsection{Rest Points and Convergence}

Under Assumption~\ref{ass:payoffs}(ii)--(iii), the exact potential of Proposition~\ref{prop:potential} satisfies $Q(B_t)-Q(A_t)=U_i(B_t)-U_i(A_t)$. If the current profile is feasible in the meeting, inertia makes every change a strict improvement. When it is infeasible, the mover's payoff and the potential can fall. Convergence therefore requires an argument that allows such decreases.

\begin{theorem}[Convergence of the deterministic $k$-team consensual dynamic]
\label{thm:deterministic-team-convergence}
Suppose Assumption~\ref{ass:payoffs} holds, $N$ is finite and nonempty, and $V(X_i,\{X_i\})=0$. From every finite initial profile and for every $k\geq1$, the dynamic \eqref{eq:deterministic-team-meetings}--\eqref{eq:deterministic-team-update} reaches a $k$-team stable equilibrium in finite time almost surely and remains there. Its absorbing states are exactly the equilibria of Definition~\ref{def:equilibrium}.
\end{theorem}

Despite this non-monotonicity, instability is temporary and the dynamic converges almost surely to a $k$-team stable equilibrium. The key is that a dissatisfied team member can leave, or leave and then start a solo project, ending with a higher payoff than in the unstable configuration. Although exit alone may lower her payoff, the complete adjustment raises it and hence the potential. Each step involves just one project and has positive probability under every consideration bound $k\geq1$.

Formally, Proposition~\ref{prop:deterministic-path-bounds} makes the reachable state space finite. Consider a closed recurrent class and a state with maximal potential within it. If that state contains an unstable membership, the adjustment just described leads with positive probability to a state with higher potential. Closedness requires the endpoint to remain in the class, contradicting maximality. If all memberships are feasible but the state is not an equilibrium, a profitable feasible revision gives the same contradiction. Every closed recurrent class must therefore contain an equilibrium. Since equilibria are absorbing, each such class consists of a single equilibrium, which the finite Markov chain reaches in finite time almost surely. Appendix~\ref{app:deterministic-team-dynamics} gives the full proof and an example of a potential decrease.

The limiting equilibrium can depend on the initial profile and meeting sequence and need not maximize $Q$ globally (Example~\ref{ex:stable-not-unconstrained}). Full support over consideration sets of size at most $k$ ensures that profitable feasible revisions involving several projects can occur alongside individual decisions to leave a team or start a project alone. The theorem holds for each $k$, with the corresponding equilibrium family $\mathcal E_k$ nested as $k$ increases (Lemma~\ref{lem:team-stability-nesting}).

\subsection{Perturbed (Logit) $k$-Team Dynamics}
\label{sec:perturbed-dynamics}

In the spirit of random utility models \citep{mcfadden1974conditional,train2009discrete}, I add random shocks to the payoffs of alternative participation choices, replacing deterministic maximization with probabilistic choice. Revisions yielding larger utility gains relative to the current profile are more likely to be chosen, although payoff-reducing changes remain possible. I first derive the stationary distribution relaxing the stability constraints, then examine its project-count tails and convergence rate. Section~\ref{sec:logit-probabilistic} reintroduces the stability constraints in the adaptive dynamic in a way that is compatible with the probabilistic choice framework.

\subsubsection{Unconstrained Logit $k$-Team Dynamics}
\label{sec:unconstrained-logit}

For a reference distribution without participation constraints, fix a finite universe of labeled project slots $\mathcal P=\{1,\ldots,\overline P\}$, with $\overline P\geq1$, and retain empty slots. The state space is $\mathcal A=\{0,1\}^{N\times\overline P}$. At each date draw a meeting $(i,R)$, where $\varnothing\ne R\subseteq\mathcal P$ and $|R|\leq\min\{k,\overline P\}$, from a fixed full-support distribution independent of the current state. Individual $i$ can choose any participation vector on $R$, holding all other entries fixed. Write
\[
\mathcal B_i(A;R):=
\{B\in\mathcal A:b_{j,r}=a_{j,r}\text{ whenever }j\ne i\text{ or }r\notin R\}.
\]
With logit scale $\tau>0$, the update rule is
\[
\Pr(A_{t+1}=B\mid A_t=A,i,R)
=\frac{\exp\{U_i(B)/\tau\}}
{\displaystyle\sum_{D\in\mathcal B_i(A;R)}\exp\{U_i(D)/\tau\}},
\qquad B\in\mathcal B_i(A;R).
\]
Equivalently, $i$ maximizes payoff plus independent type-I extreme-value shocks of scale $\tau$, one per candidate participation vector. No individual slack or consent test is imposed, so a selected revision can lower payoffs and leave teams unstable.

The exact potential $Q$ established in Proposition~\ref{prop:potential} and given by \eqref{potential} yields the following result for the unconstrained logit dynamic.

\begin{proposition}[Stationary distribution of the unconstrained logit dynamic]
\label{prop:unconstrained-logit-stationary}
Suppose $N$ is finite and nonempty, $\overline P\geq1$, and Assumption~\ref{ass:payoffs}(ii)--(iii) holds. For every $\tau>0$ and integer $k\geq1$, the unconstrained logit dynamic on $\mathcal A$, with the state-independent full-support meetings defined above, is irreducible, aperiodic, and reversible. Its unique stationary distribution is
\begin{equation}
\label{eq:unconstrained-logit-stationary}
\pi_\tau^0(A)=\frac{\exp\{Q(A)/\tau\}}
{\displaystyle\sum_{D\in\mathcal A}\exp\{Q(D)/\tau\}},\qquad A\in\mathcal A.
\end{equation}
For fixed payoffs, $\tau$, and $\overline P$, this distribution is independent of $k$ and the state-independent meeting probabilities.
\end{proposition}

A more general version is proved in Section~\ref{sec:logit-probabilistic}, where Proposition~\ref{prop:stationary-probabilistic-constraints} allows candidate profiles to be screened by a state-dependent acceptance probability $\chi(A)$. The present proposition is the special case $\chi\equiv1$; Appendix~\ref{app:count-weighted-consent} gives the full proof.

At positive $\tau$, every state has positive stationary probability. The project-count, team-size, and degree bounds in Section~\ref{sec:equilibrium_projects_collaborations} therefore need not hold along these paths or throughout stationary support. The finite slot limit is a separate modeling restriction, not a consequence of those equilibrium bounds. As $\tau\downarrow0$, mass concentrates on global maximizers of $Q$ in $\mathcal A$. Any such maximizer with an empty slot available for a new project satisfies Definition~\ref{def:equilibrium} for every $k$ and, under the assumptions of Section~\ref{sec:equilibrium_projects_collaborations}, its bounds. We use \eqref{eq:unconstrained-logit-stationary} as the unconstrained reference distribution.

\begin{theorem}[Project-count tails and the cost of trimming]
\label{thm:logit-project-count-tail}
Suppose the assumptions of Theorem~\ref{thm:existence} hold with $C_P(p)=\frac{\psi_P}{2}(p-p_0)_+^2$, $\psi_P>0$, $p_0\geq0$, on the labeled-slot universe $\mathcal A$ with $\overline P$ slots. Let $\chi:\mathcal A\to(0,1]$ satisfy $\chi(\varnothing)=1$ at the empty network, and let
\[
\pi_{\tau,\chi}(A)\propto\exp\{Q(A)/\tau\}\,\chi(A),\qquad A\in\mathcal A.
\]
This covers the unconstrained distribution \eqref{eq:unconstrained-logit-stationary} ($\chi\equiv1$) and the constrained distributions of Section~\ref{sec:logit-probabilistic}, for any finite $\kappa\geq0$ with count or depth weights (see the appendix). Write $p(A)=|P(A)|$ for the number of active projects, and let $D=2^n-1$. There are constants $K,\gamma\geq0$, independent of $\overline P$ and $\chi$, such that $Q(A)\leq K+\gamma\sum_ip_i-\sum_iC_P(p_i)$. With
\[
\alpha:=\frac{\psi_P}{2n\tau},\qquad
\ell_{\overline P}:=\log(\overline P D)+\frac{n\gamma+\psi_Pp_0}{\tau},
\]
the following hold.
\begin{enumerate}[label=(\alph*)]
\item \emph{Gaussian tails.} For every $m\in\mathbb N_0$,
\begin{equation}
\label{eq:logit-project-count-gaussian-tail}
\begin{aligned}
\Pr_{\pi_{\tau,\chi}}\{p(A)\geq m\}
&\leq\varepsilon_m:=\min\{1,C_{\overline P}e^{-\alpha m^2/2}\},\\
C_{\overline P}
&:=\left(1+\sqrt{\frac{\pi}{2\alpha}}\right)
\exp\left\{\frac K\tau+\frac{\ell_{\overline P}^2}{2\alpha}\right\}.
\end{aligned}
\end{equation}
\item \emph{Effect of trimming.} Let $\mathcal A_T\subseteq\mathcal A$ be a trimmed state space that contains every state with $p(A)<m$, for example $\{A:p(A)<m\}$ or the per-person cap $\{A:\max_ip_i\leq\bar p\}$ with $m=\bar p+1$, since $p_i\leq p(A)$. Let $\pi^T_{\tau,\chi}$ be $\pi_{\tau,\chi}$ conditioned on $\mathcal A_T$. Then
\[
\bigl\|\pi_{\tau,\chi}-\pi^T_{\tau,\chi}\bigr\|_{\mathrm{TV}}
=\pi_{\tau,\chi}(\mathcal A\setminus\mathcal A_T)\leq\varepsilon_m,
\]
so the expectation of any statistic bounded by one in absolute value changes by at most $2\varepsilon_m$, and for every state $A\in\mathcal A_T$, in particular every observed network,
\[
0\leq\log\pi^T_{\tau,\chi}(A)-\log\pi_{\tau,\chi}(A)=-\log\pi_{\tau,\chi}(\mathcal A_T)\leq-\log(1-\varepsilon_m).
\]
\end{enumerate}
\end{theorem}

The intuition comes from the exponential form of the stationary distribution, familiar from exponential-family models. In the potential \eqref{potential}, the productivity returns $\sum_i\beta_i p_i$, team values $\sum_r\Phi(N_r)$, and bilateral returns $\sum_{i<j}g_{ij}w(X_i,X_j)$ are bounded above by a linear function of total participation for a fixed population. The lower bound on $C_G$ likewise limits the contribution of relationship costs to at most linear growth, while the degree term remains bounded because $d_i\leq n-1$. The negative quadratic term $-\sum_i C_P(p_i)$ therefore eventually dominates, making the upper bound on $Q$ fall quadratically with project count. The number of configurations contributes at most an exponential factor within a fixed slot universe; the negative quadratic term in the exponent dominates this growth, yielding Gaussian-type tails.

Part (b) translates excluded probability mass into a likelihood-error bound within a fixed slot universe. At fixed parameters, let $\delta=\pi_{\tau,\chi}(\mathcal A\setminus\mathcal A_T)$. Conditioning on $\mathcal A_T$ raises an observed network's log likelihood by exactly $-\log(1-\delta)$, which is approximately $\delta$ when the excluded mass is small. Assessing the effect on the posterior additionally requires controlling this error across the relevant parameter values.

The same upper bound applies under stochastic consent. The restriction $\chi(A)\leq1$ bounds unnormalized tail weights by their unconstrained counterparts, while $\chi(\varnothing)=1$ preserves the unit lower bound on the normalizing constant. These properties preserve the envelope used in the proof; normalized tail probabilities can increase or decrease as other states are downweighted. The argument also applies within any restricted state space that contains the empty profile.

The bound displays the dependence on the slot universe: $\ell_{\overline P}$ grows as $\log\overline P$, and $\log C_{\overline P}=O((\log\overline P)^2)$. Appendix~\ref{app:logit-project-count-tail} gives the proof and sharper finite-sum and geometric bounds; for general superlinear $C_P$, it also gives $\Pr\{p\geq m\}\leq C_s(\overline P)e^{-sm}$ for every $s>0$, with explicit $\overline P$ dependence. These are bounds within the same labeled-slot model, with no change of reference measure: reducing the number of labeled slots instead changes multiplicities and defines a different reference distribution, and no tail bound uniform in $\overline P$ is asserted.

The analytical bound can be uninformative at empirically relevant congestion costs. The appendix reports boundary diagnostics for the capped simulations; these describe behavior near the cap rather than directly measuring probability outside it.

\subsubsection{Speed of Convergence of Unconstrained Logit Dynamics}
\label{sec:unconstrained-logit-speed}

The stationary distribution is independent of $k$, while convergence speed can vary with the consideration bound. I generalize the convergence comparison of \citet[Theorem~2]{Badev2021} beyond the flat-payoff benchmark: for general potential payoffs, larger nested, state-independent meetings weakly increase the spectral gap (Theorem~\ref{thm:unconstrained-logit-speed-comparison}). This result guides the choice of meetings in the inner sampler, which repeatedly simulates auxiliary project configurations at proposed parameter values (Section~\ref{sec:bayesian-estimation}). Meetings that reconsider several projects let an individual leave one project and join another in the same update, avoiding an intermediate change in her project load. Larger meetings also cost more to evaluate. I first derive the exact rate in the flat-potential benchmark (Theorem~\ref{thm:unconstrained-logit-speed}) and then establish the comparison for general potentials.

Keep the labeled-slot state space, logit rule, and stationary distribution $\pi_\tau^0$ of Section~\ref{sec:unconstrained-logit}. Write $\rho_k(i,R)$ for its state-independent meeting probabilities and $K_k$ for the transition operator. We measure convergence per meeting by the spectral gap
\[
\operatorname{gap}(K_k):=1-\lambda_2(K_k),
\qquad t_{\mathrm{rel}}(K_k):=\frac{1}{\operatorname{gap}(K_k)}.
\]
Each update draws from a Gibbs conditional. Consequently, $K_k$ is reversible with nonnegative eigenvalues, and its gap governs asymptotic $L^2(\pi_\tau^0)$ relaxation. An exact rate based only on meeting probabilities is available in the flat-potential benchmark; general payoffs require a separate comparison.

\begin{theorem}[Flat-potential convergence rate]
\label{thm:unconstrained-logit-speed}
Suppose $Q$ is constant on $\mathcal A$, as when $U_i(A)=0$ for every $i,A$, and retain the meeting assumptions of Section~\ref{sec:unconstrained-logit}. Let
\[
c_{i,r}^{(k)}:=\sum_{R\ni r}\rho_k(i,R)
\]
be the probability that coordinate $(i,r)$ is reconsidered in a meeting. Then
\begin{equation}
\label{eq:unconstrained-logit-gap}
\lambda_2(K_k)=1-\min_{i,r}c_{i,r}^{(k)},
\qquad
\operatorname{gap}(K_k)=\min_{i,r}c_{i,r}^{(k)}.
\end{equation}
In particular, select $i$ uniformly, independently draw the size $\ell=|R|$ with probabilities $w_{\ell,k}>0$ on $\{1,\ldots,m_k\}$, where $m_k=\min\{k,\overline P\}$, and then choose $R$ uniformly among sets of that size. Then
\begin{equation}
\label{eq:unconstrained-logit-mean-size-gap}
\operatorname{gap}(K_k)=\frac{\mathbb E_k[\ell]}{n\overline P},
\qquad
\mathbb E_k[\ell]=\sum_{\ell=1}^{m_k}\ell w_{\ell,k}.
\end{equation}
\end{theorem}

For the uniform size rule $w_{\ell,k}=1/m_k$, this gives
\[
\operatorname{gap}(K_k)=\frac{m_k+1}{2n\overline P},
\qquad
 t_{\mathrm{rel}}(K_k)=\frac{2n\overline P}{m_k+1}.
\]
Thus, for $k\leq\overline P$, larger consideration bounds strictly improve the benchmark relaxation rate under this rule. The expression $k/(n\overline P)$ applies instead to the exact-size variant $|R|=k$, not to a uniform draw over sizes $1,\ldots,k$. With arbitrary meetings, the slowest coordinate's update probability determines the benchmark gap; increasing the allowed bound alone does not order those probabilities.

For general potentials, the convergence rate depends on the payoff structure as well as the meeting law. The Gibbs conditional-update structure nevertheless yields a comparison without requiring an explicit formula for the rate.

\begin{theorem}[Spectral comparison for general potentials]
\label{thm:unconstrained-logit-speed-comparison}
Fix payoffs admitting an exact potential $Q$, a logit scale $\tau>0$, and the slot count $\overline P$. Suppose two state-independent meeting laws can be coupled as $(i,R)$ and $(i,R')$, with the same selected individual and $R\subseteq R'$ almost surely. Their unconstrained logit operators satisfy
\[
\operatorname{gap}(K_{\rho'})\geq\operatorname{gap}(K_\rho).
\]
This includes uniform individuals and uniform project sets conditional on size when size is drawn independently of the individual and the second size distribution first-order stochastically dominates the first. In particular, under the uniform size rule on $1,\ldots,\min\{k,\overline P\}$, the spectral gap is nondecreasing in $k$ for all payoffs admitting the potential representation.
\end{theorem}

The intuition is that a larger meeting allows the dynamic to redraw related memberships jointly. Each logit update draws from the stationary distribution conditional on the memberships outside the meeting, preserving all dependence induced by the potential. For any statistic of the profile, holding fewer memberships fixed weakly increases its expected conditional variance. This quantity is the update's Dirichlet form, which measures the expected squared change in the statistic, up to a factor of two. Nested larger meetings therefore increase this form for every statistic. The spectral gap is the smallest ratio of this form to the statistic's stationary variance, so it also weakly increases. The argument uses the conditional-variance identity under $\pi_\tau^0$ and does not require a flat potential or independent membership decisions. Appendix~\ref{app:unconstrained-logit-speed} defines the Dirichlet form, gives the formal proof, and illustrates the result with two project slots in Example~\ref{ex:two-slot-switching}.

The comparison is weak for general payoffs and does not supply a rate independent of $Q$ and $\tau$. Larger nested blocks improve the spectral gap per meeting, while enumerating their $2^{|R|}$ choices can increase computational cost. These comparisons concern the unconstrained Gibbs updates defined here; the constrained, targeted Metropolis updates used in estimation require separate performance checks. Appendix~\ref{app:unconstrained-logit-speed} gives the proofs, including the complete benchmark spectrum.

\subsection{$k$-Team Dynamics with Stochastic Consent}
\label{sec:logit-probabilistic}

This section reintroduces the stability constraints into the adaptive dynamic in a way that is compatible with the probabilistic choice framework. I do so through stochastic consent: an individual first draws a candidate revision using the logit rule, and the resulting profile then undergoes a separate acceptance test based on the participation conditions. Failed checks reduce the probability of implementation, and rejected proposals leave the current profile unchanged. The test covers all team memberships in the candidate profile, including those outside the consideration set, making acceptance a function of the candidate state alone. This yields an explicit stationary distribution combining the payoff potential with participation penalties.

\subsubsection{Construction}
\label{sec:stochastic-consent-construction}

The proposal stage uses the labeled-slot state space $\mathcal A$, exact potential $Q$, payoff-noise scale $\tau>0$, and state-independent full-support meetings of Section~\ref{sec:unconstrained-logit}. At meeting $\mu=(i,R)$, individual $i$ draws a candidate revision from $F_\mu(A):=\mathcal B_i(A;R)$ using the unconstrained logit probabilities $L_\mu(A,B)$, with $L_\mu(A,B)=0$ outside this choice set. Writing $\rho_\mu:=\rho_k(i,R)$ for the meeting probability, the proposal kernel is
\[
P_k^0(A,B)=\sum_\mu\rho_\mu L_\mu(A,B).
\]

The acceptance stage determines whether the proposed profile is implemented. Let $\chi:\mathcal A\to(0,1]$ give the probability of accepting each candidate state. After $B$ is proposed, it is implemented with probability $\chi(B)$; otherwise, the current profile $A$ is retained. Conditional on $B$, the acceptance draw is independent of the payoff shocks and past draws. There is one proposal per date, so a rejection ends the update rather than triggering another draw. For $B\neq A$, the transition probability is therefore
\[
P_k^\chi(A,B)=P_k^0(A,B)\chi(B),
\]
while the probability of remaining at $A$ includes all rejected proposals:
\[
P_k^\chi(A,A)
=
P_k^0(A,A)
+
\sum_{B\neq A}P_k^0(A,B)\bigl[1-\chi(B)\bigr].
\]

\begin{proposition}[Stationary distribution with probabilistic constraints]
\label{prop:stationary-probabilistic-constraints}
The process with transition kernel $P_k^\chi$ is irreducible,
aperiodic, and reversible. Its unique stationary distribution is
\begin{equation}
\label{eq:stationary-probabilistic-constraints}
\pi_{\tau,\chi}(A)
=
\frac{\exp\{Q(A)/\tau\}\chi(A)}
{\displaystyle\sum_{D\in\mathcal A}\exp\{Q(D)/\tau\}\chi(D)}.
\end{equation}
If $\chi$ does not depend on $k$, neither does this distribution.
\end{proposition}

This proposition generalizes Proposition~\ref{prop:unconstrained-logit-stationary}, which is recovered by setting $\chi\equiv1$. The stationary distribution has a simple interpretation: each profile's unconstrained logit weight is multiplied by its probability of acceptance. Among profiles with the same potential, those less likely to receive consent therefore receive less stationary weight. Formally, the exact potential makes the unconstrained proposals satisfy detailed balance, because $U_i(B)-Q(B)$ is constant over a meeting's choice set. Multiplying the stationary weight at $A$ by $\chi(A)$ and the transition probability to $B$ by $\chi(B)$ introduces the same factor $\chi(A)\chi(B)$ in both directions, preserving detailed balance. Appendix~\ref{app:count-weighted-consent} gives the proof. This argument requires the same candidate-state function $\chi(B)$ at every meeting; acceptance rules that also depend on the current state or the selected meeting require a separate argument.

\subsubsection{Count-Weighted Consent and the Empirical Rule}
\label{sec:count-weighted-stochastic-consent}
\label{par:counting-rule}

To connect acceptance to the stability constraints, I use the participation slack $\sigma_{j,r}(A)$ in \eqref{eq:participation-slack}. As in Definition~\ref{def:equilibrium}, this slack compares remaining in team $r$ with leaving it or replacing the membership with a solo project. A negative slack means that the member prefers at least one of these outside options; a zero slack satisfies the participation condition. The memberships subject to these checks are
\[
\mathcal M(A):=\{(j,r):a_{j,r}=1,\ n_r(A)>1\}.
\]
The count rule measures how many of these individual conditions fail:
\begin{equation}
\label{eq:individual-count-violations}
\mathcal V(A):=\sum_{(j,r)\in\mathcal M(A)}
\mathbf 1\{\sigma_{j,r}(A)<0\}.
\end{equation}
Each membership counts once if its member prefers either outside option. Violations on different projects count separately, even when they involve the same individual. Gains to other members cannot offset a failed check, and the size of the payoff shortfall does not affect the count.

Fix $\tau>0$ and a finite enforcement weight $\kappa\geq0$, and set $q=e^{-\kappa}\in(0,1]$. Here $q$ is the probability that a failed participation check nevertheless permits implementation. At each date, a state-independent meeting $\mu=(i,R)$ is drawn with full support over individuals and nonempty project sets of size at most $\min\{k,\overline P\}$. Individual $i$ then draws one candidate $B\in F_\mu(A)$ using the logit probabilities $L_\mu(A,B)$; only her memberships in $R$ can change. Every membership in $\mathcal M(B)$ is then checked. A nonnegative slack passes surely, while each negative slack passes with probability $q$. Conditional on $B$, these draws are independent across checks and independent of the payoff shocks. The candidate is implemented only if all checks pass. Thus its acceptance probability is
\begin{equation}
\label{eq:counting-rule}
\chi_\kappa(B)
=\prod_{(j,r)\in\mathcal M(B)}q^{\mathbf 1\{\sigma_{j,r}(B)<0\}}
=q^{\mathcal V(B)}=e^{-\kappa\mathcal V(B)}.
\end{equation}
The empty product equals one, so a profile with no failed checks is accepted surely. Each additional failure multiplies the probability of acceptance by $q$, and larger $\kappa$ makes enforcement stricter. For $B\ne A$, the resulting transition kernel is
\[
P_k^\kappa(A,B)=P_k^0(A,B)e^{-\kappa\mathcal V(B)},
\qquad
P_k^\kappa(A,A)=1-\sum_{B\ne A}P_k^\kappa(A,B).
\]
One interpretation is that each failed check can block a pending proposal at an independent Poisson rate $\lambda$ over a period $T$. The probability that it does not block implementation is then $q=e^{-\lambda T}$, giving $\kappa=\lambda T$. Applying Proposition~\ref{prop:stationary-probabilistic-constraints} yields the following stationary distribution.

\begin{corollary}[Count-weighted stationary distribution]
\label{cor:count-weighted-stationary}
Under the meeting and potential assumptions above, for every finite $\kappa\geq0$ the count-weighted process is irreducible, aperiodic, and reversible, with unique stationary distribution
\begin{equation}
\label{eq:counting-rule-stationary}
\pi_{\tau,\kappa}(A)
=\frac{\exp\{Q(A)/\tau-\kappa\mathcal V(A)\}}
{\displaystyle\sum_{D\in\mathcal A}\exp\{Q(D)/\tau-\kappa\mathcal V(D)\}}.
\end{equation}
For fixed payoffs, $\tau$, and $\kappa$, this distribution is independent of $k$ and of the state-independent meeting probabilities.
\end{corollary}

The exponent separates two forces: higher potential raises a profile's stationary weight, while failed participation checks reduce it. Holding the potential fixed, each additional failed check multiplies the unnormalized weight by $e^{-\kappa}$. The result follows directly by setting $\chi=\chi_\kappa$ in Proposition~\ref{prop:stationary-probabilistic-constraints}; Appendix~\ref{app:count-weighted-consent} provides the detailed verification.

Although the individual participation test is the same as in Definition~\ref{def:equilibrium}, the set of memberships screened is broader. At meeting $(i,R)$, that definition requires checks only for
\[
\mathcal M_{i,R}(B)
:=\{(j,r)\in\mathcal M(B):r\in R,\ b_{i,r}=1\}.
\]
The stochastic rule instead screens all memberships in $\mathcal M(B)$, including those outside the meeting. This global enforcement makes acceptance depend on the candidate profile alone, allowing the same acceptance factor to enter the stationary law regardless of which meeting produced the proposal. Restricting the checks to $\mathcal M_{i,R}(B)$ makes acceptance meeting dependent, and \eqref{eq:counting-rule-stationary} generally fails. The appendix gives a counterexample, characterizes compatibility with a Gibbs law, and treats a team-size-only case.

The enforcement weight connects the unconstrained model to strict participation screening. At $\kappa=0$, every proposal is accepted and \eqref{eq:counting-rule-stationary} reduces to the unconstrained distribution $\pi_\tau^0$. As $\kappa\to\infty$, stationary probability concentrates on $\Omega=\{A:\mathcal V(A)=0\}$, which contains the empty profile, and the limiting law is $\pi_\tau^0$ conditioned on this set. This limit enforces the participation conditions for current memberships. A $k$-team stable equilibrium additionally requires that no individual prefer another feasible project choice. At finite $\kappa$, every profile retains positive stationary probability.

\begin{remark}[Participation checks used in estimation]
\label{rem:exit-scale}
At the unit payoff-noise scale used in estimation, the slot adjustment expresses stability on a project payoff scale. Write $U_i$ for the per-slot payoff index and define $U_i^{\mathrm{proj}}(A)=U_i(A)+p_i(A)\log\bar P$. Withdrawal reduces the project count by one, while replacement by a solo project preserves it. The ordinary leave-or-replace slack for $U_i^{\mathrm{proj}}$ is therefore
\begin{equation}
\label{eq:slot-scale-slack}
e_{j,r}(A):=\Delta_{a_{j,r}}U_j(A)-\max\bigl\{-\log\overline P,\ \beta_j-\Delta C_P(p_j)\bigr\}
=\min\bigl\{s_{j,r}(A),\ v_{j,r}(A)+\log\overline P\bigr\},
\end{equation}
where $v_{j,r}=\Delta_{a_{j,r}}U_j$. The empirical violation count is $\bar{\mathcal V}(A)=\sum_{(j,r)\in\mathcal M(A)}\1\{e_{j,r}(A)<0\}$, as in \eqref{eq:practical-slot-adjusted-slack} and \eqref{eq:penalty-exit}, with $\beta_j$ replaced by $b_{jt}$ in the application. Relative to a test applied directly to the unshifted slot index, $e_{j,r}\geq\sigma_{j,r}$ and $\bar{\mathcal V}\leq\mathcal V$. The economic convention is to apply the same participation requirement to project payoffs; Remark~\ref{rem:practical-slot-exit} and the appendix explain its rationale and limits.

The existing proposal probabilities admit this interpretation with explicit reference weights. Let
\[
\nu_{\bar P}(A)=\bar P^{-\sum_jp_j(A)},\qquad
Q^{\mathrm{proj}}(A)=Q(A)+\log\bar P\sum_jp_j(A).
\]
Within a meeting, all other individuals' project counts are fixed, so
\[
\nu_{\bar P}(B)\exp\{U_i^{\mathrm{proj}}(B)\}
\ \propto\ \exp\{U_i(B)\}.
\]
Thus a reference-weighted logit using project payoffs gives exactly the current per-slot logit proposal. Also $\nu_{\bar P}(A)e^{Q^{\mathrm{proj}}(A)}=e^{Q(A)}$, so the reference-weighted and per-slot expressions give identical stationary weights.

Global screening with $e^{-\kappa\bar{\mathcal V}(B)}$ consequently preserves the same stationary-distribution argument. Proposition~\ref{prop:stationary-probabilistic-constraints} and Corollary~\ref{cor:count-weighted-stationary} apply directly to the slot representation with $\bar{\mathcal V}$ in place of $\mathcal V$; Appendix~\ref{app:count-weighted-consent} verifies the extension. As enforcement becomes strict, the limiting law is supported on $\bar\Omega=\{A:\bar{\mathcal V}(A)=0\}$. These profiles pass all current-membership checks on the project payoff scale. The inclusion $\bar\Omega\supseteq\Omega$ compares them with checks evaluated at the unshifted slot index.
\end{remark}

An alternative to counting failed checks is to weight them by the size of the payoff shortfall. Using the shortfalls $[-\sigma_{j,r}]_+$ gives $\pi_{\tau,\kappa_\ell}\propto\exp\{Q/\tau-\kappa_\ell\mathcal V_\ell\}$, so profiles with larger total shortfalls receive lower weight at a given potential. The appendix develops this alternative and its limits and examines observed shortfalls. The estimation uses count weighting.

\subsubsection{Discussion}
\label{sec:stochastic-consent-discussion}

Stochastic consent changes how the dynamic responds to a failed participation condition. In the deterministic dynamic, the individual cannot retain her current choices if they fail the participation checks for that meeting---for example, because a teammate would prefer to leave. She must choose a feasible revision, even if it lowers her own payoff. Theorem~\ref{thm:deterministic-team-convergence} establishes convergence to a $k$-team stable equilibrium. Here the individual draws an unconstrained logit proposal, and a separate test determines whether it is implemented. A rejection retains the current profile, even if it contains participation violations. At finite $\kappa$, such violations can therefore persist temporarily and receive positive weight in the stationary distribution.

Stationary weights do not identify the calendar timing of this process. Global screening can produce long holding times because failures in teams outside the meeting also reduce acceptance. Within the maintained dynamic, $e^{-\kappa}$ is the probability that an individual failed check permits implementation; stationary observations do not independently identify the meeting rate or establish equilibration within a year. Estimation instead uses kernels that preserve the same stationary law while moving more efficiently. Their computational performance does not validate the literal transition rates of global screening.

As payoff noise becomes small, individuals are less likely to make revisions that lower their utility. Consider a configuration that maximizes the potential $Q$. No individual can improve her payoff by changing her projects, because that would also increase $Q$. If an empty project slot is available, leaving a team or replacing a membership with a solo project is among these possible changes. Every member is therefore willing to stay, and both the count and depth penalties are zero. If every potential maximizer has an empty slot, both stationary distributions concentrate on these configurations as $\tau\downarrow0$, whether or not consent noise also vanishes (see the appendix). These configurations are $k$-team stable for every $k$; other stable equilibria may have lower potential and receive no limiting weight. The appendix gives a sufficient number of slots to ensure that the solo alternative remains available at potential maximizers. Finite-noise simulations still require checking how often the chosen project limits are approached.

For estimation, I use $\overline P=400$ labeled project slots per year and a per-person cap of $\bar p=12$ projects, compared with an observed maximum of 8. The sampler's meeting law determines how many project slots are reconsidered in each update and how they are selected. I use $(\xi,\bar K,\eta_{\mathrm{own}})=(0.25,3,0.9)$. The parameter $\xi$ controls the relative frequency of larger meetings: each additional slot makes a meeting one quarter as likely, up to the maximum of $\bar K=3$ slots. For meetings involving several slots, $\eta_{\mathrm{own}}=0.9$ gives a 90\% chance of drawing one of the individual's current projects uniformly and the remaining slots uniformly from the other slots. Otherwise, or if the individual has no current projects, all slots are drawn uniformly. This targeting helps the sampler move an individual from one project to another in a single update.

Three results connect the analysis to this empirical implementation. Proposition~\ref{prop:stationary-probabilistic-constraints} and Corollary~\ref{cor:count-weighted-stationary}, with the practical checks of Remark~\ref{rem:exit-scale}, provide the stationary likelihood of each year's participation matrix in Section~\ref{sec:full-likelihood-constrained}. Because this law is independent of $k$ and the state-independent meeting probabilities, meeting size can be chosen for computational performance in stationary estimation. Section~\ref{sec:bayesian-estimation} uses both Gibbs updates and targeted Metropolis proposals with the appropriate correction. Theorem~\ref{thm:logit-project-count-tail} bounds the likelihood error from imposing a participation cap within the fixed slot universe in terms of excluded probability mass. Finally, Theorems~\ref{thm:unconstrained-logit-speed} and~\ref{thm:unconstrained-logit-speed-comparison} compare convergence per meeting for unconstrained Gibbs updates, motivating the use of larger blocks. The performance of the constrained, targeted sampler is assessed numerically.

\section{Data and Identification}
\label{sec:data}

The empirical analysis connects individual research activity to the teams economists form and their willingness to remain in them. I first describe the data and a productivity benchmark, then specify the collaboration model and discuss the variation that separates its payoff and consent mechanisms.

\subsection{Data and Measurement}

The estimation sample covers 449 Federal Reserve Board economists in 2007--2019, with 3{,}522 economist-years, 1{,}918 research papers, and 2{,}691 Board authorships. A project is a paper, and its team consists of the Board authors employed in its year; collaborators outside the Board are not modeled as team members. Each year's observation is the participation matrix linking economists to papers. An economist remains in the yearly risk set when she produces no paper, provided that she is employed at the Board and belongs to the sample of economists who publish at least once during their tenure. Estimation and counterfactual experiments describe research activity and collaboration within this cohort of observed researchers. Entry into the research population is outside the model.

Bibliographic records come from OpenAlex \citep{priem2022openalex} and are matched to archived Board staff rosters, which supply employment histories and organizational affiliations. Author names are reconciled across sources, duplicate versions of a paper are merged, and incorrect records are corrected. Board tenure runs from one year before an author's first appearance on the rosters to her last appearance. The sample excludes years of extended absence from the Board and papers without an author in tenure. These rules distinguish a year without research output from a year in which the economist was not at risk of producing Board research.

I also exclude institutional output, including official reports, statistical documentation, policy-analysis notes, and large assigned institutional projects. Such output responds directly to the institution's policy responsibilities, whereas the model describes economists choosing research projects given their other demands. Including it would mix these different activities when assessing how policy workload affects research and collaboration. The appendix documents the sources, matching procedures, and exclusions, and records the sample at each step.

Policy workload is measured by the Flesch--Kincaid grade level, $\mathrm{FKGL}_t$, of Board staff remarks in FOMC transcripts. The measure uses words, sentences, and syllables to capture the linguistic complexity of staff briefings: denser briefings proxy for heavier policy work. It uses Board staff remarks only, since remarks by staff of the Federal Reserve Bank of New York reflect a different workload, and is not adjusted for the composition of presenters. I center it at its 2007--2019 mean, so the model's other parameters describe a year of average workload.

Table~\ref{tab:sample-stats} describes the broader bibliographic panel before the extended-absence exclusions. Research activity expanded over the sample: the number of papers rose from 103 in 2007 to 189--219 in 2017--2019, while output per active economist remained around 1.55 papers a year. Teams are small: about 70\% of papers have a single Board author, and mean Board team size rose from about 1.35 in 2007--2013 to 1.45 in 2014--2019. About half of papers with several Board authors include two members from the same section. These patterns motivate separate roles for individual productivity, team size, and organizational composition. The workload measure peaks in 2010 and is lowest in 2018--2019.

\begin{table}[ht]
\caption{Sample Statistics by Year}
\label{tab:sample-stats}
\begin{center}
{\footnotesize\setlength{\tabcolsep}{2.8pt}
\begin{tabular}{lrrrrrrrrrrrr}
\hline\hline
 & \multicolumn{2}{c}{Economists} & & & Papers & Mean & \multicolumn{4}{c}{Board team size} & Same & \\
Year & Rosters & Active & Papers & Particip. & /active & team & 1 & 2 & 3 & 4+ & section & FKGL \\
\hline
2007 & 175 & 87 & 103 & 139 & 1.60 & 1.35 & 72 & 26 & 5 & 0 & 0.55 & 13.23 \\
2008 & 179 & 81 & 112 & 145 & 1.79 & 1.29 & 84 & 23 & 5 & 0 & 0.46 & 13.46 \\
2009 & 193 & 79 & 89 & 117 & 1.48 & 1.31 & 67 & 18 & 2 & 2 & 0.50 & 13.77 \\
2010 & 199 & 96 & 102 & 150 & 1.56 & 1.47 & 65 & 26 & 11 & 0 & 0.46 & 14.26 \\
2011 & 249 & 111 & 117 & 163 & 1.47 & 1.39 & 85 & 21 & 9 & 2 & 0.50 & 14.08 \\
2012 & 248 & 125 & 136 & 185 & 1.48 & 1.36 & 99 & 27 & 8 & 2 & 0.62 & 13.68 \\
2013 & 291 & 122 & 122 & 152 & 1.25 & 1.25 & 96 & 22 & 4 & 0 & 0.46 & 13.92 \\
2014 & 309 & 152 & 167 & 236 & 1.55 & 1.41 & 117 & 38 & 7 & 5 & 0.48 & 13.38 \\
2015 & 340 & 158 & 183 & 267 & 1.69 & 1.46 & 118 & 52 & 8 & 5 & 0.45 & 13.89 \\
2016 & 367 & 179 & 196 & 274 & 1.53 & 1.40 & 143 & 34 & 14 & 5 & 0.47 & 13.44 \\
2017 & 368 & 193 & 219 & 310 & 1.61 & 1.42 & 152 & 49 & 14 & 4 & 0.58 & 13.31 \\
2018 & 381 & 192 & 199 & 307 & 1.60 & 1.54 & 125 & 52 & 13 & 9 & 0.49 & 13.00 \\
2019 & 389 & 177 & 189 & 267 & 1.51 & 1.41 & 129 & 49 & 5 & 6 & 0.47 & 12.99 \\
\hline
2007--2019 & 493 & 449 & 1{,}934 & 2{,}712 & 1.55 & 1.40 & 1{,}352 & 437 & 105 & 40 & 0.50 & 13.57 \\
\hline\hline
\end{tabular}}
\end{center}
{\RaggedRight\floatnotesfont Notes. The broader bibliographic panel, before extended-absence exclusions. Roster counts cover economists matched to the bibliographic file; active counts and paper statistics require authorship in tenure. The estimation sample is described in the text and the appendix. Rosters:
on the Board's staff rosters in the year; active: with at least one paper in tenure.
Participations: pairs of a paper and a Board author in tenure. Papers/active: mean number
of papers of active economists; team size: number of Board members of a paper; mean team: mean number of Board members of a paper. Same section: share of
papers with two or more Board members that have two members from the same section. FKGL:
Flesch--Kincaid grade level of Board staff remarks at FOMC meetings. Total row: distinct
economists over 2007--2019, and means over all papers or active person-years.\par}
\end{table}

\FloatBarrier

\subsection{A Productivity Benchmark}
\label{sec:baseline}

I begin with differences in research activity, abstracting from the benefits of teams and collaborators. For economist $i$ employed in year $t$, let $p_{it}$ be her number of projects. The own-project payoff is
\begin{equation}
\label{eq:payoffs_baseline_1}
U_{it}(p_{it})=\beta_i p_{it}-C_P(p_{it}),
\qquad C_P(p)=\frac{\psi_P}{2}(p-p_0)_+^2,
\end{equation}
where $(x)_+=\max\{x,0\}$. The individual effect $\beta_i$ raises the return to each project, while $\psi_P>0$ governs congestion once activity exceeds the threshold $p_0$. I fix $p_0=2$ and normalize the payoff-noise scale to one, as in the estimated collaboration model.

An economist chooses which of $\overline P=400$ labeled project slots to join, subject to a limit of $\bar p=12$ projects. There are $\binom{\overline P}{p}$ ways to participate in $p$ slots, all with the same payoff in this benchmark. Her project-count distribution is therefore
\begin{equation}
\label{eq:baseline_prob_slots}
\Pr(p_{it}=p\mid\beta_i)
=\frac{\binom{\overline P}{p}\exp\{\beta_i p-C_P(p)\}}
{\sum_{m=0}^{\bar p}\binom{\overline P}{m}\exp\{\beta_i m-C_P(m)\}},\qquad p=0,\ldots,\bar p.
\end{equation}
This is a restriction of the full model to individual effects and project costs, with no team benefits or consent penalty. The fixed slot universe keeps the productivity parameters on the same scale across specifications.

Conditional on the individual effects, counts are independent across economists and years, so their probabilities give an exactly evaluable full likelihood. Repeated observations inform productivity, and the decline in the odds of additional projects informs congestion. Individual effects share a normal hierarchy, $\beta_i\sim\mathcal N(\mu_\beta,s_\beta^2)$, allowing economists with short histories to draw information from the rest of the sample. The appendix gives the identification argument and estimates, together with a counts-only alternative that assigns no multiplicity to project labels.

The benchmark explains project counts but has no preferences over team sizes or collaborators, so productivity can absorb omitted collaboration returns. I next allow team size, collaborator characteristics, and past relationships to affect participation. The online appendix isolates the contribution of team size in an intermediate specification.

\FloatBarrier

\subsection{The Empirical Collaboration Model}
\label{sec:structural}

The empirical model applies the preferences of Section~\ref{sec:intro-payoffs} to the observed characteristics and histories of Board economists. Individual returns govern research activity; team size and section composition affect the value of participation; and pair characteristics and past collaboration affect the value of working together. In year $t$, let $P_{it}$ be $i$'s projects, $N_r$ the members of project $r$, $n_r=|N_r|$, and $g_{ij}$ the number of projects shared by $i$ and $j$. Suppressing the year index on these team and pair variables, the per-slot payoff index used in the labeled likelihood is
\begin{align}
\label{eq:payoffs_structural1}
U_{it}(A_t)={}&b_{it}p_{it}-C_P(p_{it})\notag\\
&+\sum_{r\in P_{it}}V_t(X_i,\{X_j\}_{j\in N_r})\notag\\
&+\sum_{j\ne i}\bigl[g_{ij}w_t(X_i,X_j)-C_{G,t}(g_{ij};H_{ijt})\bigr].
\end{align}

The per-slot return combines persistent productivity with time-varying demands and responsibilities:
\begin{equation}
\label{eq:own_return_structural}
\begin{aligned}
b_{it}={}&\beta_i+\bigl(\beta^{\mathrm{FKGL}}+\beta_S^{\mathrm{FKGL}}\1\{z_{it}=S\}+\beta_{\mathrm{MA}}^{\mathrm{FKGL}}\mathrm{MA}_{it}\bigr)\mathrm{FKGL}_t\\
&+\beta_{\mathrm{M\&O}}^{\mathrm{title}}\1\{\mathrm{Ti}_{it}=\mathrm{M\&O}\}+\beta_{\mathrm{Adv}}^{\mathrm{title}}\1\{\mathrm{Ti}_{it}=\mathrm{Adv}\}.
\end{aligned}
\end{equation}
Here $z_{it}\in\{E,M,S\}$ denotes entrant, mid-career, or senior tenure: up to one year, two to seven years, or eight or more years on the Board. The workload response is common to entrants and mid-career economists, with an additional shift for senior economists. The indicator $\mathrm{MA}_{it}$ allows a different response in the Monetary Affairs division. Title effects distinguish managers and officers (M\&O) and advisers (Adv) from the reference group of economists and senior or principal economists, including those with no recorded title.

Team value separates size from organizational composition:
\[
V_t(X_i,\{X_j\}_{j\in N_r})=V_{\mathrm{size},t}(n_r)+V_{\mathrm{section}}(m_{irt}),
\]
where $m_{irt}$ counts members of $r$ from $i$'s section, including $i$. Pair value permits sorting by productivity and tenure, and the relationship cost governs repeated projects within a year. The estimated functions are
\begin{align}
\label{eq:team_value_structural}
V_{\mathrm{size},t}(n)&=\1\{n\geq2\}
\left[v_{\mathrm{size},2}\rho_{\mathrm{size}}^{n-2}
+v_{\mathrm{size},2+}^{\mathrm{FKGL}}\mathrm{FKGL}_t\right],\\
\label{eq:payoffs_structural3}
V_{\mathrm{section}}(m)&=v_{\mathrm{section},2}\1\{m=2\}+v_{\mathrm{section},3+}\1\{m\geq3\},\\
\label{eq:payoffs_structural4}
w_t(X_i,X_j)&=\omega^{\mathrm{gap}}(\beta_i-\beta_j)^2+\omega_{z_{it}z_{jt}}^{\mathrm{tenure}}+\omega_{EM}^{\mathrm{FKGL}}\mathrm{FKGL}_t\1\{\{z_{it},z_{jt}\}=\{E,M\}\},\\
\label{eq:payoffs_structural5}
C_G(g)&=\psi_{G,2}\1\{g=2\}+\psi_{G,3+}(g-2)_+^2.
\end{align}
Solo team values and the cost of an inactive relationship are zero. The factor $\rho_{\mathrm{size}}>0$ governs how team value changes with size; with $v_{\mathrm{size},2}<0$, a value above one makes it fall more rapidly in larger teams. Section values permit organizational sorting, while a negative $\omega^{\mathrm{gap}}$ rewards productivity similarity. Tenure-pair values are symmetric, with the mid-career pair coefficient fixed at zero. Under the geometric size profile, this is a substantive restriction rather than a pure normalization. Workload can shift the value of team participation and of entrant--mid-career collaboration separately from its effect on individual activity.

Past collaboration provides an additional benefit once per active relationship:
\begin{equation}
\label{eq:collaboration_history_cost}
C_{G,t}(g;H_{ijt})=C_G(g)-\omega^{\mathrm{past}}H_{ijt}[g>0],
\end{equation}
where $H_{ijt}=1$ if $i$ and $j$ shared a project before year $t$. The initial history includes the 201 pairs of sample economists who coauthored before 2007. This benefit rewards maintaining a previous relationship, while $w_t$ applies to every shared project and $C_G$ governs repetition within the year. I fix $\psi_{G,3+}=0.5$, retain $p_0=2$, and set degree congestion to zero. The appendix collects the parameter vector and explains the maintained restrictions.

The population $N_t$ contains the sample economists at risk in year $t$. Both estimation and simulation use the fixed labeled state space
\[
\mathcal A_t=\cA_t:=\{A\in\{0,1\}^{|N_t|\times\bar P}:p_{it}\leq\bar p\text{ for all }i\},
\qquad \bar P=400,\quad\bar p=12.
\]
The project cost and symmetric collaboration terms preserve the exact potential. Writing $\Phi_t(S)=\sum_{h=1}^{|S|}V_{\mathrm{size},t}(h)+\sum_c\sum_{h=1}^{m_{S,c}}V_{\mathrm{section}}(h)$, where $m_{S,c}$ counts members of section $c$, gives
\begin{align}
\label{eq:potential_structural}
Q_t(A_t)={}&\sum_{i\in N_t}\bigl[b_{it}p_{it}-C_P(p_{it})\bigr]
+\sum_{r=1}^{\bar P}\Phi_t(N_r)\notag\\
&+\sum_{i<j}\bigl[g_{ij}w_t(X_i,X_j)-C_{G,t}(g_{ij};H_{ijt})\bigr].
\end{align}

Consent adds each member's willingness to remain. Let $B_{j,r,t}$ replace $j$'s membership in team $r$ with a new solo project. Her surplus from the team relative to that alternative is
\begin{equation}
\label{eq:team_surplus}
s_{j,r,t}(A_t)=U_{jt}(A_t)-U_{jt}(B_{j,r,t}).
\end{equation}
Solo replacement leaves her project count unchanged, so the own-project return and congestion cost cancel from this comparison. Exit also changes her project count and therefore gives a separate participation test.

\begin{remark}[Stability checks on the project payoff scale]
\label{rem:practical-slot-exit}
The slot adjustment is intended to preserve the economic content of stability when project opportunities are represented by labeled slots. With many available labels, fitting sparse participation pushes the per-slot return downward; using that index directly for withdrawal can make failures reflect the counting of opportunities rather than members' incentives to remain. At unit payoff-noise scale, I distinguish the slot index in \eqref{eq:payoffs_structural1} from project-scale utility:
\begin{equation}
\label{eq:project-scale-payoff}
U_{it}^{\mathrm{proj}}(A)
=U_{it}(A)+p_{it}(A)\log\bar P,
\qquad
b_{it}^{\mathrm{proj}}=b_{it}+\log\bar P.
\end{equation}
The implemented margin is the ordinary leave-or-replace slack for $U_{jt}^{\mathrm{proj}}$:
\begin{equation}
\label{eq:practical-slot-adjusted-slack}
\begin{aligned}
e_{j,r,t}(A_t)
&=\min\left\{s_{j,r,t}(A_t),\ \Delta_{a_{j,r}}U_{jt}(A_t)+\log\bar P\right\}\\
&=\Delta_{a_{j,r}}U_{jt}(A_t)
-\max\left\{-\log\bar P,\ b_{jt}-\Delta C_P(p_{jt})\right\}.
\end{aligned}
\end{equation}
Solo replacement preserves the project count, so its surplus is unchanged by the normalization. Withdrawal reduces the count by one, giving the additional $\log\bar P$ in its margin. Requiring $e_{j,r,t}\geq0$ therefore retains both participation requirements on the project payoff scale. Relative to the unshifted slot index, $e_{j,r,t}\geq\sigma_{j,r,t}$; this comparison uses different payoff normalizations. The project payoff convention is maintained explicitly, with the offset carried by reference weights so the implemented likelihood is unchanged. The appendix develops the argument and its limits, and Remark~\ref{rem:exit-scale} gives the equivalent reference-weighted proposal.
\end{remark}

Count each failed nonsingleton membership once:
\begin{equation}
\label{eq:penalty-exit}
\bar{\mathcal V}_t(A\mid G^U_{<t};\theta)=\sum_{r:\,n_r\geq2}\ \sum_{j\in N_r}\1\bigl\{e_{j,r}(A)<0\bigr\}.
\end{equation}
The checks use each member's own outside options, so gains to other members cannot offset a failure. By Corollary~\ref{cor:count-weighted-stationary} and Remark~\ref{rem:exit-scale}, the empirical stationary law, at unit payoff-noise scale, is
\begin{equation}
\label{eq:pi-t-constrained}
\pi_t^{\kappa}(A\mid G^U_{<t};\theta)=
\frac{\exp\{Q_t(A\mid G^U_{<t};\theta)-\kappa \bar{\mathcal V}_t(A\mid G^U_{<t};\theta)\}}{Z_t(\theta,\kappa)},
\end{equation}
where $Z_t(\theta,\kappa)$ sums the numerator over $\mathcal A_t$, $\theta$ collects payoff parameters, and $\kappa\geq0$ measures the penalty on failed checks. Characteristics and the past collaboration network $G^U_{<t}$ are conditioned on throughout. Setting $\kappa=0$ gives the unconstrained model; positive values discount profiles with failed empirical participation checks. This law supports the identification analysis below and the likelihood in Section~\ref{sec:full-likelihood}. It retains positive probability on every state at finite $\kappa$, including states that are not stable equilibria.

\FloatBarrier

\subsection{Identification}
\label{sec:structural-identification}

Identification asks whether different parameters can produce the same conditional distributions of project assignments. I first describe the variation that separates activity incentives, collaborator preferences, and participation penalties, then state the conditions for local identification and for distinguishing the penalty from broader additive payoffs.

\subsubsection{Intuition for Identification}
\label{sec:identification-intuition}

Individual project counts and their variation over time inform productivity and congestion. Solo configurations are particularly useful because they contain no team benefits, pair terms, or consent penalties. Conditional on these activity incentives, changes in team size inform $V_{\mathrm{size},t}$, and differences in section composition among equally sized teams inform $V_{\mathrm{section}}$. Conditional membership odds in the size-only model make this distinction explicit (see the online appendix). These arguments require variation over the relevant project loads and team sizes, as detailed in the appendix.

Partner choice adds information beyond project counts and team sizes. Productivity gaps and tenure-pair composition inform $w_t$; repeated projects within a year inform $C_G$, while the continuation of past relationships informs $\omega^{\mathrm{past}}$. Some payoff components can overlap: a common pair value or relationship cost can be absorbed by sufficiently flexible team-size values. Specification restrictions and variation in repeated relationships are therefore necessary to distinguish these channels (see the appendix).

Sorting and consent can favor similar teams but affect their probabilities differently. Conditional on the individual effects, productivity sorting has coefficient $\gamma:=\omega^{\mathrm{gap}}$ multiplying $G_t(A):=\sum_{i<j}g_{ij}(\beta_i-\beta_j)^2$ in the potential. Comparing assignments with different $G_t$ and the same violation count isolates this contribution once other payoffs are accounted for. Such comparisons can involve teams whose participation checks all pass, so the consent penalty is zero.

The participation penalty is instead informed by changes in the outside-option comparisons. Holding team surplus and collaborator characteristics fixed, a higher project load raises congestion and can make the adjusted exit margin negative. A change in the own-project return can likewise move a membership across its empirical exit threshold. One informative case is a membership that passes the solo-replacement check but fails the adjusted exit check. After accounting for the own-project payoff, these contrasts reveal the penalty on failed checks. Conditional on payoffs, $\kappa$ compares observed failures with those predicted by the model. Separating it jointly from sorting requires variation that the other payoff terms cannot reproduce. Workload variation can contribute, provided its direct effects on activity and collaboration are also accounted for.

Participation checks can also be distinguished from a broader class of additive payoffs. They link a member's benefits within a particular team to opportunities elsewhere in her portfolio. Proposition~\ref{prop:participation-nonabsorption} uses two configurations with the same individual project counts and weighted collaboration network, then adds the same solo projects to each. Even with arbitrary portfolio costs, project-composition values, and network payoffs, the difference between the two potentials is unchanged by these additions. The difference in violation counts can change, however, because portfolio congestion moves particular memberships across their participation thresholds. When it does, no unconstrained model in that additive class can reproduce the penalized stationary distribution.

Sorting coefficients describe preferences conditional on the included characteristics, history, and payoff specification. Unobserved project opportunities, common interests, or persistent pair affinity can produce similar matching patterns, and assigned team formation is not modeled explicitly. The online appendix provides additional analysis of sorting and participation constraints and discusses identification in large networks.

\subsubsection{Formal Results}
\label{sec:identification-formal}

Fix the slot universe, population, payoff scale, and reference normalizations above. For $\eta=(\theta,\kappa)$, write the log stationary weight as $F_t(A;\eta)=Q_t(A;\theta)-\kappa\bar{\mathcal V}_t(A;\theta)$ and the law in \eqref{eq:pi-t-constrained} as $\bar\pi_t(A;\eta)$. Comparing two states in the same year eliminates the normalizing constant:
\begin{equation}
\label{eq:sorting-consent-log-ratio}
\log\frac{\bar\pi_t(A;\eta)}{\bar\pi_t(B;\eta)}
=Q_t(A;\theta)-Q_t(B;\theta)
-\kappa\bigl[\bar{\mathcal V}_t(A;\theta)-\bar{\mathcal V}_t(B;\theta)\bigr].
\end{equation}
Two parameter vectors are observationally equivalent exactly when their $F_t$ functions differ only by a state-independent constant in each year. For a fixed collection of statistics, the exponential-family representation identifies their coefficients when no nonzero linear combination of the statistics is constant within every year \citep{WainwrightJordan2008}. This is linear independence across configurations; the statistics may be correlated. Under this condition, a penalty statistic cannot be absorbed by reweighting the payoff statistics. The empirical specification maps structural parameters nonlinearly into the potential, and violation classifications can change with payoffs. The local rank condition below accounts for these features and requires independent variation after all nuisance payoff scores are included.

The following criterion makes this requirement precise for count weighting. It concerns the conditional population distributions and does not establish precision from one observed network per year.
\begin{proposition}[Local identification of payoffs and participation constraints]
\label{prop:sorting-consent-local-identification}
Fix the normalizations and state spaces above. Consider an interior parameter vector $\eta_0=(\theta_0,\kappa_0)$ at which $Q_t$ is continuously differentiable and every violation indicator is locally constant as a function of $\theta$, for all candidate states. Define the column vector
\begin{equation}
\label{eq:sorting-consent-information}
D_t(A)=
\begin{pmatrix}\nabla_\theta Q_t(A;\theta_0)\\-\bar{\mathcal V}_t(A;\theta_0)\end{pmatrix},
\qquad
\mathcal I(\eta_0)=\sum_t\operatorname{Cov}_{\bar\pi_t(\cdot;\eta_0)}\!\bigl(D_t(A)\bigr).
\end{equation}
If $\mathcal I(\eta_0)$ is positive definite, the collection of conditional stationary distributions locally identifies all free components of $\eta$. Conditional on the other payoff parameters, separating $\gamma$ and $\kappa$ requires independent variation in $G_t$ and $\bar{\mathcal V}_t$ up to year-specific constants. With nuisance payoff parameters unknown, the regular rank criterion requires these two statistics to remain independent after their variation explained by the nuisance score columns is removed.
\end{proposition}

The proof and probability contrasts separating sorting from participation penalties are in Appendix~\ref{app:sorting-consent-identification}. The condition is local to a region with unchanged violation indicators; it does not establish identification across different violation patterns. If failures do not vary across states, $\kappa$ is uninformative, and dependence between the sorting and violation statistics can prevent their separation. Depth weighting requires a separate assessment. Stationary observations also do not identify the consideration bound $k$, since the stationary law is independent of it.

The second result distinguishes participation penalties from a broader class of additive payoff specifications. Fix a year and suppress its index. Consider the class
\begin{equation}
\label{eq:additive-payoff-class}
\widetilde Q(A)
=\sum_i f_i\bigl(p_i(A)\bigr)
+\sum_r\widetilde\Phi\bigl(N_r(A)\bigr)
+\Psi\bigl(G^W(A)\bigr),
\end{equation}
where the portfolio functions $f_i$, project-composition function $\widetilde\Phi$, and network function $\Psi$ are unrestricted, with $\widetilde\Phi(\varnothing)=0$. Characteristics and past collaboration are held fixed and may enter these functions. This class contains the potential in \eqref{eq:potential_structural}, while allowing arbitrary project-composition and network values in place of the estimated functions.

For four feasible configurations, define the linear contrast
\begin{equation}
\label{eq:participation-four-profile-contrast}
\mathcal R h
:=
h(A_2)-h(B_2)-h(A_0)+h(B_0).
\end{equation}
\begin{proposition}[Participation penalties beyond additive payoffs]
\label{prop:participation-nonabsorption}
Fix the labeled state space, unit payoff-noise scale, and parameters $(\theta_0,\kappa_0)$ with $\kappa_0>0$ and $Q(\cdot;\theta_0)$ in the class \eqref{eq:additive-payoff-class}. Let $A_0$ and $B_0$ have the same individual project counts and weighted collaboration network. Form $A_2$ and $B_2$ by adding two distinct fresh solo projects for the same individual to each, leaving all existing memberships fixed. Suppose all four configurations are feasible and
\[
\mathcal R\bar{\mathcal V}(\cdot;\theta_0)\ne0.
\]
Then no unconstrained logit model with a potential $\widetilde Q$ in \eqref{eq:additive-payoff-class} generates the stationary law proportional to $\exp\{Q(\cdot;\theta_0)-\kappa_0\bar{\mathcal V}(\cdot;\theta_0)\}$.
\end{proposition}

The proof is in Appendix~\ref{app:sorting-consent-identification}, where Example~\ref{ex:portfolio-team-consent} constructs this contrast with six economists. The argument also allows network payoffs for cross-section pairs with no prior collaboration: such terms cannot generally absorb the participation penalty, although they can resemble it on a restricted set of dyads. The example establishes structural nonredundancy, rather than verifying the condition at the estimated parameters or establishing precision in this sample. Appendix~\ref{app:identification-verification} reports the checks at the fitted parameters. An unrestricted payoff expansion can reproduce the weights by adding the term $-\kappa\bar{\mathcal V}_t(A;\theta)$ itself. Under the proposition's condition, that term introduces dependence beyond the additive class. The identification claim therefore distinguishes the penalty from the admitted payoff components; its interpretation as stochastic consent follows from the behavioral assumptions.

\FloatBarrier

\section{Estimation and Empirical Results}
\label{sec:estimation}

First, I introduce the estimation method. I then present the parameter estimates and assess model fit, using a comparison with a model re-estimated without participation penalties to evaluate the role of stability constraints. Finally, I examine how policy workload affects research activity and collaboration, providing the starting point for the counterfactual experiments in Section~\ref{sec:counterfactuals}.

\subsection{Bayesian Estimation}
\label{sec:full-likelihood}
\label{sec:full-likelihood-constrained}
\label{sec:bayesian-estimation}
\label{sec:structural-estimation}

I use likelihood-based Bayesian estimation implemented through a double Metropolis--Hastings algorithm. At each proposed parameter value, the algorithm simulates project assignments for each year using Markov chains that approximate draws from the model's stationary distribution. It evaluates the observed and simulated assignments at the current and proposed parameters to decide whether to accept the proposal, avoiding direct evaluation of the likelihood's intractable normalizing constants. The appendix provides the full algorithm and implementation details.

The likelihood uses the stationary law \eqref{eq:pi-t-constrained} for each year's participation matrix, including zero-output years for sample economists at risk. It conditions on observed rosters, characteristics, workload, and collaboration history before the year. The log-likelihood is
\begin{equation}
\label{eq:full-likelihood-constrained}
\ell(\theta,\kappa)=\sum_{t=2007}^{2019}\Bigl[Q_t(A_t\mid G^U_{<t};\theta)
-\kappa \bar{\mathcal V}_t(A_t\mid G^U_{<t};\theta)-\log Z_t(\theta,\kappa)\Bigr].
\end{equation}
Each year has its own normalizing constant because its population, workload, and history differ. Individual effects are common across years and are learned jointly from research activity, matching, and participation checks.

I use the hierarchical normal prior for individual effects described in Section~\ref{sec:baseline}, with a flat prior on $\mu_\beta$ and an inverse-gamma$(2,1)$ prior on $s_\beta^2$. Common parameters have independent normal priors with the required support restrictions; the appendix gives their distributions. Write $\eta=(\theta,\kappa)$, with $\theta$ including the individual effects, and $F_t(A;\eta)=Q_t(A;\theta)-\kappa\bar{\mathcal V}_t(A;\theta)$. Conditional on the hierarchy, the posterior is proportional to the prior times $\exp\{\ell(\eta)\}$. Direct evaluation is infeasible because each $Z_t$ sums over an enormous number of assignments.

The exchange algorithm avoids evaluating these constants \citep{murray2006exchange,liang2010double}. After proposing $\eta'$, it independently draws an auxiliary network $A'_t$ for each year from that year's stationary law at the proposed parameters, keeping the observed past fixed. The current and proposed parameters are then evaluated on both the observed and auxiliary networks. Writing $p(\eta)$ for the prior conditional on the hierarchy and $q$ for the proposal density, the exchange ratio is
\begin{equation*}
\bar a=\frac{p(\eta')q(\eta\mid\eta')}{p(\eta)q(\eta'\mid\eta)}\exp\!\left\{\sum_t\left[F_t(A_t;\eta')-F_t(A_t;\eta)+F_t(A'_t;\eta)-F_t(A'_t;\eta')\right]\right\}.
\end{equation*}
The proposal is accepted with probability $\min\{1,\bar a\}$. With exact auxiliary draws, this update preserves the posterior. The double Metropolis--Hastings implementation approximates those draws by finite inner chains started from the observed networks. Because payoffs are nonlinear in the parameters and failed checks change at participation thresholds, both $Q_t$ and $\bar{\mathcal V}_t$ are recomputed at each parameter value.

The inner sampler targets the penalized stationary law directly. It combines single-cell Gibbs updates with meetings that reconsider several project slots and use a Metropolis correction for state-dependent targeting. Theorem~\ref{thm:unconstrained-logit-speed-comparison} motivates larger Gibbs blocks, while the cost and performance of the targeted kernels are assessed numerically. The reported meeting law is $(\xi,\bar K,\eta_{\mathrm{own}})=(0.25,3,0.9)$, as calibrated in the appendix. The appendix records a sweep of the algorithm and explains the parameter blocks and numerical settings. These computational updates share the behavioral dynamic's stationary distribution; they need not reproduce its proposal-and-consent paths.

\FloatBarrier

\subsection{Parameter Estimates}
\label{sec:estimation-results}

Table~\ref{tab:k-posterior} reports the posterior of the collaboration model in Section~\ref{sec:structural}. Intervals quoted below contain 90\% of the posterior. Workload coefficients and their implications for simulated outcomes are examined together in Section~\ref{sec:response-workload}.

\begin{table}[!htbp]
\centering
\caption{Posterior of the Collaboration Model}
\label{tab:k-posterior}
\footnotesize
{\renewcommand{\arraystretch}{1.3}%
\begin{tabular}{@{}lrrrrrrrr@{}}
 & & & & & & & \multicolumn{2}{c@{}}{ESS} \\
 & Median & Mean & s.d. & 5\% & 95\% & $\hat R$ & Bulk & Tail \\ \hline\hline
$\psi_P$ & 0.068 & 0.072 & 0.042 & 0.011 & 0.151 & 1.03 & 217 & 367 \\
$\beta^{\mathrm{FKGL}}$ & $-$0.415 & $-$0.416 & 0.094 & $-$0.575 & $-$0.262 & 1.02 & 299 & 578 \\
$\beta^{\mathrm{FKGL}}_S$ & 0.216 & 0.215 & 0.121 & 0.014 & 0.412 & 1.02 & 497 & 1{,}132 \\
$\beta^{\mathrm{FKGL}}_{\mathrm{MA}}$ & 0.208 & 0.208 & 0.119 & 0.017 & 0.403 & 1.02 & 283 & 670 \\
$\beta^{\mathrm{title}}_{\mathrm{M\&O}}$ & $-$0.534 & $-$0.535 & 0.068 & $-$0.651 & $-$0.425 & 1.05 & 194 & 555 \\
$\beta^{\mathrm{title}}_{\mathrm{Adv}}$ & $-$0.818 & $-$0.818 & 0.159 & $-$1.082 & $-$0.558 & 1.00 & 1{,}399 & 4{,}555 \\
$v_{\mathrm{size}{,}2}$ & $-$0.112 & $-$0.116 & 0.037 & $-$0.185 & $-$0.066 & 1.04 & 250 & 277 \\
$\rho_{\mathrm{size}}$ & 2.495 & 2.508 & 0.201 & 2.199 & 2.852 & 1.03 & 293 & 381 \\
$v^{\mathrm{FKGL}}_{\mathrm{size}{,}2+}$ & 0.442 & 0.447 & 0.128 & 0.245 & 0.663 & 1.01 & 369 & 914 \\
$v_{\mathrm{section}{,}2}$ & 2.287 & 2.286 & 0.095 & 2.131 & 2.440 & 1.02 & 266 & 829 \\
$v_{\mathrm{section}{,}3+}$ & 1.611 & 1.609 & 0.166 & 1.330 & 1.877 & 1.02 & 413 & 835 \\
$\omega^{\mathrm{past}}$ & 2.355 & 2.356 & 0.063 & 2.253 & 2.457 & 1.02 & 381 & 726 \\
$\omega^{\mathrm{gap}}$ & $-$0.118 & $-$0.123 & 0.097 & $-$0.296 & 0.029 & 1.03 & 149 & 259 \\
$\omega^{\mathrm{tenure}}_{EE}$ & $-$0.157 & $-$0.170 & 0.123 & $-$0.393 & 0.007 & 1.01 & 469 & 894 \\
$\omega^{\mathrm{tenure}}_{EM}$ & $-$0.026 & $-$0.026 & 0.077 & $-$0.154 & 0.100 & 1.02 & 318 & 515 \\
$\omega^{\mathrm{tenure}}_{ES}$ & $-$0.051 & $-$0.054 & 0.096 & $-$0.221 & 0.104 & 1.04 & 267 & 513 \\
$\omega^{\mathrm{tenure}}_{MS}$ & 0.295 & 0.294 & 0.061 & 0.193 & 0.392 & 1.03 & 343 & 727 \\
$\omega^{\mathrm{tenure}}_{SS}$ & $-$0.002 & $-$0.001 & 0.085 & $-$0.141 & 0.139 & 1.02 & 277 & 729 \\
$\omega^{\mathrm{FKGL}}_{EM}$ & 0.088 & 0.087 & 0.196 & $-$0.239 & 0.406 & 1.00 & 1{,}131 & 3{,}345 \\
$\psi_{G{,}2}$ & $-$2.197 & $-$2.195 & 0.104 & $-$2.364 & $-$2.019 & 1.03 & 340 & 807 \\
$\kappa$ & 0.342 & 0.342 & 0.052 & 0.256 & 0.426 & 1.05 & 161 & 440 \\
\hline
$\mu_\beta$ & $-$6.279 & $-$6.278 & 0.045 & $-$6.351 & $-$6.204 & 1.03 & 163 & 338 \\
$s_\beta$ & 0.491 & 0.492 & 0.030 & 0.443 & 0.542 & 1.01 & 370 & 1{,}054 \\
\hline
\end{tabular}}
\par\smallskip
\begin{minipage}{0.97\textwidth}\floatnotesfont
Notes. The 21 free common parameters and the hierarchy of individual productivity. Posterior summaries pool 108{,}000 post-adaptation draws from eight chains, 13{,}500 per chain. 5\% and 95\%: posterior quantiles. $\hat R$: the larger of rank-normalized split-$\hat R$ and its folded version; ESS: bulk and tail effective sample sizes \citep{vehtari2021rank}. The appendix lists the fixed parameters and provides run details and posterior-density plots.
\end{minipage}
\end{table}

Individual productivity is dispersed, with $\mu_\beta=-6.28$ per slot and $s_\beta=0.49$. Its location reflects the 400 labeled slots, while the dispersion captures persistent differences in research activity. Congestion beyond the fixed threshold of two projects is modest, $\psi_P=0.07$ $[0.01,0.15]$. Managers and officers have a lower own-project return in the years they hold the title, $\beta_{\mathrm{M\&O}}^{\mathrm{title}}=-0.53$ $[-0.65,-0.43]$, and advisers a lower one still, $\beta_{\mathrm{Adv}}^{\mathrm{title}}=-0.82$ $[-1.08,-0.56]$. Title shifts are learned primarily from economists who change title; persistent differences can be absorbed by the individual effects.

The baseline value of a team falls with its size. Evaluating the size profile at the posterior means $v_{\mathrm{size},2}=-0.12$ and $\rho_{\mathrm{size}}=2.51$ gives values of about $-0.12$, $-0.29$, $-0.73$, and $-1.8$ for teams of two to five at average workload. Same-section teammates can offset these costs: one adds $v_{\mathrm{section},2}=2.29$ $[2.13,2.44]$, while two or more give $v_{\mathrm{section},3+}=1.61$ $[1.33,1.88]$.

Repeated relationships are especially valuable. Past collaboration adds $\omega^{\mathrm{past}}=2.36$ $[2.25,2.46]$ once per active pair, and a second common project within the year is rewarded, with $\psi_{G,2}=-2.20$ $[-2.36,-2.02]$. The clearest tenure complementarity is between mid-career and senior economists, $\omega_{MS}^{\mathrm{tenure}}=0.30$ $[0.19,0.39]$, relative to the mid-career pair reference. Entrant-pair values are close to zero or negative. Evidence of productivity sorting is weaker: $\omega^{\mathrm{gap}}=-0.12$ $[-0.30,0.03]$, with posterior probability 0.92 of a negative coefficient.

The participation penalty is $\kappa=0.34$ $[0.26,0.43]$. Holding the potential fixed, an additional failed membership check multiplies a profile's stationary weight by $e^{-\kappa}\approx0.71$. This parameter has a role beyond the payoff terms if it accounts for patterns that the model cannot reproduce by changing those terms. The following comparison evaluates that contribution.

\FloatBarrier

\subsection{Model Fit and Participation Constraints}
\label{sec:model-fit-consent}

Posterior simulations reproduce research output, its distribution across economists and careers, larger teams, and much of the collaboration network. The data lie inside the 90\% posterior predictive intervals for 34 of 42 statistics (see the appendix). The principal misses are the split between papers with one and two Board authors and the output of entrants. External collaborators are dropped from the modeled network: 71\% of papers with a single Board author have coauthors outside the Board. Papers classified as solo can therefore be collaborative, making discrepancies in the split between solo papers and small Board teams less concerning as evidence about collaboration more broadly. Large teams of entrants may reflect assignment that the model does not represent.

To assess consent, I compare the estimated model with the same model re-estimated at $\kappa=0$, allowing all other parameters to adjust. The sample, controls, and priors of those parameters are held fixed. Both models reproduce participation, and the unconstrained model places 38 of the 42 conventional statistics inside their predictive intervals. It also fits the split between one and two Board authors better. Aggregate activity and team composition therefore provide limited discrimination between the two mechanisms.

\begin{table}[!htbp]
\centering
\caption{Model Fit with and without Participation Penalties}
\label{tab:consent-fit-main}
\footnotesize
\setlength{\tabcolsep}{3.5pt}
\begin{tabular}{@{}lrrr@{}}
Statistic & Data & Estimated $\kappa$ & $\kappa=0$, re-estimated \\ \hline\hline
Failed memberships: observed & --- & 409 & 451 \\
Failed memberships: simulated & --- & 406 [344, 477] & 569 [511, 639] \\
Simulated minus observed & --- & $-$1 [$-$62, 70] & 118 [57, 191] \\
Authorships & 2{,}691 & 2{,}696 [2{,}581, 2{,}832] & 2{,}711 [2{,}580, 2{,}839] \\
Papers & 1{,}918 & 1{,}991 [1{,}911, 2{,}066] & 1{,}959 [1{,}870, 2{,}034] \\
Economist-years with a paper (\%) & 49.3 & 49.6 [47.8, 51.2] & 49.6 [47.9, 51.4] \\
Statistics inside 90\% interval (of 42) & --- & 34 & 38 \\
\hline
\end{tabular}
\par\smallskip
\begin{minipage}{0.97\textwidth}\floatnotesfont
Notes. Selected results from a fuller comparison in the appendix; medians and 90\% posterior predictive intervals from 500 joint states per model, conditional on observed past collaboration. Failures in observed networks are evaluated at each model's own payoffs and averaged over its states, so their counts differ between columns. Simulated-minus-observed differences are computed within each state before summarizing. Other data counts are common to both models. The final row counts the conventional fit statistics reported in the appendix that lie inside their predictive intervals.
\end{minipage}
\end{table}

The models differ in the frequency of failed empirical participation checks. The estimated model generates 406 failed memberships, compared with 409 in the observed networks evaluated at its payoffs. Conditional on payoffs, the likelihood score for $\kappa$ is expected minus observed failures, summed across years; this agreement is therefore fit to the statistic directly informing $\kappa$, rather than independent validation of consent. The re-estimated unconstrained model generates 569 failures, compared with 451 observed at its own payoffs; its paired excess is 118 $[57,191]$. Violation classifications depend on payoffs, so each simulation is compared with the observed network evaluated at the same parameter draw. Table~\ref{tab:consent-fit-main} summarizes these comparisons, and the appendix reports the full results and diagnostics.

With $\kappa$ fixed at zero, the re-estimated model relies more on same-section ties, past collaboration, and productivity similarity, while lowering the base team value. These adjustments fit activity and several composition moments but leave an excess of failed empirical checks. The restricted-model comparison supports a separate contribution of the participation penalty within the maintained payoff specification, consistent with the identification argument in Section~\ref{sec:structural-identification}. It does not distinguish every unrestricted payoff expansion that could generate the same weights. Fitting the participation statistic also accounts for part of the shortfall of two-author papers.

Removing the penalty while holding all payoff parameters fixed gives 752 simulated members in violation, compared with 406 at the estimated $\kappa$. This is a mechanism comparison rather than re-estimation; the appendix examines the associated changes in output and collaboration. The count rule targets the number of failed checks and does not ensure a fit to their depths, which are examined separately in the appendix.

\FloatBarrier

\subsection{Response to Policy Workload}
\label{sec:response-workload}

The clearest response to higher policy workload is a shift from solo projects toward teams. Raising centered FKGL from $-0.6$ to $+0.6$ in every year lowers solo projects by 24\% and raises team memberships by 22\%, with the latter increasing in 93\% of the 500 joint posterior states. The team share rises from 0.41 to 0.52, about 12 percentage points, with a 90\% posterior interval of 6 to 18 points. Total participations fall by 5.5\% on average, but their interval ranges from a 17\% fall to a 6\% rise. These simulations hold the observed collaboration history fixed in each year.

Three channels explain the response. Workload changes the return to every project through the own-project terms (D), the value of teams and entrant--mid-career pairs (T), and the resulting participation margins (C). To separate their contributions, I vary the workload input to the checks separately from the payoff potential and average contributions over the orders in which the three channels are switched from low to high workload. These Shapley contributions sum to the total effect. Table~\ref{tab:fkgl-decomp} reports means and 90\% intervals over the 500 joint posterior states; the appendix gives the margin adjustments, simulation protocol, and additional results.

\begin{table}[!htbp]
\centering
\caption{Response to Higher Policy Workload and Its Channels}
\label{tab:fkgl-decomp}
\footnotesize
\setlength{\tabcolsep}{3.5pt}
\begin{tabular}{l rr r rrr}
\hline
 & \multicolumn{2}{c}{Level} & & \multicolumn{3}{c}{Channel} \\
 & FKGL $-0.6$ & $+0.6$ & Total & D: direct & T: team value & C: cut-offs \\
\hline
Participations & 2{,}843 & 2{,}686 & $-$157 & $-$1{,}210 & 984 & 69 \\
 & & & $[-482,169]$ & $[-1{,}964,-599]$ & $[527,1{,}627]$ & $[38,100]$ \\[2pt]
Solo projects & 1{,}688 & 1{,}278 & $-$410 & $-$241 & $-$151 & $-$17 \\
 & & & $[-566,-252]$ & $[-325,-156]$ & $[-241,-84]$ & $[-26,-9]$ \\[2pt]
Team memberships & 1{,}156 & 1{,}408 & 253 & $-$969 & 1{,}135 & 87 \\
 & & & $[-20,554]$ & $[-1{,}658,-446]$ & $[620,1{,}876]$ & $[48,126]$ \\[2pt]
Team share & 0.406 & 0.523 & 0.117 & $-$0.115 & 0.213 & 0.019 \\
 & & & $[0.058,0.183]$ & $[-0.173,-0.055]$ & $[0.129,0.311]$ & $[0.011,0.027]$ \\[2pt]
Members in violation & 442 & 342 & $-$100 & $-$266 & 270 & $-$105 \\
 & & & $[-169,-41]$ & $[-409,-142]$ & $[153,416]$ & $[-147,-64]$ \\[2pt]
\quad of the solo condition & 432 & 111 & $-$321 & $-$182 & 200 & $-$338 \\
 & & & $[-425,-209]$ & $[-263,-111]$ & $[115,296]$ & $[-451,-210]$ \\[2pt]
\quad of the exit condition & 348 & 323 & $-$25 & $-$239 & 246 & $-$32 \\
 & & & $[-59,10]$ & $[-376,-125]$ & $[135,389]$ & $[-84,15]$ \\[2pt]
\hline
\end{tabular}
\par\smallskip
\begin{minipage}{0.97\textwidth}\floatnotesfont
Notes. Centered FKGL is raised from $-0.6$ to $+0.6$ in each year, conditional on observed past collaboration. Outcomes are summed over 2007--2019. First line: mean across 500 joint posterior states; second line: 5\% and 95\% quantiles. D changes own-project returns, T team and pair values, and C the participation margins. Shapley contributions average over channel orderings and sum to the total. Team share is the fraction of participations in projects with two or more Board authors. The appendix gives the simulation and decomposition details.
\end{minipage}
\end{table}

The direct and team-value channels largely offset. For an entrant or mid-career economist outside Monetary Affairs, higher workload lowers the own-project return, $\beta^{\mathrm{FKGL}}=-0.41$ $[-0.58,-0.26]$, while raising the value of team participation, $v_{\mathrm{size},2+}^{\mathrm{FKGL}}=0.44$ $[0.24,0.66]$. Their joint effect on the return to a team membership is close to zero, $0.03$ $[-0.13,0.21]$, while solo work loses value. The direct channel alone reduces participation by about 30\%; the team-value channel offsets most of this loss and draws activity toward teams. The two contributions interact and are uncertain individually, with correlation $-0.89$ across the posterior states. Their sum is better determined than their separate magnitudes.

Responses differ across economists. Senior tenure and Monetary Affairs each add a positive shift to the own-project workload coefficient, so the total response is about $-0.20$ for either group separately and $0.01$ $[-0.25,0.26]$ for senior economists in Monetary Affairs (Table~\ref{tab:fkgl-groups}). For groups whose own return falls less, the positive team-value effect also raises the return to team participation. The near-zero own-return response of Monetary Affairs seniors is the sum of two estimated shifts, since the specification has no separate senior-by-division interaction. Its interpretation therefore depends on this additive restriction and the limited information for that group. The workload interaction for entrant--mid-career pairs is imprecise, $\omega_{EM}^{\mathrm{FKGL}}=0.09$ $[-0.24,0.41]$.

\begin{table}[!htbp]
\centering
\caption{Response of the Own-Project Return to Workload, by Group}
\label{tab:fkgl-groups}
\footnotesize
\setlength{\tabcolsep}{3.3pt}
\begin{tabular}{@{}lccrrr@{}}
 & Person-years & Model: total & \multicolumn{3}{c}{Data: within-person slope on FKGL$_t$} \\
Group & (economists) & coefficient & All projects & Team & Solo \\ \hline\hline
Entrant/mid-career, not MA & 1{,}957 (335) & $-0.41$ $[-0.58,-0.26]$ & $-0.23$ (0.07) & $-0.21$ (0.05) & $-0.01$ (0.05) \\
Senior, not MA & \phantom{1{,}}924 (142) & $-0.20$ $[-0.41,\phantom{-}0.01]$ & $-0.00$ (0.08) & $-0.03$ (0.07) & $0.02$ (0.04) \\
Entrant/mid-career, MA & \phantom{1{,}}436 \phantom{0}(80) & $-0.20$ $[-0.44,\phantom{-}0.01]$ & $0.10$ (0.11) & $-0.10$ (0.06) & $0.20$ (0.10) \\
Senior, MA & \phantom{1{,}}205 \phantom{0}(37) & $\phantom{-}0.01$ $[-0.25,\phantom{-}0.26]$ & $0.19$ (0.22) & $0.31$ (0.18) & $-0.12$ (0.11) \\ \hline
\end{tabular}
\par\smallskip
\begin{minipage}{0.97\textwidth}\floatnotesfont
Notes. Model: posterior median and 90\% interval of $\beta^{\mathrm{FKGL}}+\beta_S^{\mathrm{FKGL}}\1\{z_{it}=S\}+\beta_{\mathrm{MA}}^{\mathrm{FKGL}}\mathrm{MA}_{it}$, the coefficient on $\mathrm{FKGL}_t$ in the own-project return \eqref{eq:own_return_structural}, from the draws of Table~\ref{tab:k-posterior}; entrants share the mid-career coefficient, $\beta_E^{\mathrm{FKGL}}=0$. Data: slope of $i$'s number of projects in year $t$ (all, teams of two or more Board economists, solo) on $\mathrm{FKGL}_t$, with person fixed effects, in the person-years of the group; standard errors clustered by person in parentheses. Groups by tenure in year $t$ (senior: eight or more years on the Board) and division in year $t$ (MA: Monetary Affairs). The standard deviation of $\mathrm{FKGL}_t$ across the thirteen years is 0.38.
\end{minipage}
\end{table}

The consent-margin channel makes an additional contribution. Higher team value raises the solo margin; the exit margin also depends on the economist's own-project workload response. In the decomposition, changing the margins adds 87 team memberships and 1.9 percentage points to the team share, about 16\% of its total increase, with an interval of 9--30\%. This is the effect of changing willingness to remain while holding the payoff potential fixed for that channel, rather than a change in $\kappa$.

The workload input varies across only thirteen years. The descriptive within-person slopes in Table~\ref{tab:fkgl-groups} include changes in partners and team sizes, and the unadjusted cross-year relationship between workload and team share has the opposite sign from the model's conditional response. The positive simulated response depends on accounting for changing rosters, tenure, and past relationships. Entrants' large teams remain a limitation of this interpretation. Section~\ref{sec:counterfactuals} next allows simulated collaborations to affect future years and asks how the timing and allocation of workload change these outcomes.

\FloatBarrier

\section{Counterfactual Experiments}
\label{sec:counterfactuals}

I use the estimated model to examine how the timing and allocation of policy workload shape research output and collaboration over 2007--2019. I first vary the sequence of high- and low-workload years, allowing collaborations formed in one year to affect subsequent choices. I then shift policy work toward more experienced economists while holding its total over the period fixed.

\subsection{Experimental Design}
\label{sec:cf-setup}
\label{sec:cf-paths}

I simulate alternative workload paths while holding staffing at its observed values and allowing collaboration histories to evolve. In each year, economists retain their observed sections, tenure, Monetary Affairs affiliations, and titles. Workload replaces observed FKGL in own-project returns, team and pair values, and the resulting participation margins. Past collaboration includes ties formed before 2007 and in earlier simulated years of the same history. This allows workload to affect future research through the value of repeated collaboration.

\begin{figure}[!htbp]
\centering
\includegraphics[width=\textwidth]{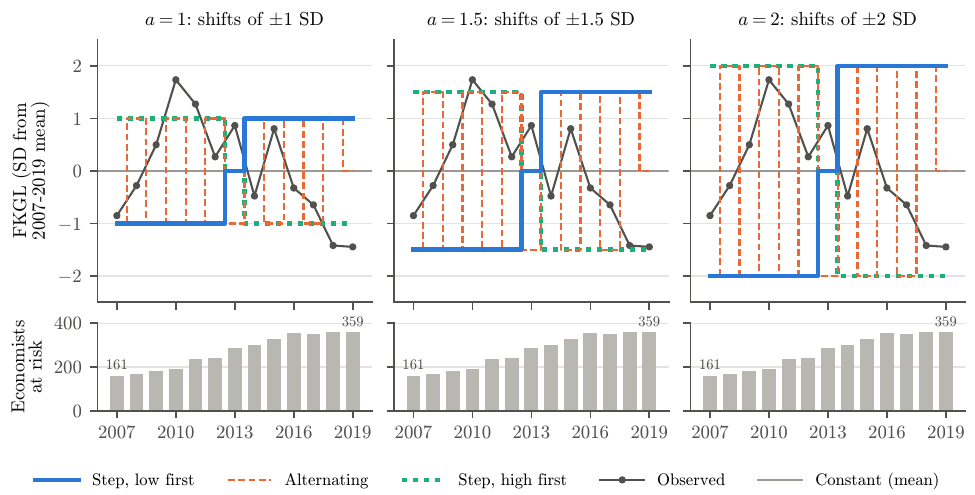}
\caption{Workload paths with amplitudes $a=1,1.5,2$ standard deviations from the 2007--2019 mean. Low-first and high-first paths reverse the order of six quiet and six busy years; alternating paths switch each year. Every synthetic path has mean zero. The observed and constant paths are repeated in each panel. Lower panels show the observed roster, which grows from 161 to 359 economists. Colors match Figure~\ref{fig:cf-totals}.}
\label{fig:cf-paths}
\end{figure}

The timing experiments compare a constant workload at the sample mean with paths containing six high-workload years, six low-workload years, and one year at the mean. Let $a\in\{1,1.5,2\}$ denote the amplitude in standard deviations, with $\sigma_F=0.400$ grade levels. The \emph{low-first} path puts workload at $-a\sigma_F$ in 2007--2012, at the mean in 2013, and at $+a\sigma_F$ in 2014--2019; the \emph{high-first} path reverses this order. The \emph{alternating} path switches between low and high workload from 2007 to 2018, with 2019 at the mean. These paths have the same distribution of workload over calendar years and differ only in its sequence. Figure~\ref{fig:cf-paths} also shows the observed path, and the appendix lists all paths year by year. Its range is $[-1.45,1.73]\sigma_F$: $a=1$ lies within it, $a=1.5$ extends slightly beyond its lower end, and $a=2$ extends beyond both ends.

Equal means across calendar years do not imply equal exposure across economists. The roster grows from 161 economists in 2007 to 359 in 2019, so high workload late in the period reaches more economists. Under the low-first path, average workload across economist-years is $+0.25a\sigma_F$, despite a zero calendar-year mean; the high-first path gives the opposite exposure. Alternating paths are almost balanced. A variant holding the roster and attributes at their 2013 values removes changes in staff size and composition.

Outcomes are summed over 2007--2019. Papers are distinct projects with at least one Board economist, and team size counts Board authors. Coauthoring pairs are counted in each year; repeat pairs have collaborated before that year. Each scenario uses 500 joint posterior draws and two simulated histories per draw, with common random numbers across scenarios. Effects are differences from the constant path, except in the tenure experiment, which uses its own flat-workload benchmark. Reported 90\% intervals approximate posterior uncertainty after adjustment for simulation noise; the appendix describes the simulation protocol, the uncertainty calculations, and the fit on the observed workload path.

\FloatBarrier

\subsection{Timing, Research Output, and Collaboration}
\label{sec:cf-efficiency}

\begin{figure}[!htbp]
\centering
\includegraphics[width=\textwidth]{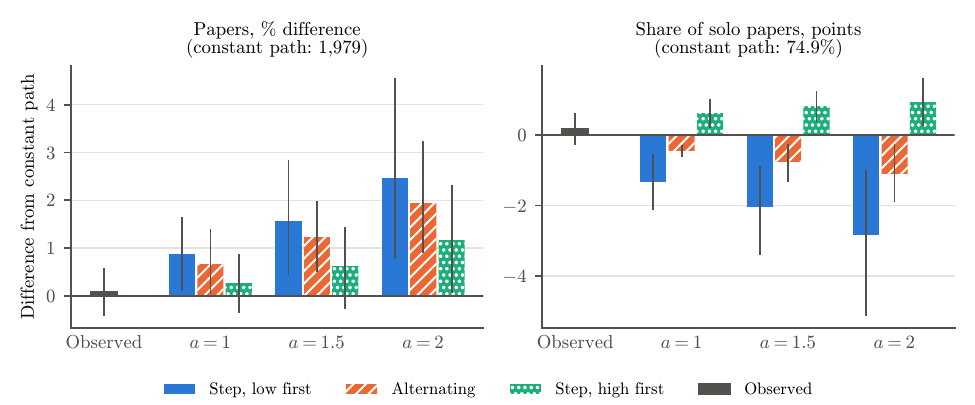}
\caption{Effects of workload timing on paper counts and the solo share over 2007--2019, relative to constant workload. Bars are mean effects across 500 joint posterior draws and two histories per draw; lines give approximate 90\% posterior intervals adjusted for simulation noise. Workload amplitudes $a$ are in standard deviations; paths are defined in Figure~\ref{fig:cf-paths}.}
\label{fig:cf-totals}
\end{figure}

Rearranging policy workload changes team composition much more than total research output. Figure~\ref{fig:cf-totals} shows that bursts raise paper counts by 0.3--0.9\% for $a=1$, 0.6--1.6\% for $a=1.5$, and 1.2--2.5\% for $a=2$, relative to constant workload. These are posterior mean effects; the smallest gain, under high-first workload at $a=1$, has an uncertain sign. The largest gain, 2.5\% under low-first workload at $a=2$, has a 90\% interval of 0.8--4.6\%. The gain persists when past ties have no value and when staffing is fixed. It reflects an asymmetric response within the year: without the value of past ties, a year at $-2\sigma_F$ produces about 15\% more papers than the same year at mean workload, while a year at $+2\sigma_F$ produces about 11\% fewer. The appendix reports these within-year comparisons and the effects in the low- and high-workload years of each path.

At $a=2$, placing low-workload years first lowers the solo share by 2.8 percentage points (90\% interval $-5.1$ to $-1.0$), raises the share of papers with four or more Board authors from 2.3\% to 3.7\%, and increases repeat coauthoring pairs by 38\%. The last effect is much less precisely determined, with an interval of 7--83\%. Alternating workload shifts composition in the same direction, with the solo share falling by 1.1 points. Placing high-workload years first raises it by 0.9 points and reduces coauthoring pairs by about 5\%. The composition ranking is the same at the smaller amplitudes. The appendix reports all outcomes, including comparisons with the observed path.

Staff growth explains part of this ranking because late high-workload years expose more economists to the incentive to collaborate. Turning off the value of past ties, $\omega^{\mathrm{past}}=0$, leaves this channel in place: at $a=2$, low-first workload still lowers the solo share by 1.9 points. Most of the shift is toward pairs, whose paper share rises by 1.0 point; the share of papers with four or more authors rises by only 0.2 points, compared with 1.4 points in the estimated model. Past collaboration therefore amplifies the response and supports larger teams.

Holding staffing fixed reverses the ranking of the two step paths. With the 2013 roster and attributes in every year, an early high-workload episode lowers the solo share by 1.9 points and raises coauthoring pairs by 17\% at $a=2$, compared with 0.6 points and 9\% when high workload comes late. Ties formed early have more years in which to be reused. The effect persists after workload returns to normal: in 2013, when both step paths are at the mean, the path following six high-workload years has 44 more coauthoring pairs than the path following six low-workload years, including 37 repeat pairs. This comparison isolates the role of prior collaboration within the model.

The largest collaboration responses occur among senior economists, economists in Monetary Affairs, managers, and advisers. Under low-first workload at $a=2$, these groups' authorships rise by 11--33\%, compared with falls of 2--12\% under high-first workload. Entrants and mid-career economists outside Monetary Affairs gain about 3--4\% of authorships on all three paths. The appendix reports the group comparisons. Predictions for entrants' collaboration require particular care because the model does not account for the apparent assignment of their large teams.

\FloatBarrier

\subsection{Allocating Policy Workload by Tenure}
\label{sec:cf-tenure}

Allocating more policy work to experienced economists shifts research toward entrants and produces a modest increase in total paper counts. The \emph{graded} rule sets workload at $0$, $1\sigma_F$, $1.5\sigma_F$, and $2\sigma_F$ for economists with 0--1, 2--4, 5--8, and nine or more years at the Board. The \emph{flat} rule gives everyone the graded rule's average over all economist-years: $1.12\sigma_F$, or 0.45 grade levels above the sample mean. Total workload is therefore equal over 2007--2019. It is not equal separately in each year: the graded rule's yearly average ranges from $0.93\sigma_F$ to $1.39\sigma_F$ as staff composition changes.

Individual workload enters each economist's own-project return. Pair values use the partners' average workload, while team values use the year's average across economists. These choices extend coefficients estimated from common annual workload to workloads that differ across economists, an assumption of this counterfactual. Since the yearly mean remains near the flat level, the graded rule primarily changes own-project returns. The appendix provides the workload profile, payoff formulas, and an alternative calculation holding team and pair values at the flat level.

The graded rule produces 1{,}923 papers, compared with 1{,}875 under flat workload, an increase of 2.6\% (90\% interval 1.0--4.3\%). Table~\ref{tab:cf-tenure} shows that aggregate composition changes little: the solo share is nearly unchanged, while the share with four or more Board authors falls from 3.6\% to 3.4\%. Repeat coauthoring pairs fall by 7\% on average, but the posterior probability of a decline is only 74\%.

\begin{table}[!htb]
\centering
\caption{Policy workload rising with tenure against flat workload at the same average, totals over 2007--2019.}
\label{tab:cf-tenure}
\footnotesize
\setlength{\tabcolsep}{5pt}
\begin{tabular}{l r r r r}
\hline
\multicolumn{5}{l}{\emph{A. All economists, totals over 2007--2019}} \\
 & Flat at average & Graded & Difference & (s.e.) \\
\hline
Papers & 1{,}875 & 1{,}923 & $+2.6\%$ & (0.03) \\
\quad share solo & 70.7\% & 70.8\% & $+0.1^{\dagger}$ & (0.03) \\
\quad share 2 authors & 19.5\% & 19.7\% & $+0.2$ & (0.03) \\
\quad share 3 authors & 6.2\% & 6.1\% & $-0.1$ & (0.02) \\
\quad share 4+ authors & 3.6\% & 3.4\% & $-0.3$ & (0.02) \\
Coauthoring pairs & 1{,}145 & 1{,}135 & $-0.9\%^{\dagger}$ & (0.29) \\
\quad repeat & 397 & 369 & $-7.0\%^{\dagger}$ & (0.56) \\
\hline
\end{tabular}

\medskip

\begin{tabular}{l rr rr rr rr}
\hline
\multicolumn{9}{l}{\emph{B. By years at the Board, totals over 2007--2019}} \\
 & \multicolumn{2}{c}{0--1 years} & \multicolumn{2}{c}{2--4 years} & \multicolumn{2}{c}{5--8 years} & \multicolumn{2}{c}{9+ years} \\
 & Flat & Diff. & Flat & Diff. & Flat & Diff. & Flat & Diff. \\
\hline
Papers per economist-year & 0.70 & $+18\%$ & 0.80 & $+3\%$ & 0.82 & $-5\%$ & 0.78 & $-7\%$ \\
\quad share solo & 65.5\% & $-1.0$ & 51.8\% & $-1.3$ & 45.6\% & $-0.6^{\dagger}$ & 36.0\% & $+0.6^{\dagger}$ \\
\quad share 2 authors & 24.6\% & $+0.8$ & 28.2\% & $+0.6$ & 27.8\% & $+0.5$ & 27.2\% & $+0.5$ \\
\quad share 3 authors & 7.5\% & $+0.2^{\dagger}$ & 12.3\% & $+0.3$ & 13.9\% & $+0.3$ & 16.6\% & $+0.2$ \\
\quad share 4+ authors & 2.4\% & $0.0^{\dagger}$ & 7.6\% & $+0.3^{\dagger}$ & 12.8\% & $-0.2^{\dagger}$ & 20.1\% & $-1.3$ \\
Coauthors per economist-year & 0.32 & $+20\%$ & 0.58 & $+6\%$ & 0.75 & $-4\%$ & 0.95 & $-9\%$ \\
\hline
\end{tabular}
\par\smallskip
\begin{minipage}{0.97\textwidth}\floatnotesfont
Notes. Graded workload is $0,1,1.5,2\,\sigma_F$ for 0--1, 2--4, 5--8, and nine or more years at the Board; flat workload is $1.12\sigma_F$. Pair values use mean partner workload and team values use mean yearly workload. Total workload is equal over the period. Panel A counts distinct papers and annual coauthoring pairs, summed over years; repeat pairs collaborated before the year. Changes are percentages for counts and percentage points for shares; s.e. is the simulation standard error. Panel B reports papers and coauthors per economist-year and shares of each group's authorships by Board team size. Group sizes average 72, 67, 55, and 77 economists at risk. Results average 500 posterior draws and two histories per draw. $^{\dagger}$ Sign probability below 75\% or within two simulation standard errors of zero; uncertainty calculations are in the appendix.
\end{minipage}
\end{table}

The redistribution across economists is substantially larger than the aggregate gain. Entrants author 18\% more papers per economist-year (90\% interval 10--26\%), and economists with 2--4 years gain 3\%. Economists with 5--8 years author 5\% fewer papers; those with nine or more years author 7\% fewer, with an 86\% posterior probability of a decline. The most experienced economists lose mainly large-team papers: the share of their authorships with four or more Board authors falls from 20.1\% to 18.8\%. Since entrants author a larger fraction of their papers alone, shifting research toward them helps explain why distinct papers rise more than total authorships, which increase by 1.9\%.

Collaboration also shifts toward entrants. Table~\ref{tab:cf-tenure-pairs} shows that the most experienced economists have 9\% fewer coauthors on average, although the 90\% interval, $-27$ to $+6\%$, includes an increase. The mean decline is concentrated among experienced partners: 14\% fewer coauthors with nine or more years and 11\% fewer with five to eight. Collaboration with entrants rises by 11\% (interval 1--24\%), increasing entrants' share of these economists' coauthors from 7.5\% to 9.3\%. Entrants gain coauthors in every tenure group, with a total increase of 20\% (interval 7--34\%) and the largest proportional gain among other entrants, 40\%.

These changes operate mainly through own-project returns and the availability of partners. Experienced economists facing heavier policy work give up marginal projects, while entrants with lighter workloads start more projects, some with experienced partners. Team values move with average yearly workload, so heavier individual workload does not directly raise the value of that individual's teams. Holding team and pair values at the flat level gives a similar output gain, 2.9\%, and the same collaboration pattern. The entrants' gains, especially their collaboration with one another, remain conditional on a model that does not represent the apparent assignment of large entrant teams.

\begin{table}[!htb]
\centering
\caption{Policy workload rising with tenure: coauthors by the partner's years at the Board, totals over 2007--2019.}
\label{tab:cf-tenure-pairs}
\footnotesize
\setlength{\tabcolsep}{5pt}
\begin{tabular}{l rrrrr}
\hline
 & & \multicolumn{4}{c}{Partner's years at the Board} \\ \cline{3-6}
 & All & 0--1 & 2--4 & 5--8 & 9+ \\
\hline
\multicolumn{6}{l}{\emph{A. Economists with 9 or more years at the Board}} \\
Coauthors per economist-year, flat & 0.95 & 0.07 & 0.18 & 0.23 & 0.47 \\
\quad graded, change (\%) & $-$9 & +11 & $-$1$^{\dagger}$ & $-$11 & $-$14 \\
\quad 90\% interval & [$-$27, +6] & [+1, +24] & [$-$14, +11] & [$-$26, +4] & [$-$37, +5] \\
Share of coauthors, flat (\%) & 100 & 7.5 & 19.6 & 24.2 & 48.7 \\
\quad graded, change (points) & & +1.8 & +1.8 & $-$0.4 & $-$3.2 \\[3pt]
\multicolumn{6}{l}{\emph{B. Entrants, 0--1 years at the Board}} \\
Coauthors per economist-year, flat & 0.32 & 0.07 & 0.10 & 0.07 & 0.07 \\
\quad graded, change (\%) & +20 & +40 & +18 & +11 & +11 \\
\quad 90\% interval & [+7, +34] & [+20, +69] & [0, +38] & [0, +22] & [+1, +24] \\
Share of coauthors, flat (\%) & 100 & 23.7 & 30.5 & 22.6 & 23.2 \\
\quad graded, change (points) & & +3.8 & $-$0.4$^{\dagger}$ & $-$1.7 & $-$1.7 \\[3pt]
\hline
\end{tabular}
\par\smallskip
\begin{minipage}{0.97\textwidth}\floatnotesfont
Notes. Graded workload relative to the flat benchmark of Table~\ref{tab:cf-tenure}. Coauthors are Board economists sharing a paper in the year, counted per economist-year of the focal group. Partner tenure is measured in that year. Changes are percentages for counts and percentage points for partner shares. Results average 500 posterior draws and two histories per draw. Intervals approximate 90\% posterior uncertainty after adjustment for simulation noise, as described in the appendix. $^{\dagger}$ Sign probability below 75\% or within two simulation standard errors of zero.
\end{minipage}
\end{table}

\FloatBarrier

\subsection{Interpretation and Limitations}
\label{sec:cf-caveats}

These experiments compare research activity under specified workload rules, with staffing and the estimated behavioral relationships held fixed. Paper counts and team sizes do not measure research quality, and the comparisons do not rank economists' payoffs or welfare. FKGL proxies for policy work, so interpreting a change in the measure as a change in workload requires that relationship to remain valid. The model also omits staffing responses and external collaborators.

The simulated responses depend on extending relationships estimated from thirteen years of common workload variation. In particular, the small output gains from bursts reflect the model's nonlinear response to workload entering payoffs linearly. The tenure exercise additionally places economists with nine or more years at $2\sigma_F$ in every year, above the sample's highest annual workload, $1.73\sigma_F$. Both tenure rules have the same average workload above the sample mean; their comparison concerns its allocation and does not establish a preferred workload level.

Collaboration history and the fit of entrants' teams limit the interpretation of the results. Past ties do not depreciate, so simulated networks continue to accumulate historical relationships; extending the exercise beyond 2019 would require addressing this persistence. Within the sample, simulations produce too many papers with a single Board author and too many coauthoring pairs late in the period. Apparent assignment of large entrant teams also limits predictions of entrants' responses to lighter policy work. The appendix reports these fit checks, the behavior beyond the sample, and the uncertainty around the additional results.

\FloatBarrier

\clearpage

\section{Concluding Remarks}
\label{sec:conclusion}

This paper develops and estimates an equilibrium model of team formation in which interactions among team members extend beyond bilateral relationships. Teams form with consent, interactions reflect team composition, and payoffs feature externalities and several forms of congestion. Individuals choose their projects and collaborators, jointly determining the number of projects and their membership. Stability requires each team member to be willing to remain rather than withdraw or pursue a project alone. Logit choice and stochastic consent connect the theory to a stationary likelihood, allowing the empirical analysis to assess the role of participation constraints alongside the payoffs to collaboration.

Research collaborations among economists at the Federal Reserve Board support a distinct role for these constraints. A model re-estimated without participation penalties reproduces overall research activity and team composition but predicts too many memberships that economists would prefer to leave. Large estimated violations are concentrated in teams of entrants that appear to be assigned, suggesting where institutional assignment may matter for understanding observed collaborations.

The estimates and counterfactuals suggest that policy workload affects team composition more than total research output. Higher common workload shifts research toward teams, partly compensating for lower own-project returns. The timing of busy periods matters through the number of economists exposed and the reuse of earlier collaborations. Allocating more policy work to experienced economists raises entrants' output and collaboration and modestly increases total paper counts, while experienced economists produce less research and shift their collaborations toward entrants. Several effects on collaboration remain imprecisely estimated.

Further work could combine voluntary participation with institutional assignment to study how organizations shape the teams that form. Allowing past relationships to depreciate would support counterfactuals over longer horizons, while modeling learning within teams could connect current assignments to future productivity and collaboration. Data on research quality and the internal allocation of tasks and effort could also help assess the gains from team formation beyond paper counts.

\clearpage
\phantomsection
\addcontentsline{toc}{section}{\refname}
\bibliographystyle{plainnat}
\bibliography{references}

\begin{thebibliography}{49}
\providecommand{\natexlab}[1]{#1}
\providecommand{\url}[1]{\texttt{#1}}
\expandafter\ifx\csname urlstyle\endcsname\relax
  \providecommand{\doi}[1]{doi: #1}\else
  \providecommand{\doi}{doi: \begingroup \urlstyle{rm}\Url}\fi

\bibitem[Allocca(2023)]{allocca2023teams}
Alessandra Allocca.
\newblock ``{No} man is an island'': An empirical study on team formation and
  performance.
\newblock Discussion Paper 389, Collaborative Research Center Transregio 190:
  Rationality and Competition, 2023.
\newblock URL
  \url{https://rationality-and-competition.de/wp-content/uploads/discussion_paper/389.pdf}.

\bibitem[Anderson and
  Richards-Shubik(2022)]{andersonrichardsshubik2022collaborative}
Katharine~A. Anderson and Seth Richards-Shubik.
\newblock Collaborative production in science: An empirical analysis of
  coauthorships in economics.
\newblock \emph{The Review of Economics and Statistics}, 104\penalty0
  (6):\penalty0 1241--1255, 2022.
\newblock \doi{10.1162/rest_a_01025}.

\bibitem[Badev(2021)]{Badev2021}
Anton Badev.
\newblock Nash equilibria on (un)stable networks.
\newblock \emph{Econometrica}, 89\penalty0 (3):\penalty0 1179--1206, 2021.

\bibitem[Badev and Bardales(2026)]{Badev_Bardales_2026}
Anton Badev and Arturo Bardales.
\newblock Collaboration cycles: Evidence from the {Federal Reserve}.
\newblock Working paper, 2026.

\bibitem[Battaglini et~al.(2020)Battaglini, {Leone Sciabolazza}, and
  Patacchini]{battaglinileonepatacchini2020effectiveness}
Marco Battaglini, Valerio {Leone Sciabolazza}, and Eleonora Patacchini.
\newblock Effectiveness of connected legislators.
\newblock \emph{American Journal of Political Science}, 64\penalty0
  (4):\penalty0 739--756, 2020.

\bibitem[Becker and Murphy(1992)]{beckermurphy1992division}
Gary~S. Becker and Kevin~M. Murphy.
\newblock The division of labor, coordination costs, and knowledge.
\newblock \emph{The Quarterly Journal of Economics}, 107\penalty0 (4):\penalty0
  1137--1160, 1992.
\newblock \doi{10.2307/2118383}.

\bibitem[Bendor et~al.(2026)Bendor, Bolte, Immorlica, and
  Jackson]{bendor2026multiple}
Jonathan Bendor, Lukas Bolte, Nicole Immorlica, and Matthew~O. Jackson.
\newblock Interactions across multiple games: Cooperation, corruption, and
  organizational design.
\newblock Working paper, arXiv:2507.03030, February 2026.
\newblock URL \url{https://arxiv.org/abs/2507.03030}.

\bibitem[Bhamidi et~al.(2011)Bhamidi, Bresler, and Sly]{bhamidi2011mixing}
Shankar Bhamidi, Guy Bresler, and Allan Sly.
\newblock Mixing time of exponential random graphs.
\newblock \emph{The Annals of Applied Probability}, 21\penalty0 (6):\penalty0
  2146--2170, 2011.
\newblock \doi{10.1214/10-AAP740}.

\bibitem[Blume et~al.(2011)Blume, Brock, Durlauf, and
  Ioannides]{blume2011identification}
Lawrence~E. Blume, William~A. Brock, Steven~N. Durlauf, and Yannis~M.
  Ioannides.
\newblock Identification of social interactions.
\newblock In Jess Benhabib, Alberto Bisin, and Matthew~O. Jackson, editors,
  \emph{Handbook of Social Economics}, volume~1B, chapter~18, pages 853--964.
  North-Holland, 2011.
\newblock \doi{10.1016/B978-0-444-53707-2.00001-3}.

\bibitem[Boerma et~al.(2021)Boerma, Tsyvinski, and Zimin]{boerma2021sorting}
Job Boerma, Aleh Tsyvinski, and Alexander~P. Zimin.
\newblock Sorting with team formation.
\newblock Working Paper 29290, National Bureau of Economic Research, 2021.
\newblock URL \url{https://www.nber.org/papers/w29290}.

\bibitem[Boncinelli et~al.(2022)Boncinelli, Muscillo, and
  Pin]{boncinelli2022teams}
Leonardo Boncinelli, Alessio Muscillo, and Paolo Pin.
\newblock Efficiency and stability in a process of teams formation.
\newblock \emph{Dynamic Games and Applications}, 12:\penalty0 1101--1129, 2022.
\newblock \doi{10.1007/s13235-022-00438-y}.

\bibitem[Bonhomme(2026)]{bonhomme2026teams}
St{\'e}phane Bonhomme.
\newblock Teams: Heterogeneity, sorting, and complementarity.
\newblock In Victor Chernozhukov, Johannes H{\"o}rner, Eliana {La Ferrara}, and
  Iv{\'a}n Werning, editors, \emph{Advances in Economics and Econometrics:
  Twelfth World Congress}, volume~2, pages 49--89. Cambridge University Press,
  2026.
\newblock \doi{10.1017/9781009589727.004}.
\newblock First published online in November 2025.

\bibitem[Bramoull{\'e} et~al.(2020)Bramoull{\'e}, Djebbari, and
  Fortin]{bramoulle2020peer}
Yann Bramoull{\'e}, Habiba Djebbari, and Bernard Fortin.
\newblock Peer effects in networks: A survey.
\newblock \emph{Annual Review of Economics}, 12:\penalty0 603--629, 2020.
\newblock \doi{10.1146/annurev-economics-020320-033926}.

\bibitem[Brock and Durlauf(2001)]{brockdurlauf2001interactions}
William~A. Brock and Steven~N. Durlauf.
\newblock Interactions-based models.
\newblock In James~J. Heckman and Edward~E. Leamer, editors, \emph{Handbook of
  Econometrics}, volume~5, chapter~54, pages 3297--3380. Elsevier, 2001.
\newblock \doi{10.1016/S1573-4412(01)05007-3}.

\bibitem[Brown(2024)]{brown2024team}
Christopher~L. Brown.
\newblock Team production in endogenous networks.
\newblock \emph{Journal of Economic Behavior \& Organization}, 217:\penalty0
  560--580, 2024.
\newblock \doi{10.1016/j.jebo.2023.11.018}.

\bibitem[Canen et~al.(2023)Canen, Jackson, and Trebbi]{canen2023social}
Nathan Canen, Matthew~O. Jackson, and Francesco Trebbi.
\newblock Social interactions and legislative activity.
\newblock \emph{Journal of the European Economic Association}, 21\penalty0
  (3):\penalty0 1072--1118, 2023.
\newblock \doi{10.1093/jeea/jvac051}.

\bibitem[Carlier and Ekeland(2010)]{carlierekeland2010teams}
Guillaume Carlier and Ivar Ekeland.
\newblock Matching for teams.
\newblock \emph{Economic Theory}, 42\penalty0 (2):\penalty0 397--418, 2010.
\newblock \doi{10.1007/s00199-008-0415-z}.

\bibitem[Chalkiadakis et~al.(2010)Chalkiadakis, Elkind, Markakis, Polukarov,
  and Jennings]{chalkiadakis2010overlapping}
Georgios Chalkiadakis, Edith Elkind, Evangelos Markakis, Maria Polukarov, and
  Nicholas~R. Jennings.
\newblock Cooperative games with overlapping coalitions.
\newblock \emph{Journal of Artificial Intelligence Research}, 39:\penalty0
  179--216, 2010.
\newblock \doi{10.1613/jair.3075}.

\bibitem[de~Paula(2020)]{depaula2020formation}
{\'A}ureo de~Paula.
\newblock Econometric models of network formation.
\newblock \emph{Annual Review of Economics}, 12:\penalty0 775--799, 2020.
\newblock \doi{10.1146/annurev-economics-093019-113859}.

\bibitem[Ductor et~al.(2014)Ductor, Fafchamps, Goyal, and van~der
  Leij]{ductor2014social}
Lorenzo Ductor, Marcel Fafchamps, Sanjeev Goyal, and Marco~J. van~der Leij.
\newblock Social networks and research output.
\newblock \emph{The Review of Economics and Statistics}, 96\penalty0
  (5):\penalty0 936--948, 2014.
\newblock \doi{10.1162/REST_a_00430}.

\bibitem[Fowler(2006)]{fowler2006connecting}
James~H. Fowler.
\newblock Connecting the {Congress}: A study of cosponsorship networks.
\newblock \emph{Political Analysis}, 14\penalty0 (4):\penalty0 456--487, 2006.

\bibitem[Goyal et~al.(2006)Goyal, van~der Leij, and
  Moraga-Gonz{\'a}lez]{goyal2006economics}
Sanjeev Goyal, Marco~J. van~der Leij, and Jos{\'e}~Luis Moraga-Gonz{\'a}lez.
\newblock Economics: An emerging small world.
\newblock \emph{Journal of Political Economy}, 114\penalty0 (2):\penalty0
  403--412, 2006.
\newblock \doi{10.1086/500990}.

\bibitem[Graham(2020)]{graham2020network}
Bryan~S. Graham.
\newblock Network data.
\newblock In \emph{Handbook of Econometrics}, volume~7A, chapter~2, pages
  111--218. Elsevier, 2020.
\newblock \doi{10.1016/bs.hoe.2020.05.001}.

\bibitem[Herkenhoff et~al.(2024)Herkenhoff, Lise, Menzio, and
  Phillips]{herkenhoff2024production}
Kyle Herkenhoff, Jeremy Lise, Guido Menzio, and Gordon~M. Phillips.
\newblock Production and learning in teams.
\newblock \emph{Econometrica}, 92\penalty0 (2):\penalty0 467--504, 2024.
\newblock \doi{10.3982/ECTA16748}.

\bibitem[Hsieh et~al.(2020)Hsieh, Lee, and
  Boucher]{hsiehleeboucher2020specification}
Chih-Sheng Hsieh, Lung-Fei Lee, and Vincent Boucher.
\newblock Specification and estimation of network formation and network
  interaction models with the exponential probability distribution.
\newblock \emph{Quantitative Economics}, 11\penalty0 (4):\penalty0 1349--1390,
  2020.
\newblock \doi{10.3982/QE944}.

\bibitem[Hsieh et~al.(2022)Hsieh, K{\"o}nig, and
  Liu]{hsiehkonigliu2022coevolution}
Chih-Sheng Hsieh, Michael~D. K{\"o}nig, and Xiaodong Liu.
\newblock A structural model for the coevolution of networks and behavior.
\newblock \emph{The Review of Economics and Statistics}, 104\penalty0
  (2):\penalty0 355--367, 2022.
\newblock \doi{10.1162/rest_a_00958}.

\bibitem[Hsieh et~al.(2025{\natexlab{a}})Hsieh, K{\"o}nig, and
  Liu]{hsiehkonigliu2025spillovers}
Chih-Sheng Hsieh, Michael~D. K{\"o}nig, and Xiaodong Liu.
\newblock Endogenous technology spillovers in {R\&D} collaboration networks.
\newblock \emph{The RAND Journal of Economics}, 56\penalty0 (4):\penalty0
  419--443, 2025{\natexlab{a}}.
\newblock \doi{10.1111/1756-2171.70010}.

\bibitem[Hsieh et~al.(2025{\natexlab{b}})Hsieh, K{\"o}nig, Liu, and
  Zimmermann]{hsieh2025collaboration}
Chih-Sheng Hsieh, Michael~D. K{\"o}nig, Xiaodong Liu, and Christian Zimmermann.
\newblock Collaboration in bipartite networks.
\newblock \emph{American Economic Journal: Microeconomics}, 2025{\natexlab{b}}.
\newblock URL
  \url{https://research.vu.nl/ws/portalfiles/portal/393362186/paper.pdf}.
\newblock Forthcoming.

\bibitem[Jackson(2008)]{jackson2008networks}
Matthew~O. Jackson.
\newblock \emph{Social and Economic Networks}.
\newblock Princeton University Press, Princeton, NJ, 2008.

\bibitem[Jackson and Wolinsky(1996)]{Jackson_Wolinsky_1996}
Matthew~O. Jackson and Asher Wolinsky.
\newblock {A Strategic Model of Social and Economic Networks}.
\newblock \emph{Journal of Economic Theory}, 71\penalty0 (1):\penalty0 44--74,
  1996.
\newblock \doi{10.1006/jeth.1996.0108}.

\bibitem[Jackson et~al.(2017)Jackson, Rogers, and
  Zenou]{jacksonrogerszenou2017consequences}
Matthew~O. Jackson, Brian~W. Rogers, and Yves Zenou.
\newblock The economic consequences of social-network structure.
\newblock \emph{Journal of Economic Literature}, 55\penalty0 (1):\penalty0
  49--95, 2017.
\newblock \doi{10.1257/jel.20150694}.

\bibitem[Jarosch et~al.(2021)Jarosch, Oberfield, and
  Rossi-Hansberg]{jarosch2021learning}
Gregor Jarosch, Ezra Oberfield, and Esteban Rossi-Hansberg.
\newblock Learning from coworkers.
\newblock \emph{Econometrica}, 89\penalty0 (2):\penalty0 647--676, 2021.
\newblock \doi{10.3982/ECTA16915}.

\bibitem[Kinateder(2009)]{kinateder2009team}
Markus Kinateder.
\newblock Team formation in a network.
\newblock Working Paper 36.2009, Fondazione Eni Enrico Mattei, 2009.
\newblock URL
  \url{https://www.feem.it/publications/team-formation-in-a-network/}.

\bibitem[Kremer(1993)]{kremer1993oring}
Michael Kremer.
\newblock The {O-Ring} theory of economic development.
\newblock \emph{The Quarterly Journal of Economics}, 108\penalty0 (3):\penalty0
  551--575, 1993.
\newblock \doi{10.2307/2118400}.

\bibitem[Liang(2010)]{liang2010double}
Faming Liang.
\newblock A double {Metropolis--Hastings} sampler for spatial models with
  intractable normalizing constants.
\newblock \emph{Journal of Statistical Computation and Simulation}, 80\penalty0
  (9):\penalty0 1007--1022, 2010.

\bibitem[Lo et~al.(2023)Lo, Olivella, and Imai]{loolivellaimai2023bipartite}
Adeline Lo, Santiago Olivella, and Kosuke Imai.
\newblock A statistical model of bipartite networks: Application to
  cosponsorship in the {United States Senate}.
\newblock arXiv:2305.05833, 2023.

\bibitem[McFadden(1974)]{mcfadden1974conditional}
Daniel~L. McFadden.
\newblock Conditional logit analysis of qualitative choice behavior.
\newblock In Paul Zarembka, editor, \emph{Frontiers in Econometrics}, pages
  105--142. Academic Press, New York, 1974.
\newblock URL \url{https://eml.berkeley.edu/reprints/mcfadden/zarembka.pdf}.

\bibitem[Mele(2017)]{mele2017structural}
Angelo Mele.
\newblock A structural model of dense network formation.
\newblock \emph{Econometrica}, 85\penalty0 (3):\penalty0 825--850, 2017.
\newblock \doi{10.3982/ECTA10400}.

\bibitem[Mele(2020)]{mele2020desegregation}
Angelo Mele.
\newblock Does school desegregation promote diverse interactions? an
  equilibrium model of segregation within schools.
\newblock \emph{American Economic Journal: Economic Policy}, 12\penalty0
  (2):\penalty0 228--257, 2020.
\newblock \doi{10.1257/pol.20170604}.

\bibitem[Monderer and Shapley(1996)]{monderer1996potential}
Dov Monderer and Lloyd~S. Shapley.
\newblock Potential games.
\newblock \emph{Games and Economic Behavior}, 14\penalty0 (1):\penalty0
  124--143, 1996.

\bibitem[Murray et~al.(2006)Murray, Ghahramani, and MacKay]{murray2006exchange}
Iain Murray, Zoubin Ghahramani, and David J.~C. MacKay.
\newblock {MCMC} for doubly-intractable distributions.
\newblock In \emph{Proceedings of the 22nd Conference on Uncertainty in
  Artificial Intelligence}, pages 359--366, 2006.

\bibitem[Nakajima(2007)]{nakajima2007}
Ryo Nakajima.
\newblock Measuring peer effects on youth smoking behaviour.
\newblock \emph{Review of Economic Studies}, 74\penalty0 (3):\penalty0
  897--935, 2007.
\newblock \doi{10.1111/j.1467-937X.2007.00448.x}.

\bibitem[O'Donnell(2014)]{odonnell2014analysis}
Ryan O'Donnell.
\newblock \emph{Analysis of Boolean Functions}.
\newblock Cambridge University Press, 2014.

\bibitem[Priem et~al.(2022)Priem, Piwowar, and Orr]{priem2022openalex}
Jason Priem, Heather Piwowar, and Richard Orr.
\newblock {OpenAlex}: A fully-open index of scholarly works, authors, venues,
  institutions, and concepts.
\newblock arXiv:2205.01833, 2022.

\bibitem[Rosenthal(1973)]{rosenthal1973}
Robert~W. Rosenthal.
\newblock A class of games possessing pure-strategy {Nash} equilibria.
\newblock \emph{International Journal of Game Theory}, 2\penalty0 (1):\penalty0
  65--67, 1973.

\bibitem[Train(2009)]{train2009discrete}
Kenneth~E. Train.
\newblock \emph{Discrete Choice Methods with Simulation}.
\newblock Cambridge University Press, second edition, 2009.
\newblock \doi{10.1017/CBO9780511805271}.

\bibitem[Vehtari et~al.(2021)Vehtari, Gelman, Simpson, Carpenter, and
  B{\"u}rkner]{vehtari2021rank}
Aki Vehtari, Andrew Gelman, Daniel Simpson, Bob Carpenter, and Paul-Christian
  B{\"u}rkner.
\newblock Rank-normalization, folding, and localization: An improved
  $\widehat{R}$ for assessing convergence of {MCMC} (with discussion).
\newblock \emph{Bayesian Analysis}, 16\penalty0 (2):\penalty0 667--718, 2021.

\bibitem[Wainwright and Jordan(2008)]{WainwrightJordan2008}
Martin~J. Wainwright and Michael~I. Jordan.
\newblock Graphical models, exponential families, and variational inference.
\newblock \emph{Foundations and Trends in Machine Learning}, 1\penalty0
  (1--2):\penalty0 1--305, 2008.
\newblock \doi{10.1561/2200000001}.

\bibitem[Walsh(1923)]{walsh1923closed}
J.~L. Walsh.
\newblock A closed set of normal orthogonal functions.
\newblock \emph{American Journal of Mathematics}, 45\penalty0 (1):\penalty0
  5--24, 1923.

\end{thebibliography}

\clearpage
\appendix

\section{Proofs}
\label{app:proofs}

\subsection{Potential and Existence}
\label{app:potential-existence}

An exact potential links unilateral payoff changes to a common function and supplies the existence argument. The first part verifies this identity; the second uses superlinear project costs to show that the potential attains a maximum despite an endogenous number of projects.

\subsubsection{Potential Representation}

\begin{proposition}[Exact potential representation]
\label{prop:potential}
For the payoffs in \eqref{eq:payoffs-project}, suppose Assumption~\ref{ass:payoffs}(ii)--(iii) holds. Then $Q$ in \eqref{potential} is an exact potential.
The representation also holds
when each individual's choices are restricted to at most $\bar p$ projects, $p_i\leq\bar p$:
restricting strategy sets preserves an exact potential \citep{monderer1996potential}.
\end{proposition}

\begin{proof}
Consider the incremental change related to $i$ joining project $r$, i.e. adding the link $a_{i,r}$. Recall that $N_r=\{i: a_{i,r}=1\}$ denotes the set of collaborators in project $r$ with $n_r=|N_r|$. Further, let $N_r^{(i)}$ denote the identities and $n_r^{(i)}$ the number of new collaborators associated with $i$ joining project $r$, where $N_r^{(i)}\subsetneq N_r$ and $n_r^{(i)}\leq d_i$ (note that $i \in N_r$ but $i\notin N_r^{(i)}$). We need to establish $\Delta_{a_{i,r}} U_i=\Delta_{a_{i,r}} Q $. Indeed, write
\begin{align}
\Delta_{a_{i,r}}Q
={}&\beta_i-\Delta C_P(p_i)+V(X_i,\{X_j\}_{j\in N_r})\notag\\
&+\sum_{j\in N_r\setminus\{i\}}\Bigl[w(X_i,X_j)-\Delta C_G(g_{ij}-1)\Bigr]\notag\\
\label{eq:Da_Dp}
&-\psi_D\left[d_i n_r^{(i)}-\frac{n_r^{(i)}(n_r^{(i)}-1)}{2}
+\sum_{j\in N_r^{(i)}}d_j\right],
\end{align}
where $p_i$, $N_r$, $g_{ij}$ and the degrees count $i$'s membership in $r$; for a new collaborator, $g_{ij}=1$ and the cost is $\Delta C_G(0)$. The terms in $p_i$ change as $i$'s own terms. Adding $a_{i,r}$ changes only the membership of $r$, so $\sum_{r'}\Phi(N_{r'})$ changes by $\Phi(N_r)-\Phi(N_r\setminus\{i\})=V(X_i,\{X_j\}_{j\in N_r})$ by Assumption~\ref{ass:payoffs}(ii), which equals the change in $i$'s own payoff term $\sum_{r'}a_{i,r'}V(X_i,\{X_j\}_{j\in N_{r'}})$; the
payoffs of $r$'s other members also change, but the potential only needs to
track the payoff of the individual who moves. The pair terms are symmetric by Assumption~\ref{ass:payoffs}(iii): adding $a_{i,r}$ raises $g_{ij}$ by one for every $j\in N_r\setminus\{i\}$ and changes no other pair, so $\sum_{i<j}\bigl(g_{ij}w(X_i,X_j)-C_G(g_{ij})\bigr)$ changes by $\sum_{j\in N_r\setminus\{i\}}\bigl[w(X_i,X_j)-\Delta C_G(g_{ij}-1)\bigr]$, as does $i$'s payoff. Finally, the $n_r^{(i)}$ new relationships raise $d_i$ by $n_r^{(i)}$ and each $d_j$, $j\in N_r^{(i)}$, by one, which changes $-\frac{\psi_D}{2}\sum_k(d_k+d_k^2)$ by the last term of \eqref{eq:Da_Dp}, as it changes $i$'s degree term. The project term is the only one that needs (ii): a value of membership that is not a marginal contribution to a common project-level function would make the change in the potential depend on the order in which members join.

The identity extends to any unilateral change in several memberships by summing the single-membership identities along a sequence of additions and deletions. If the two profiles have different project sets, first add empty project slots to put them on a common set of labels; empty slots contribute zero to both payoff and potential. Hence
\[
Q(\widetilde a_i,a_{-i})-Q(a_i,a_{-i})=U_i(\widetilde a_i,a_{-i})-U_i(a_i,a_{-i}).
\]
\end{proof}

\subsubsection{Existence}

\begin{proof}[Proof of Theorem~\ref{thm:existence}]
Let $n=|N|<\infty$. If $n=0$, the empty profile suffices, so suppose $n\geq1$. Proposition~\ref{prop:potential} gives an exact potential under Assumption~\ref{ass:payoffs}(ii)--(iii), for any $\psi_D\in\mathbb R$. We first show that it attains a global maximum even though the number of projects is endogenous.

Let $a,b\geq0$ be the constants in the theorem, and define the finite constants
\begin{align*}
B_\Phi&=\max_{\varnothing\ne S\subseteq N}\frac{\max\{\Phi(S),0\}}{|S|},\\
B_w&=\max\Bigl(\{0\}\cup\{w(X_i,X_j)+a:i<j\}\Bigr).
\end{align*}
Then $\sum_r\Phi(N_r)\leq B_\Phi\sum_i p_i$, while $C_G(g)\geq-ag-b$ gives
\[
\sum_{i<j}\bigl[g_{ij}w(X_i,X_j)-C_G(g_{ij})\bigr]
\leq B_w\sum_{i<j}g_{ij}+b\binom n2.
\]
Moreover,
\[
\sum_{i<j}g_{ij}=\sum_{r\in P}\frac{n_r(n_r-1)}{2}
\leq\frac{n-1}{2}\sum_i p_i.
\]
Since $0\leq d_i\leq n-1$, the degree term is bounded above for either sign of $\psi_D$:
\[
-\frac{\psi_D}{2}\sum_i(d_i+d_i^2)
\leq K_D,
\qquad
K_D=\frac{\max\{-\psi_D,0\}}{2}\,n^2(n-1).
\]
Consequently, with $K=b\binom n2+K_D<\infty$,
\begin{equation}
\label{eq:potential-existence-bound}
Q(A)\leq K+\sum_i\left[\left(\beta_i+B_\Phi+\frac{n-1}{2}B_w\right)p_i-C_P(p_i)\right].
\end{equation}
Because $C_P(p)/p\to\infty$, each summand on the right of \eqref{eq:potential-existence-bound} tends to $-\infty$ as $p_i\to\infty$ and is bounded above over the nonnegative integers. Since $N$ is finite and $K$ is constant, there is a finite $M$ such that $Q(A)<0$ whenever $\max_i p_i>M$. The network with no active projects has $Q=0$, so a global maximizer must lie among profiles with $p_i\leq M$ for every $i$. Such profiles have at most $nM$ nonempty projects. Since project labels have no payoff consequences, only finitely many profiles remain after identifying project relabelings. Hence a global maximizer $(P^\ast,A^\ast)$ exists.

For any unilateral change by $i$, pad both profiles with empty project slots if necessary. Exact potentiality gives
\[
U_i(\widetilde a_i,a_{-i}^\ast)-U_i(A^\ast)
=Q(\widetilde a_i,a_{-i}^\ast)-Q(A^\ast)\leq0.
\]
Thus $A^\ast$ is optimal without stability constraints for every consideration set, including every set of at most $\max\{k,2\}$ projects. Deleting a membership cannot improve its member's payoff, and neither can replacing it with a new solo project. Lemma~\ref{lem:optimality_implies_stability} therefore implies that $(P^\ast,A^\ast)$ is $\max\{k,2\}$-team stable and hence $k$-team stable. This proves existence for every $k\geq1$.
\end{proof}

\subsection{Participation Conditions and Equilibrium Properties}
\label{app:equilibrium-properties}

Individual participation slacks determine which memberships can be retained and which deviations are feasible. The following arguments derive these slacks and their response to entry, establish the equilibrium bounds and nesting, and verify the example separating two-team from three-team stability.

\subsubsection{Participation Slacks and Entry}
\label{app:slacks-entry}

This subsection derives the participation slack in Remark~\ref{rem:slack-equilibrium} and tracks how entry changes it within the entered team and in other projects.

\begin{proof}[Proof of Remark~\ref{rem:slack-equilibrium}]
Adding a solo project to $A^{-jr}$ changes neither team nor relationship terms and raises $j$'s project count from $p_j-1$ to $p_j$. This proves \eqref{eq:participation-slack} and the equivalence between its nonnegativity and the constraint in \eqref{stability_constraint}. At an equilibrium the current choice is feasible in every consideration set, so every current member's constraint holds, and by Definition~\ref{def:equilibrium} a change is feasible exactly when the constraints required at the resulting profile hold.
\end{proof}

Let $A^-$ be the profile before $i$ joins $r$ and $A^+$ the profile after entry, with all other memberships fixed. Write $N_r$ and $n_r$ for the team and its size after entry, and $g_{ij}^-$ for shared projects before entry. For each incumbent $j\in N_r\setminus\{i\}$, define her payoff externality and an adjustment for her outside option:
\begin{align*}
E_{j;i,r}&:=U_j(A^+)-U_j(A^-),\\
\xi_{j;i,r}&:=
\left.
\sum_{\ell\ne j}[g_{j\ell}>0]\,\Delta_{a_{i,r}}d_\ell
\right|_{a_{j,r}=0}\geq0.
\end{align*}
The degrees and relationship indicators in $\xi_{j;i,r}$ are evaluated with $j$ removed from $r$. Her relationships then remain fixed when $i$ enters, while the degrees of some of her collaborators may rise.

\begin{lemma}[Entry and individual participation slacks]
\label{lem:participation-stability}
Suppose Assumption~\ref{ass:payoffs}(ii)--(iii) holds. For every incumbent $j\in N_r\setminus\{i\}$ and any $\psi_D\in\mathbb R$,
\begin{equation}
\label{eq:entry-individual-slack}
\begin{aligned}
\sigma_{j,r}(A^+)-\sigma_{j,r}(A^-)
&=\Delta_{a_{i,r}}\Delta_{a_{j,r}}U_j\\
&=E_{j;i,r}+\psi_D\xi_{j;i,r}.
\end{aligned}
\end{equation}
Consequently, the incumbents' constraints after entry hold if and only if
\begin{equation}
\label{eq:entry-feasibility-slacks}
\sigma_{j,r}(A^-)+E_{j;i,r}+\psi_D\xi_{j;i,r}\geq0
\qquad\text{for every }j\in N_r\setminus\{i\}.
\end{equation}
When $\psi_D=0$, the change in each incumbent's slack is
\begin{equation}
\label{eq:entry-slack-zero-degree}
\begin{aligned}
E_{j;i,r}
={}&V(X_j,\{X_\ell\}_{\ell\in N_r})
-V(X_j,\{X_\ell\}_{\ell\in N_r\setminus\{i\}})\\
&+w(X_i,X_j)-\Delta C_G(g_{ij}^-).
\end{aligned}
\end{equation}
In particular, initially satisfied incumbent constraints remain satisfied if these changes are nonnegative for every incumbent.
\end{lemma}

The entry test is individual. A favorable effect on one incumbent cannot compensate another for a negative slack. Moreover, \eqref{eq:entry-feasibility-slacks} checks only the incumbents in $r$: the entrant's own constraint, any other constraints required by Definition~\ref{def:equilibrium}, and the profitability of her full choice must also be checked. Example~\ref{ex:stable-not-unconstrained} illustrates how an incumbent's negative slack can block privately profitable entry.
The aggregate identities below compare the incumbents with the entrant. Under Assumption~\ref{ass:payoffs}(ii), the incumbents' total change in team value can differ from the entrant's team value, and it falls short when a positive value of team size is constant or declining in size. When $\psi_D=0$ and there is no such shortfall, an entrant who prefers the team to a solo project raises the sum of the incumbents' slacks, but not necessarily each of them. Entry into $r$ can also change the slacks of members of other projects $s\neq r$, through the cost of repeated collaboration with the entrant and through degree congestion (Lemma~\ref{lem:cross-participation} below). These spillovers are network externalities; they add no constraints beyond those of Definition~\ref{def:equilibrium}.

\begin{proof}[Proof of Lemma~\ref{lem:participation-stability}]
Entry by $i\neq j$ leaves $p_j$ unchanged, so the outside-option threshold $\max\{0,\beta_j-\Delta C_P(p_j)\}$ in \eqref{eq:participation-slack} is the same at $A^-$ and $A^+$. The slack change therefore equals the mixed payoff difference in \eqref{eq:entry-individual-slack}. With $a_{j,r}=0$, $i$'s entry changes neither $j$'s team benefits, shared-project counts, nor her own degree. It changes only the degrees of her existing collaborators, giving
\[
\left.\Delta_{a_{i,r}}U_j\right|_{a_{j,r}=0}
=-\psi_D\xi_{j;i,r}.
\]
Subtracting this outside-state payoff change from $E_{j;i,r}$ proves the identity. Degrees can only rise when a membership is added, so $\xi_{j;i,r}\geq0$. Equation~\eqref{eq:entry-feasibility-slacks} is the resulting nonnegativity test. When $\psi_D=0$, the payoff externality consists only of the change in $j$'s team value and the additional pair value net of collaboration cost, proving \eqref{eq:entry-slack-zero-degree}.
\end{proof}

 For an aggregate interpretation, define, for $i\in S$,
\[
\begin{aligned}
G_V(i,S)&:=\sum_{j\in S\setminus\{i\}}\Bigl[V(X_j,\{X_\ell\}_{\ell\in S})-V(X_j,\{X_\ell\}_{\ell\in S\setminus\{i\}})\Bigr],\\
D_V(i,S)&:=G_V(i,S)-V(X_i,\{X_j\}_{j\in S}).
\end{aligned}
\]
Here $G_V$ is the incumbents' total change in team value and $D_V$ is its difference from the entrant's team value. Under Assumption~\ref{ass:payoffs}(ii), $D_V(i,S)=0$ in a team of two, because both members' team value is $\Phi(S)$ and solo team value is zero. For a value of team size,
\[
D_V(i,S)=(n-1)[V(n)-V(n-1)]-V(n),\qquad n=|S|\geq2.
\]
For $n\geq3$, this is nonnegative if and only if $V(n)\geq\frac{n-1}{n-2}V(n-1)$. A positive team value that is constant or declining therefore gives a negative aggregate wedge: the entrant receives a team benefit while incumbents receive little or lose from its dilution.

Let $m=n_r^{(i)}$ be the number of $i$'s new collaborators in $r$, and put $\Xi_{i,r}=\sum_{j\in N_r\setminus\{i\}}\xi_{j;i,r}$. The aggregate payoff externality satisfies
\begin{equation}
\label{eq:D_irU_j}
\begin{aligned}
\sum_{j\in N_r\setminus\{i\}}\Delta_{a_{i,r}}U_j
={}&\Delta_{a_{i,r}}U_i^{\mathrm{new}}
-\Delta_{a_{i,r}}U_i^{\mathrm{alone}}+D_V(i,N_r)\\
&-\psi_D\frac{m}{2}(4n_r-m-7),
\end{aligned}
\end{equation}
and summing the individual slack changes gives
\begin{equation}
\label{eq:entry-aggregate-slack}
\begin{aligned}
\sum_{j\in N_r\setminus\{i\}}\Delta_{a_{i,r}}\sigma_{j,r}
&=\sum_{j\in N_r\setminus\{i\}}\Delta_{a_{i,r}}U_j+\psi_D\Xi_{i,r}\\
&=\Delta_{a_{i,r}}\left(\sum_{j\ne i}a_{j,r}\Delta_{a_{j,r}}U_j\right).
\end{aligned}
\end{equation}
To verify \eqref{eq:D_irU_j}, evaluate degrees before entry. The degree expression multiplying $\psi_D$ in each existing collaborator's cost rises by $2m$, and that of each new collaborator $j$ rises by $d_i+d_j+2m$. Summing gives $md_i+\sum_{j\in N_r^{(i)}}d_j+2m(n_r-1)$. The entrant's corresponding increment is $md_i+\sum_{j\in N_r^{(i)}}d_j+m(m+3)/2$. Their difference is $m(4n_r-m-7)/2$; the pair terms coincide in the two totals by symmetry, and the team terms differ by $D_V$. Equation~\eqref{eq:entry-aggregate-slack} follows from \eqref{eq:entry-individual-slack} and the constancy of each incumbent's outside-option threshold.

These identities explain the sources of divergence between an entrant's gain and the effects on incumbent incentives. With $\psi_D=0$ and $D_V\geq0$, preferring the team to a solo project implies a nonnegative aggregate externality and a nonnegative change in the sum of incumbent slacks. Neither conclusion implies \eqref{eq:entry-feasibility-slacks}: Definition~\ref{def:equilibrium} requires every individual slack to be nonnegative, rather than their sum or its change.

Entry into $r$ may also change the slack of a member $j\neq i$ in another project $s\neq r$. As before, $i$'s entry leaves $p_j$ unchanged, so
\[
\Delta_{a_{i,r}}\sigma_{j,s}
=\Delta_{a_{i,r}}\Delta_{a_{j,s}}U_j,
\qquad j\in N_s\setminus\{i\}.
\]
This comparison isolates the collaboration and degree channels: neither the composition of $s$ nor $j$'s own project count changes.

\begin{lemma}[Participation slacks in other projects]
\label{lem:cross-participation}
Suppose Assumption~\ref{ass:payoffs}(ii)--(iii) holds. Let $i$ join $r$, hold all other memberships fixed, and consider $s\neq r$ and $j\in N_s\setminus\{i\}$.
\begin{enumerate}[label=(\alph*)]
\item If $i\notin N_s$ and $\psi_D\geq0$, then $\Delta_{a_{i,r}}\sigma_{j,s}\leq0$. If $\psi_D=0$, the change is zero.
\item If $i\in N_s$, $\psi_D=0$, and $\Delta C_G(h+1)\leq\Delta C_G(h)$ for all $h\geq0$, then $\Delta_{a_{i,r}}\sigma_{j,s}\geq0$. For $C_G(g)=c_N[g>0]$ with $c_N\geq0$,
\[
\Delta_{a_{i,r}}\sigma_{j,s}
=c_N\,[j\in N_r,\ g_{ij}^-=1],
\]
where $g_{ij}^-$ counts shared projects before $i$ joins $r$.
\item If $i\in N_s$ and $\psi_D>0$, the change cannot be signed in general, even when $C_G(g)=c_N[g>0]$ with $c_N\geq0$.
\end{enumerate}
\end{lemma}

\begin{proof}
Fix $j\in N_s\setminus\{i\}$. Let $A^{xy,j}$ set $a_{i,r}=x$ and $a_{j,s}=y$, holding all other memberships fixed, and denote the corresponding shared-project counts and degrees by $g_{uv}^{xy,j}$ and $d_v^{xy,j}$. Define
\[
e_j=[a_{i,s}=1,\ a_{j,r}=1],\qquad
h_j=g_{ij}^{00,j},\qquad c_j=e_j[h_j=0].
\]
The only relationship that both memberships can create is $\{i,j\}$. The other degree increments are nonnegative and specific to one membership:
\begin{align*}
\alpha_v^{(j)}&:=d_v^{10,j}-d_v^{00,j}-c_j[v\in\{i,j\}],\\
\eta_v^{(j)}&:=d_v^{01,j}-d_v^{00,j}-c_j[v\in\{i,j\}].
\end{align*}
Consequently,
\[
d_v^{xy,j}=d_v^{00,j}+\alpha_v^{(j)}x+\eta_v^{(j)}y
+c_j[v\in\{i,j\}](x+y-xy).
\]
By exact potentiality, the mixed payoff difference equals $\Delta_x\Delta_y Q$. Because the two memberships concern distinct individuals and projects, the own-project, team, and linear pair-value terms of $Q$ have zero mixed difference. Only the collaboration cost of $\{i,j\}$ and the degree terms contribute. Write
\[
D^2C_G(h):=C_G(h+2)-2C_G(h+1)+C_G(h)
=\Delta C_G(h+1)-\Delta C_G(h).
\]
The pair cost contributes $-e_jD^2C_G(h_j)$. For $f(d)=(d^2+d)/2$, the degree representation gives
\[
\Delta_x\Delta_y f(d_v^{xy,j})
=\alpha_v^{(j)}\eta_v^{(j)}
-c_j[v\in\{i,j\}](d_v^{00,j}+1).
\]
Thus, for any $\psi_D\in\mathbb R$,
\begin{equation}
\label{eq:cross-participation-individual}
\begin{aligned}
\Delta_{a_{i,r}}\sigma_{j,s}
={}&-e_jD^2C_G(h_j)
+c_j\psi_D(d_i^{00,j}+d_j^{00,j}+2)\\
&-\psi_D\sum_{v\in N}\alpha_v^{(j)}\eta_v^{(j)}.
\end{aligned}
\end{equation}

If $i\notin N_s$, then $e_j=c_j=0$, leaving only the last term. This proves (a), and also shows that its inequality reverses when $\psi_D<0$. If $\psi_D=0$, only $-e_jD^2C_G(h_j)$ remains, proving the first claim in (b). For a maintenance cost, $D^2C_G(0)=-c_N$ and $D^2C_G(h)=0$ for $h\geq1$. When $e_j=1$, the actual pre-entry count is $g_{ij}^-=h_j+1$, so the slack increases by $c_N$ exactly when the pair previously shared only $s$.

For (c), take a profile with only $N_s=\{i,j\}$ and $N_r=\{j\}$. Entry gives $\Delta_{a_{i,r}}\sigma_{j,s}=c_N+2\psi_D>0$. Instead, take $N_s=\{i,j\}$ and $N_r=\{h\}$, with $h\notin\{i,j\}$. Entry then gives $\Delta_{a_{i,r}}\sigma_{j,s}=-\psi_D<0$.
\end{proof}

For an aggregate interpretation, sum over the other members of $s$:
\begin{equation}
\label{eq:delta2}
\begin{aligned}
\Delta^*&:=\sum_{j\in N_s\setminus\{i\}}\Delta_{a_{i,r}}\sigma_{j,s}\\
&=\Delta_{a_{i,r}}\left(\sum_{j\ne i}a_{j,s}\Delta_{a_{j,s}}U_j\right).
\end{aligned}
\end{equation}
Summing \eqref{eq:cross-participation-individual} yields the aggregate identity
\begin{equation}
\label{eq:cross-participation-explicit}
\begin{aligned}
\Delta^*
={}&\sum_{j\in N_s\setminus\{i\}}\Bigl[-e_jD^2C_G(h_j)
+c_j\psi_D(d_i^{00,j}+d_j^{00,j}+2)\Bigr]\\
&-\psi_D\sum_{j\in N_s\setminus\{i\}}\sum_{v\in N}
\alpha_v^{(j)}\eta_v^{(j)}.
\end{aligned}
\end{equation}
When $\psi_D=0$ and $C_G(g)=c_N[g>0]$ with $c_N\geq0$, this reduces to
\begin{equation}
\label{eq:cross-participation-zero-convex-cost}
\Delta^*=c_N\,\mathbf 1\{i\in N_s\}\left|\{j\in N_s\cap N_r:j\ne i,\ g_{ij}^-=1\}\right|\geq0.
\end{equation}
In particular, the aggregate change is zero when $i\notin N_s$, as Lemma~\ref{lem:cross-participation}(a) states.
The identities summarize the total change in other members' slacks, while the lemma's signs apply separately to each of them. Repeated collaboration can strengthen participation in an existing team when it reduces the marginal cost of that membership. With $\psi_D=0$, convex collaboration costs reverse this effect on the other members' slacks. Degree congestion can instead weaken participation in projects whose members share collaborators with the entrant. To assess feasibility, each affected slack required by Definition~\ref{def:equilibrium} must be evaluated at the candidate profile. Spillovers on projects outside that constraint set describe network externalities, without supplying an additional veto. Together with the mover's payoff comparison and the participation slacks of Remark~\ref{rem:slack-equilibrium}, these calculations explain which deviations can be blocked in equilibrium; they do not substitute aggregate consent for individual participation.

\subsubsection{Equilibrium Bounds and Nesting}
\label{app:bounds-nesting}

The participation conditions bound individual project loads, team sizes, and collaboration degrees. Nesting then follows from the inclusion of consideration sets.

This appendix proves Propositions~\ref{prop:number_projects}--\ref{prop:degree_bounds} and Lemma~\ref{lem:team-stability-nesting} of Sections~\ref{sec:equilibrium_projects_collaborations} and~\ref{sec:team-stability-discussion}, and illustrates that equilibrium does not require nonnegative aggregate entry externalities.

\begin{proof}[Proof of Proposition~\ref{prop:number_projects}]
Fix an equilibrium and write $p_i=p_i^*$. The extrema defining the bounds exist because $\Delta C_P$ is nondecreasing and tends to infinity, while $W_i$ is finite.

\emph{Lower bound.} Adding a new solo project gives
\[
\Delta U_i^{\mathrm{alone}}=\beta_i-\Delta C_P(p_i+1).
\]
If this gain is positive, the addition is feasible. It leaves all team compositions, shared-project counts, and degrees unchanged, so every other individual's slack is unchanged. If $i$ already has a team membership, nondecreasing marginal project costs imply that her solo-project return is positive both before and after the addition. Equation~\eqref{eq:slack-team-surplus} therefore gives $\sigma_{i,r}=s_{i,r}$ at both profiles for every such membership, and these slacks are unchanged as well. The new project is singleton and has no team stability constraint. Thus no required participation constraint blocks this profitable deviation. Equilibrium requires $\Delta C_P(p_i+1)\geq\beta_i$, which gives $p_i\geq\underline p_i$.

\emph{Upper bound.} Suppose $p_i>0$ and take $r\in P_i$. If $r$ is nonsingleton, Definition~\ref{def:equilibrium} and \eqref{eq:participation-slack} give
\[
\Delta_{a_{i,r}}U_i(A^*)
\geq\max\{0,\beta_i-\Delta C_P(p_i)\}\geq0.
\]
This conclusion follows from current feasibility; it does not require dropping a team membership to be a feasible deviation. If $r$ is singleton, dropping it leaves all collaboration and team terms unchanged. It weakly increases $i$'s slacks in any remaining teams by \eqref{eq:slack-team-surplus} and nondecreasing $\Delta C_P$, and it leaves all other members' slacks unchanged. The deletion is feasible, so equilibrium again requires $\Delta_{a_{i,r}}U_i(A^*)\geq0$.

For either kind of project, the own-project return and team surplus give
\[
\begin{aligned}
0\leq\Delta_{a_{i,r}}U_i(A^*)
&=\beta_i-\Delta C_P(p_i)+s_{i,r}(A^*)\\
&\leq\beta_i-\Delta C_P(p_i)+\overline s_i(n_r^*)
\leq\beta_i-\Delta C_P(p_i)+W_i.
\end{aligned}
\]
The surplus bound uses \eqref{eq:new-vs-alone}: collaboration costs are bounded below by $c_G$, and the degree-cost contribution is nonpositive when $\psi_D\geq0$. Hence $\Delta C_P(p_i)\leq\beta_i+W_i$, proving the upper bound. The case $p_i=0$ is included in its definition. Finally, each active project has at least one membership, so $|P^*|\leq\sum_i p_i^*$, giving \eqref{eq:total-project-count-bound}.
\end{proof}

\begin{proof}[Proof of Proposition~\ref{prop:project_size}]
Every current member of a nonsingleton project has nonnegative slack by Remark~\ref{rem:slack-equilibrium}. The payoff decomposition gives the exact expression
\begin{equation}
\label{eq:individual-slack-team-bound}
\begin{aligned}
\sigma_{i,r}(A^*)
={}&V(X_i,\{X_j\}_{j\in N_r^*})
+\sum_{j\in N_r^*\setminus\{i\}}
\bigl[w(X_i,X_j)-\Delta C_G(g_{ij}^*-1)\bigr]\\
&-\psi_DD^{\mathrm{degree}}_{i,r}(A^*)-\lambda_i(p_i^*).
\end{aligned}
\end{equation}
Rearranging and dividing by $n_r^*-1>0$ proves \eqref{eq:average-benefit-participation} and its equivalence to the member's constraint. In particular, the conclusion follows from current feasibility for any $k\geq1$; it does not rely on a two-project replacement deviation or on an aggregate consent condition.

Since $\Delta C_G\geq c_G$ and $\psi_DD^{\mathrm{degree}}_{i,r}\geq0$,
\[
0\leq\sigma_{i,r}(A^*)
\leq\overline s_i(n_r^*)-\lambda_i(p_i^*).
\]
Thus $n_r^*$ passes the size test defining $\overline n_i(p_i^*)$ for every member. Singleton projects satisfy the bound by the inclusion of $1$. Because $\lambda_i(p)\geq0$ and $\overline s_i(1)=0$, $\overline n_i(p)\leq\overline n_i$, proving \eqref{eq:individual-project-size-bound}.
\end{proof}

\begin{proof}[Proof of Proposition~\ref{prop:degree_bounds}]
The identity in \eqref{eq:weighteddegree} and the individual size bounds give
\[
d_i^{W,*}=\sum_{r\in P_i^*}(n_r^*-1)
\leq p_i^*(\overline n_i-1)
\leq\overline p_i(\overline n_i-1).
\]
For $p_i^*>0$, the same argument uses $\overline n_i(p_i^*)$ to obtain the sharper first bound. Every distinct collaborator contributes at least once to weighted degree, so $d_i^*\leq d_i^{W,*}$, and a finite population gives $d_i^*\leq n-1$. Conversely, any given collaborator shares at most $p_i^*$ projects with $i$. Hence $d_i^{W,*}=\sum_{j\ne i}g_{ij}^*\leq p_i^*d_i^*$, which yields the lower bound on $d_i^*$ by division and integrality. If $\overline n_i=1$, no nonsingleton project can contain $i$, and both degrees are zero.
\end{proof}

\begin{proof}[Proof of Lemma~\ref{lem:team-stability-nesting}]
Take $A\in\mathcal E_k$. Every consideration set $R$ with $|R|\leq k'$ also satisfies $|R|\leq k$. The current choice is therefore feasible and optimal in that problem, with exactly the same objective and constraints as at bound $k'$. Any profitable feasible deviation at the smaller bound would therefore also contradict $k$-team stability. Thus $A\in\mathcal E_{k'}$. In Definition~\ref{def:equilibrium}, the bound enters only through $|R|\leq k$; the individual slack checks for a given problem are independent of that bound. This verifies the stated conditions.
\end{proof}

The ``at most $k$'' convention is important for this argument. A candidate feasible on $R$ need not remain feasible if $R$ is enlarged to include additional retained memberships, since those memberships may introduce further slack constraints. The proof keeps the original $R$ available and never requires such an enlargement. Under a variant allowing only sets of exactly $k$ projects, nesting would instead require an extension argument. A sufficient condition is that every smaller consideration problem can be padded with projects in which the mover remains absent, with enough such slots available: they change neither payoffs nor the required participation checks. Padding with retained memberships would require verifying their additional constraints.

 Definition~\ref{def:equilibrium} does not require aggregate entry externalities to be nonnegative. For illustration, take three individuals with $\beta_i=1$, $C_P(p)=p^2$, $V_{\mathrm{size}}(m)=[m\geq2]4(3/4)^{m-2}$, and all other payoff terms zero. One project containing all three gives each member payoff and slack $3$. Leaving or replacing it with a solo project gives payoff $0$, and retaining it while adding a solo project gives payoff $1$. These exhaust unilateral choices, so the profile is a $k$-team stable equilibrium for every $k\geq1$. Nevertheless, adding any one member to the other two lowers each incumbent's team value from $4$ to $3$, an aggregate externality of $-2$. The members remain willing to participate despite that externality. The $G_V$ and $D_V$ identities in Appendix~\ref{app:slacks-entry} interpret such changes; nonnegative aggregate entry effects are not an additional equilibrium requirement.

\subsubsection{Two-Team Stability without Three-Team Stability}
\label{app:two-versus-three-team-stability}

This subsection verifies Example~\ref{ex:two-versus-three-team-stability} of Section~\ref{sec:consideration-bound}.

\begin{proof}[Verification of Example~\ref{ex:two-versus-three-team-stability}]
These payoffs satisfy Assumption~\ref{ass:payoffs}, with $\Phi=0$, $\psi_D=0$, $\Delta C_P(p)=2p-1$, and marginal collaboration costs $4$ for the first shared project and zero thereafter.

Holding individual 2's memberships fixed, individual 1's choices over the existing projects give the following payoffs. The last column counts memberships changed relative to $A^0$.
\begin{center}
{\small\upshape\renewcommand{\arraystretch}{1.2}
\begin{tabular}{@{}lrrrr@{}}
\hline
$(a_{1,r},a_{1,s},a_{1,t})$ & $p_1$ & $g_{12}$ & $U_1$ & Changes\\
\hline
$(1,0,0)$ & 1 & 0 & 1 & 0\\
$(0,0,0)$ & 0 & 0 & 0 & 1\\
$(1,1,0)$ or $(1,0,1)$ & 2 & 1 & $-1$ & 1\\
$(0,1,0)$ or $(0,0,1)$ & 1 & 1 & 0 & 2\\
$(1,1,1)$ & 3 & 2 & $-1$ & 2\\
$(0,1,1)$ & 2 & 2 & 2 & 3\\
\hline
\end{tabular}}
\end{center}
No change of at most two memberships increases her payoff. Adding a fresh solo project to a choice with $p_1\geq1$ lowers it by $2p_1-1>0$; when she holds no existing project, a fresh solo project gives payoff $1$. Thus the new-project option supplies no profitable two-membership deviation either.

Individual 2 cannot gain from any unilateral change. Her own-project payoff satisfies $4p_2-p_2^2\leq4$, and she can share at most one project with individual 1, who participates only in $r$. Any such collaboration adds $3-4=-1$. Her current two solo projects therefore maximize her payoff even without participation constraints. Since every current project is solo, current feasibility is immediate. Hence $A^0$ is 2-team stable.

Individual 1 can nevertheless gain by leaving $r$ and joining both $s$ and $t$. These are three membership changes. After deleting the empty project $r$, the resulting profile is
\[
A^1=
\begin{array}{c|cc}
 &s&t\\\hline
1&1&1\\
2&1&1
\end{array},
\qquad (U_1(A^1),U_2(A^1))=(2,6).
\]
The gain to individual 1 is $1$. Each membership passes the individual participation test: because $g_{12}=2$, removing one membership leaves the relationship in place and saves no collaboration cost. Equation~\eqref{eq:individual-slack-team-bound} gives, for $u\in\{s,t\}$,
\[
\sigma_{i,u}(A^1)=3-[3-\beta_i]_+
=\begin{cases}
2,&i=1,\\
3,&i=2.
\end{cases}
\]
Thus the three-membership deviation is feasible, proving $A^0\notin\mathcal E_3$.

The resulting profile $A^1$ is itself stable for every $k$. Holding the other individual's memberships fixed, let $g\in\{0,1,2\}$ be the number of joint projects retained and $z\in\{0,1\}$ indicate a new solo project. The payoff from any unilateral choice is
\[
\beta_i(g+z)-(g+z)^2+3g-4[g>0].
\]
Its maximum is $2$ for individual 1 and $6$ for individual 2, in both cases attained at $(g,z)=(2,0)$. Both individuals therefore choose full unconstrained best responses at $A^1$.
\end{proof}

\subsection{Deterministic Team Dynamics}
\label{app:deterministic-team-dynamics}

The pathwise bounds give a finite reachable state space up to project relabeling. The convergence proof then considers a potential maximizer within each closed recurrent class, allowing temporary decreases along individual updates.

\begin{proof}[Proof of Proposition~\ref{prop:deterministic-path-bounds}]
The constants $M_i$ and $L_i$ are finite by Section~\ref{sec:equilibrium_projects_collaborations}, and the initial profile satisfies the asserted bounds. Consider one update by individual $i$, from a profile $A$ satisfying those bounds to a maximizing feasible candidate $B$. Empty-column deletion and relabeling do not affect any of the quantities involved.

\emph{Project loads.} Suppose $p_i(B)>M_i$. Since $p_i(A)\leq M_i$, at least one of $i$'s memberships in $B$ is newly added, in a considered project $r$. Moreover, $p_i(B)>\overline p_i$ implies
\[
\Delta C_P(p_i(B))>\beta_i+W_i.
\]
If $r$ is nonsingleton, the membership-margin bound from Proposition~\ref{prop:number_projects} gives
\[
\Delta_{a_{i,r}}U_i(B)
\leq\beta_i-\Delta C_P(p_i(B))+W_i<0.
\]
Thus $\sigma_{i,r}(B)<0$, contradicting feasibility. If $r$ is singleton, omitting that membership instead raises $i$'s payoff by
\[
\Delta C_P(p_i(B))-\beta_i>0.
\]
This omission does not change any team composition, collaboration count, or degree outside that solo project. It weakly raises $i$'s other participation slacks because $\Delta C_P$ is nondecreasing, and it leaves every other individual's slacks unchanged. The resulting candidate remains feasible on $R$, contradicting the maximizing choice of $B$. Hence $p_i(B)\leq M_i$; all other individuals' loads are unchanged.

\emph{Team sizes.} Any expanding project has a new membership by $i$ and belongs to $R$. If it is nonsingleton after entry, every member $j$ must satisfy $\sigma_{j,r}(B)\geq0$. The pointwise bound used in Proposition~\ref{prop:project_size} yields
\[
0\leq\sigma_{j,r}(B)
\leq\overline s_j(n_r(B))-\lambda_j(p_j(B)),
\]
so $n_r(B)\leq\overline n_j\leq L_j$ for every member. A new solo project satisfies the bound since $L_i\geq1$. Exits reduce team size and remove a member from its bound; other project sizes are unchanged. The size bounds therefore also propagate by induction.

Finally, each active project has a member, giving $|P_t|\leq\sum_i p_i(A_t)\leq M$. The identities
\[
d_i^W(A_t)=\sum_{r\in P_i(A_t)}(n_r(A_t)-1),
\qquad d_i(A_t)\leq\min\{n-1,d_i^W(A_t)\}
\]
give \eqref{eq:deterministic-degree-bounds}. With at most $M$ active projects and only $2^n-1$ nonempty participant sets for each project, there are finitely many multisets of project teams. Hence the reachable state space is finite modulo relabeling. None of these arguments requires the chosen update to increase the mover's payoff relative to $A$.
\end{proof}

\begin{proof}[Proof of Theorem~\ref{thm:deterministic-team-convergence}]
Proposition~\ref{prop:deterministic-path-bounds} gives a finite reachable state space after identifying project relabelings. The state-dependent meeting probabilities and fixed tie rule define a time-homogeneous finite Markov chain. All allowed meetings, including every singleton consideration set, have positive probability. Proposition~\ref{prop:potential} gives the exact potential identity for every unilateral revision.

\emph{Absorbing states.} At a $k$-team stable equilibrium, the current choice is feasible and maximizing for every meeting. Inertia therefore keeps the profile fixed. Conversely, if a profile contains a nonsingleton membership with $\sigma_{j,r}(A)<0$, the singleton meeting $(j,\{r\})$ cannot retain it. Setting $a_{j,r}=0$ is the only feasible choice, so that positive-probability meeting changes the state by reducing $j$'s project load. An absorbing state must consequently have nonnegative slacks for all current team memberships. Its current choice is then feasible in every meeting. If it were not an equilibrium, some allowed consideration set would admit a feasible strict improvement. The maximizing update would strictly raise the mover's payoff and potential, changing the state with positive probability. Thus the absorbing states are exactly the $k$-team stable equilibria.

\emph{Closed recurrent classes.} Let $\mathcal C$ be a closed communicating class and choose $A\in\mathcal C$ maximizing $Q$ over $\mathcal C$. Suppose first that every current nonsingleton membership has nonnegative slack. If $A$ is not an equilibrium, a profitable feasible deviation exists on some consideration set of size at most $k$. The corresponding meeting selects a best response with strictly higher payoff and hence higher potential. Closedness puts its successor in $\mathcal C$, contradicting maximality.

It remains to rule out a negative slack at $A$. Suppose $\sigma_{j,r}(A)<0$. Let $X=A^{-jr}$ denote exit from $r$, and let $Y$ replace that membership by a fresh solo project. By the definition of the slack,
\begin{equation}
\label{eq:deterministic-membership-repair}
\max\{U_j(X),U_j(Y)\}>U_j(A).
\end{equation}
The singleton meeting $(j,\{r\})$ forces exit to $X$, because retention violates $j$'s own participation condition. Thus $X\in\mathcal C$. If $U_j(X)>U_j(A)$, exact potentiality gives $Q(X)>Q(A)$, a contradiction. Otherwise \eqref{eq:deterministic-membership-repair} implies
\[
U_j(Y)>U_j(A)\geq U_j(X).
\]
At $X$, the singleton meeting $(j,\{r_{\mathrm{new}}\})$ offers only remaining absent and adding a solo project. The latter is feasible, faces no nonsingleton participation constraint, and is strictly better. It therefore leads to $Y$ with positive probability. Closedness gives $Y\in\mathcal C$, while exact potentiality for the same individual's combined exit and addition gives
\[
Q(Y)-Q(A)=U_j(Y)-U_j(A)>0,
\]
again a contradiction. These are successive singleton meetings, so this argument applies even when $k=1$.

Every closed communicating class therefore contains an equilibrium at its potential maximum. That state is absorbing, so the communicating class containing it must be a singleton. A finite Markov chain enters one of its closed recurrent classes in finite time almost surely. It consequently reaches a $k$-team stable equilibrium and remains there.
\end{proof}

The following example illustrates a temporary decrease in potential. Consider three individuals with $\beta_i=2$, $C_P(p)=p^2$, $V=0$, $C_G=0$, and $\psi_D=0$, and symmetric pair values
\[
w(X_1,X_2)=\tfrac12,\qquad
w(X_1,X_3)=3,\qquad
w(X_2,X_3)=-10.
\]
These primitives satisfy Assumption~\ref{ass:payoffs}. Initially let $N_r=\{1,2\}$ and $N_s=\{3\}$. Both members of $r$ have slack $1/2$. At the meeting $(1,\{s\})$, individual 1 joins $s$ and gains $2$. In the resulting team $\{1,3\}$, the two members' slacks are $2$ and $3$, so entry is feasible. However, individual 1's slack in the unconsidered project $r$ falls to $-1/2$: her payoff is $7/2$, whereas leaving $r$ would give $4$. At the subsequent meeting $(2,\{r\})$, retention is infeasible because member 1 has negative slack. Individual 2 must exit, reducing her payoff from $3/2$ to $0$ and reducing $Q$ by $3/2$. This does not contradict the theorem: the closed-class argument allows temporary decreases and uses exit, followed when needed by a new solo project, to rule out nontrivial recurrent classes.

\subsection{Stationary Distributions and Mixing}
\label{app:stationary-laws-mixing}

The stationary law follows from the exact potential and an acceptance probability that depends only on the candidate profile. Its proof gives the unconstrained and count-weighted laws together. The remaining arguments establish project-count tails and convergence-rate comparisons for the unconstrained logit dynamic.

\subsubsection{General Stationary Law and Count Weighting}
\label{app:count-weighted-consent}

\begin{proof}[Proof of Propositions~\ref{prop:stationary-probabilistic-constraints} and~\ref{prop:unconstrained-logit-stationary}]
It suffices to prove the general screened law; setting $\chi\equiv1$ gives the unconstrained case.
First consider the unconstrained proposal kernel. Exact
potentiality implies that $U_i(B)-Q(B)$ is constant over
$F_\mu(A)$. Hence
\[
L_\mu(A,B)
=
\frac{\exp\{Q(B)/\tau\}}
{\displaystyle\sum_{D\in F_\mu(A)}\exp\{Q(D)/\tau\}}
\qquad\text{for }B\in F_\mu(A).
\]
Moreover,
\[
B\in F_\mu(A)
\quad\Longrightarrow\quad
F_\mu(B)=F_\mu(A).
\]
Thus the forward and reverse conditional choice probabilities
have the same denominator, giving
\[
e^{Q(A)/\tau}L_\mu(A,B)
=
e^{Q(B)/\tau}L_\mu(B,A).
\]
Multiplication by $\rho_\mu$ and summation over meetings yield
\begin{equation}
e^{Q(A)/\tau}P_k^0(A,B)
=
e^{Q(B)/\tau}P_k^0(B,A).
\label{eq:proposal-detailed-balance}
\end{equation}

Now let
\[
Z_{\tau,\chi}
=
\sum_{D\in\mathcal A}e^{Q(D)/\tau}\chi(D).
\]
For $A\neq B$, equation
\eqref{eq:proposal-detailed-balance} implies
\begin{align*}
\pi_{\tau,\chi}(A)P_k^\chi(A,B)
&=
\frac{
e^{Q(A)/\tau}\chi(A)P_k^0(A,B)\chi(B)
}{Z_{\tau,\chi}}\\
&=
\frac{
e^{Q(B)/\tau}\chi(B)P_k^0(B,A)\chi(A)
}{Z_{\tau,\chi}}\\
&=
\pi_{\tau,\chi}(B)P_k^\chi(B,A).
\end{align*}
Detailed balance is automatic for $A=B$. Therefore
$\pi_{\tau,\chi}$ is stationary.

Full-support meetings and logit proposals permit every
single-coordinate change with positive probability: the other
selected coordinates can retain their current values.
Consequently, $P_k^0$ is irreducible on $\mathcal A$.
Since $\chi(B)>0$ for every $B$, every positive-probability
off-diagonal transition of $P_k^0$ remains possible under
$P_k^\chi$. Irreducibility is therefore preserved.

Finally, logit choice assigns positive probability to the
status quo, so
\[
P_k^\chi(A,A)\geq P_k^0(A,A)>0.
\]
The process is aperiodic. Finiteness and irreducibility imply
uniqueness of the stationary distribution. Its expression
contains no $k$ when $\chi$ is independent of $k$.
\end{proof}

\begin{proof}[Proof of Corollary~\ref{cor:count-weighted-stationary}]
Conditional on a candidate $B$, independent checks give joint acceptance probability $q^{\mathcal V(B)}=e^{-\kappa\mathcal V(B)}$. One proposal is drawn and rejection retains the current state, so this is precisely Proposition~\ref{prop:stationary-probabilistic-constraints} with $\chi(B)=e^{-\kappa\mathcal V(B)}$. For finite $\kappa$, this weight is strictly positive; the proposition therefore gives the stated stationary law, reversibility, irreducibility, aperiodicity, and independence from the meeting law.

At $\kappa=0$, the weight is one. As $\kappa\to\infty$, it converges to $\mathbf 1\{B\in\Omega\}$. The empty profile belongs to $\Omega$, so the limiting denominator is positive and the finite normalization converges to the Gibbs law conditioned on $\Omega$. This is a limit of stationary laws and does not establish irreducibility of the limiting hard-consent kernel.
\end{proof}

\label{par:slot-adjusted-stationarity}
The same construction applies to the empirical participation benchmark. For the margin $e_{j,r}$ of Remark~\ref{rem:exit-scale}, fix the slot count and the parameters. The practical count $\bar{\mathcal V}(B)$ depends only on the candidate profile, and global screening gives
\[
\bar\chi_\kappa(B)=e^{-\kappa\bar{\mathcal V}(B)}
\in\bigl[e^{-\kappa|\mathcal M(B)|},1\bigr].
\]
It is strictly positive for finite $\kappa$, shared by every meeting, and independent of the proposal shocks conditional on $B$. Proposition~\ref{prop:stationary-probabilistic-constraints} therefore applies directly. Its detailed-balance calculation multiplies the unconstrained stationary flows by the symmetric factor $\bar\chi_\kappa(A)\bar\chi_\kappa(B)$, giving the count-weighted stationary law with $\bar{\mathcal V}$ in place of $\mathcal V$. At $\kappa=0$ the unconstrained law is recovered; as $\kappa\to\infty$, the law conditions on $\bar\Omega$. The inequality $e_{j,r}\geq\sigma_{j,r}$ implies $\bar\Omega\supseteq\Omega$. The same argument covers depth weights based on $\sum_{(j,r)\in\mathcal M(A)}[-e_{j,r}(A)]_+$. Thus the implemented criterion is the ordinary participation test for $U^{\mathrm{proj}}$ in Remark~\ref{rem:practical-slot-exit}; its comparison with $\Omega$ uses the unshifted slot index. The reference-weight identity in the appendix gives the same stationary law.

Restricting checks to the current meeting can change the stationary distribution. Consider three individuals, one labeled project slot, uniform individual selection, and $k=1$. Set $\beta_i=0$, $C_P(p)=p^2$, $w=0$, $C_G=0$, and $\psi_D=0$. Let $\Phi(S)=4\mathbf 1\{|S|=3\}$ and let team value be the marginal contribution to $\Phi$, as in Assumption~\ref{ass:payoffs}(ii). These primitives satisfy Assumption~\ref{ass:payoffs}. A pair gives each member payoff and slack $-1$, while the three-person team gives each member payoff and slack $3$. Thus $\mathcal V=2$ at every pair, and $\mathcal V=0$ at the empty, singleton, and three-person profiles. The potential is $Q(S)=-|S|+4\mathbf 1\{|S|=3\}$.

Let $A^T$ denote the three-person team and suppose $0<q=e^{-\kappa}<1$. If a member leaves $A^T$, the resulting pair has two negative slacks. The global rule accepts this candidate with probability $q^2$. Under checks restricted to $\mathcal M_{i,R}(B)$, the mover is absent from the resulting project, so that check set is empty and acceptance is certain. Every other off-diagonal transition has the same acceptance probability under the two rules. Conditional on any selected individual at $A^T$, the logit probability of exit is $1/(1+e^{3/\tau})$.

Consequently, if $\widetilde P$ denotes the meeting-local kernel, the proposed global Gibbs law satisfies
\[
(\pi_{\tau,\kappa}\widetilde P)(A^T)-\pi_{\tau,\kappa}(A^T)
=-\pi_{\tau,\kappa}(A^T)
\frac{1-q^2}{1+e^{3/\tau}}<0.
\]
It is therefore not stationary for that local-screening process. This counterexample isolates the screening-scope issue while retaining exactly the individual participation criterion of Definition~\ref{def:equilibrium}.

\subsubsection{Project-Count Tails in the Labeled-Slot Model}
\label{app:logit-project-count-tail}

\begin{proof}[Proof of Theorem~\ref{thm:logit-project-count-tail}]
Choosing $q$ active slots and one of $D=2^n-1$ nonempty teams for each gives exactly
\[
H_{\overline P}(q)=\binom{\overline P}{q}D^q,
\qquad 0\leq q\leq\overline P
\]
labeled states at count $q$. Let
\[
Z_{\tau,\overline P}=\sum_{D\in\mathcal A}\exp\{Q(D)/\tau\}\chi(D)
\]
be the normalizing constant of $\pi_{\tau,\chi}$. The empty state, with $Q(\varnothing)=0$ and $\chi(\varnothing)=1$, gives $Z_{\tau,\overline P}\geq1$. Since $\chi\leq1$, the unnormalized weight of every state is at most $\exp\{Q(A)/\tau\}$, so every bound below holds for $\pi_{\tau,\chi}$ exactly as for $\pi_\tau^0$; we write $\pi_\tau^0$ for brevity. Equation~\eqref{eq:potential-existence-bound} gives $K$ and $\gamma$ in the remark by taking the maximum of zero and its linear coefficients as $\gamma$. Both are independent of $\overline P$. Set $T=\sum_i p_i$, with $q\leq T\leq nq$ at count $q$.

\emph{Superlinear costs.} For any $r>0$, define the finite constant
\[
B_r:=n\sup_{u\in\mathbb N_0}
\bigl\{(\gamma+r)u-C_P(u)\bigr\}<\infty.
\]
Then $Q(A)\leq K+B_r-rT\leq K+B_r-rq$. Using $H_{\overline P}(q)\leq(\overline P D)^q$ and $Z_{\tau,\overline P}\geq1$, for every $s>0$ choose $r_s=\tau[s+\log(\overline P D)]$ to get
\begin{equation}
\label{eq:logit-project-count-tail}
\Pr_{\pi_\tau^0}\{p\geq m\}
\leq C_s(\overline P)e^{-sm},\qquad
C_s(\overline P):=
\frac{\exp\{(K+B_{r_s})/\tau\}}{1-e^{-s}}.
\end{equation}
Sum the geometric bound from $m$ to $\overline P$ and extend it to infinity.

\emph{Quadratic costs.} Jensen's inequality gives
\[
\sum_i C_P(p_i)\geq\frac{\psi_P}{2n}(T-np_0)_+^2.
\]
Since $\gamma\geq0$ and $q\leq T\leq nq$, this implies $Q(A)\leq K+n\gamma q-\psi_P(q-np_0)_+^2/(2n)$. Therefore the exact label count gives the finite-sum tail bound
\begin{equation}
\label{eq:logit-project-count-level-bound}
\begin{aligned}
\Pr_{\pi_\tau^0}\{p\geq m\}
&\leq e^{K/\tau}\sum_{q=m}^{\overline P}
\binom{\overline P}{q}D^q
\exp\left\{\frac{n\gamma q}{\tau}-\alpha(q-np_0)_+^2\right\}.
\end{aligned}
\end{equation}
The bound is zero for $m>\overline P$, and any probability upper bound here can be capped at one. Retaining the factor $Z_{\tau,\overline P}^{-1}$ on the right would sharpen it further if that normalizing constant were available.

To obtain a closed form, use $\binom{\overline P}{q}\leq\overline P^q$ and $(q-np_0)_+^2\geq q^2-2np_0q$. The unnormalized mass at count $q$ is then at most
\[
\exp\{K/\tau-\alpha q^2+\ell_{\overline P}q\}.
\]
The inequality $\ell_{\overline P}q\leq\alpha q^2/2+\ell_{\overline P}^2/(2\alpha)$ consequently bounds the tail by
\[
e^{K/\tau+\ell_{\overline P}^2/(2\alpha)}
\sum_{q=m}^{\infty}e^{-\alpha q^2/2}.
\]
For integers $m,j\geq0$, $(m+j)^2\geq m^2+j^2$, and the decreasing Gaussian summand satisfies
\[
\sum_{j=0}^{\infty}e^{-\alpha j^2/2}
\leq1+\int_0^{\infty}e^{-\alpha x^2/2}\,dx
=1+\sqrt{\frac{\pi}{2\alpha}}.
\]
Combining these inequalities proves \eqref{eq:logit-project-count-gaussian-tail} with the stated $C_{\overline P}$.

Alternatively, retain the full quadratic coefficient. Successive terms $e^{-\alpha q^2+\ell_{\overline P}q}$ have ratio $e^{\ell_{\overline P}-\alpha(2q+1)}$. Hence, whenever $\delta_m:=\alpha(2m+1)-\ell_{\overline P}>0$,
\begin{equation}
\label{eq:logit-project-count-geometric-tail}
\Pr_{\pi_\tau^0}\{p\geq m\}
\leq\frac{\exp\{K/\tau-\alpha m^2+\ell_{\overline P}m\}}
{1-e^{-\delta_m}}.
\end{equation}
Thus the leading negative quadratic term is unchanged by labels, while their contribution to the linear coefficient is $\log\overline P$. The bounds can be conservative because they use a potential envelope and replace the normalizing constant by one.

\emph{Truncation within the fixed reference.} Let $\pi_\tau^{0,M}$ be $\pi_\tau^0$ conditioned on $p\leq M$, extended by zero outside that set. Then
\[
\left\|\pi_\tau^0-\pi_\tau^{0,M}\right\|_{\mathrm{TV}}
=\Pr_{\pi_\tau^0}\{p>M\},
\]
directly from the definition of total variation. The same holds for any $\mathcal A_T\supseteq\{p<m\}$: conditioning on $\mathcal A_T$ rescales the weights on $\mathcal A_T$ by $1/\pi_{\tau,\chi}(\mathcal A_T)$ and removes the rest, so the total variation distance is $\pi_{\tau,\chi}(\mathcal A\setminus\mathcal A_T)\leq\Pr\{p\geq m\}\leq\varepsilon_m$, and for $A\in\mathcal A_T$ the log ratio of the two probabilities is $-\log\pi_{\tau,\chi}(\mathcal A_T)\in[0,-\log(1-\varepsilon_m)]$; for a statistic $|f|\leq1$, $|\mathbb E^Tf-\mathbb Ef|\leq2\|\pi-\pi^T\|_{\mathrm{TV}}$. All probabilities and normalizations in this proof refer to the same labeled-slot universe.
\end{proof}

\subsubsection{Speed of Unconstrained Logit Dynamics}
\label{app:unconstrained-logit-speed}

The proofs of Theorems~\ref{thm:unconstrained-logit-speed} and~\ref{thm:unconstrained-logit-speed-comparison} treat the dynamic as an operator acting on functions of the profile. Fix the payoffs, $\tau$ and $\overline P$, and write $\pi=\pi_\tau^0$ for the stationary distribution \eqref{eq:unconstrained-logit-stationary}. $A\sim\pi$ denotes a profile drawn from it, and $\mathbb E_\pi$ and $\operatorname{Var}_\pi$ denote expectations and variances over that draw.

A statistic is a function $f:\mathcal A\to\mathbb R$ of the profile, such as a membership $a_{i,r}$, a project count $p_i$, or the potential $Q$. Statistics form the space $L^2(\pi)$, with inner product and norm
\[
\langle f,g\rangle_\pi:=\sum_{A\in\mathcal A}\pi(A)f(A)g(A)=\mathbb E_\pi[f(A)g(A)],
\qquad
\|f\|_\pi^2:=\langle f,f\rangle_\pi .
\]
Constants are statistics, $\langle f,1\rangle_\pi=\mathbb E_\pi f$, and $\operatorname{Var}_\pi(f)=\|f-\mathbb E_\pi f\|_\pi^2$.

A meeting $(i,R)$ lets $i$ reconsider the block of coordinates $\{(i,r):r\in R\}$. Write $A_{-(i,R)}:=(a_{j,s})_{(j,s)\notin\{i\}\times R}$ for the entries of $A$ outside the block, and $\mathcal F_{i,R}$ for the information they carry. Two profiles have the same entries outside the block exactly when one lies in the other's choice set, $B\in\mathcal B_i(A;R)$. Conditioning on $A_{-(i,R)}$ therefore restricts $\pi$ to the choice set:
\[
\pi\bigl(B\mid A_{-(i,R)}\bigr)=\frac{\pi(B)}{\pi\bigl(\mathcal B_i(A;R)\bigr)},
\qquad B\in\mathcal B_i(A;R).
\]
A larger block holds fewer entries fixed: $R\subseteq R'$ implies that $A_{-(i,R')}$ is a sub-vector of $A_{-(i,R)}$, so $\mathcal F_{i,R'}\subseteq\mathcal F_{i,R}$.

Profiles in $\mathcal B_i(A;R)$ differ only in $i$'s memberships. By the exact potential (Proposition~\ref{prop:potential}), $U_i(B)-U_i(D)=Q(B)-Q(D)$ for $B,D\in\mathcal B_i(A;R)$. Multiplying the numerator and denominator of the logit probability of Section~\ref{sec:unconstrained-logit} by $\exp\{[Q(A)-U_i(A)]/\tau\}$ gives
\[
\Pr(A_{t+1}=B\mid A_t=A,\,i,R)
=\frac{\exp\{Q(B)/\tau\}}{\sum_{D\in\mathcal B_i(A;R)}\exp\{Q(D)/\tau\}}
=\pi\bigl(B\mid A_{-(i,R)}\bigr).
\]
A meeting therefore redraws its block from the stationary conditional distribution given everything else, as in a block Gibbs sampler. Its action on statistics is the conditional-update operator
\[
(H_{i,R}f)(A):=\mathbb E\bigl[f(A_{t+1})\mid A_t=A,\,i,R\bigr]
=\sum_{B\in\mathcal B_i(A;R)}\pi\bigl(B\mid A_{-(i,R)}\bigr)f(B)
=\mathbb E_\pi\bigl[f\mid A_{-(i,R)}\bigr](A),
\]
which is the conditional expectation of $f$ given the entries outside the block. The tower property of conditional expectations gives three properties:
\begin{itemize}
\item redrawing a block twice is the same as redrawing it once, $H_{i,R}^2=H_{i,R}$;
\item $H_{i,R}$ is self-adjoint, $\langle g,H_{i,R}f\rangle_\pi=\mathbb E_\pi\bigl[\mathbb E_\pi(g\mid\mathcal F_{i,R})\,\mathbb E_\pi(f\mid\mathcal F_{i,R})\bigr]=\langle H_{i,R}g,f\rangle_\pi$;
\item $\langle f,H_{i,R}f\rangle_\pi=\|H_{i,R}f\|_\pi^2\geq0$.
\end{itemize}
Thus $H_{i,R}$ is the orthogonal projection of $L^2(\pi)$ onto the statistics that depend only on $A_{-(i,R)}$.

For a state-independent meeting law $\rho(i,R)$, one step of the dynamic acts on statistics by
\[
(K_\rho f)(A):=\mathbb E\bigl[f(A_{t+1})\mid A_t=A\bigr]=\sum_{(i,R)}\rho(i,R)\,(H_{i,R}f)(A),
\]
and $K_k=K_{\rho_k}$. $K_\rho$ is an average of orthogonal projections with weights that do not depend on the state. It is therefore self-adjoint, which is reversibility, $\pi(A)K_\rho(A,B)=\pi(B)K_\rho(B,A)$ for the transition probabilities $K_\rho(A,B)$. It is also positive semidefinite, with $\|K_\rho f\|_\pi\leq\|f\|_\pi$, so its eigenvalues lie in $[0,1]$. Constants have eigenvalue one, and by irreducibility (Proposition~\ref{prop:unconstrained-logit-stationary}) no other statistic does, so the eigenvalues are $1=\lambda_1>\lambda_2\geq\cdots\geq0$. If the meeting law depended on the state, the weights could not be taken outside the conditional expectations, and $K_\rho$ would not be an average of projections.

For a statistic $f$, the Dirichlet form of a reversible operator $K$ with stationary distribution $\pi$ is
\begin{equation}
\label{eq:dirichlet-form}
\mathcal E_K(f):=\langle f,(I-K)f\rangle_\pi
=\tfrac12\sum_{A,B\in\mathcal A}\pi(A)K(A,B)\bigl[f(B)-f(A)\bigr]^2,
\end{equation}
where $I$ is the identity and $K(A,B)$ the transition probability from $A$ to $B$. The second expression follows from expanding the square. Using $\sum_BK(A,B)=1$ and reversibility, $\tfrac12\sum_{A,B}\pi(A)K(A,B)\bigl[f(B)^2+f(A)^2\bigr]=\|f\|_\pi^2$, while $\sum_{A,B}\pi(A)K(A,B)f(A)f(B)=\langle f,Kf\rangle_\pi$.

The second expression has a direct reading. Start the dynamic at a profile drawn from $\pi$ and take one step; $\mathcal E_K(f)=\tfrac12\mathbb E\bigl[(f(A_{t+1})-f(A_t))^2\bigr]$ is half the expected squared change in the statistic $f$. It is zero for constants and unchanged when a constant is added to $f$. It is also linear in the operator: $\mathcal E_{K_\rho}(f)=\sum_{(i,R)}\rho(i,R)\,\mathcal E_{H_{i,R}}(f)$.

Since $K$ is self-adjoint, its second eigenvalue is the largest value of $\langle f,Kf\rangle_\pi/\|f\|_\pi^2$ over nonzero statistics orthogonal to the constants, that is, with $\mathbb E_\pi f=0$ (the Rayleigh--Ritz characterization). Subtracting from one and recentering an arbitrary statistic at its mean,
\begin{equation}
\label{eq:gap-variational}
\operatorname{gap}(K)=1-\lambda_2(K)
=\min_{f\ \mathrm{nonconstant}}\frac{\mathcal E_K(f)}{\operatorname{Var}_\pi(f)}.
\end{equation}
The gap is the smallest fraction of its stationary variance that any statistic moves in one step. The dynamic is slow when some statistic with substantial variance barely changes in a typical step. Because the eigenvalues are nonnegative, the gap also gives the rate at which expectations approach their stationary values. For every statistic, $\|K^tf-\mathbb E_\pi f\|_\pi\leq\bigl(1-\operatorname{gap}(K)\bigr)^t\|f-\mathbb E_\pi f\|_\pi$.

\begin{proof}[Proof of Theorem~\ref{thm:unconstrained-logit-speed}]
By the construction above, $K_k=\sum_{i,R}\rho_k(i,R)H_{i,R}$ is an average of the orthogonal projections $H_{i,R}f=\mathbb E_{\pi_\tau^0}[f\mid A_{-(i,R)}]$. It is therefore self-adjoint, with eigenvalues in $[0,1]$, for general payoffs.

When $Q$ is constant, $\pi_\tau^0$ is uniform and every considered bit is refreshed independently with probability $1/2$. Use the Walsh characters on the Boolean hypercube, as in the spectral counting argument of \citet{Badev2021}; see also \citet{walsh1923closed} and \citet[Ch.~1]{odonnell2014analysis}. For $I\subseteq N\times\mathcal P$, define
\[
e_I(A):=\prod_{(j,r)\in I}(-1)^{a_{j,r}},
\qquad I_i:=\{r:(i,r)\in I\}.
\]
These $2^{n\overline P}$ characters form an orthogonal basis: the product of two different characters contains a bit whose two values cancel in the uniform sum. Conditional averaging leaves $e_I$ unchanged when $R\cap I_i=\varnothing$ and gives zero otherwise. Thus the complete spectrum, counted with multiplicity through the sets $I$, is
\begin{equation}
\label{eq:unconstrained-logit-spectrum}
\lambda_{k,I}=\sum_{i,R}\rho_k(i,R)
\mathbf 1\{R\cap I_i=\varnothing\}.
\end{equation}
The empty set gives eigenvalue one. For nonempty $I$, choose any $(i,r)\in I$. A meeting that refreshes that coordinate necessarily touches $I$, so
\[
1-\lambda_{k,I}\geq c_{i,r}^{(k)}
\geq\min_{j,s}c_{j,s}^{(k)}.
\]
A singleton $I=\{(i,r)\}$ attaining the minimum gives equality. This proves \eqref{eq:unconstrained-logit-gap}. Full support makes the minimum positive. Under uniform individuals and uniform project sets conditional on size,
\[
c_{i,r}^{(k)}=\frac1n\sum_{\ell=1}^{m_k}
w_{\ell,k}\frac{\ell}{\overline P}
=\frac{\mathbb E_k[\ell]}{n\overline P},
\]
proving \eqref{eq:unconstrained-logit-mean-size-gap} and its uniform-size specialization. Concentrating all size mass at $k$ gives the stated exact-size variant. Since the eigenvalues are nonnegative, for every mean-zero $f$,
\[
\|K_k^t f\|_{L^2(\pi_\tau^0)}
\leq\bigl(1-\operatorname{gap}(K_k)\bigr)^t
\|f\|_{L^2(\pi_\tau^0)},
\]
which explains the relaxation-rate interpretation.
\end{proof}

\begin{proof}[Proof of Theorem~\ref{thm:unconstrained-logit-speed-comparison}]
By \eqref{eq:gap-variational}, the gap is set by the statistic that moves least in one step relative to its variance. The proof shows that a larger nested meeting moves every statistic at least as much. A meeting redraws its block given everything outside it, and the less it holds fixed, the more the statistic can vary in the redraw. The argument needs no formula for the gap, so it holds for any potential. It has four steps:
\begin{enumerate}
\item the Dirichlet form of one meeting is an expected conditional variance;
\item a conditional variance is larger when less is conditioned on;
\item averaging over the coupled meeting laws carries the comparison to the transition operators and their gaps;
\item the required coupling exists for uniform project sets.
\end{enumerate}

First, fix $(i,R)$ and write $H=H_{i,R}$ and $\mathcal F=\mathcal F_{i,R}$. Since $H$ is an orthogonal projection, $\langle f,Hf\rangle_\pi=\|Hf\|_\pi^2$, so
\[
\mathcal E_H(f)=\|f\|_\pi^2-\|Hf\|_\pi^2
=\mathbb E_\pi\Bigl[\mathbb E_\pi\bigl(f^2\mid\mathcal F\bigr)-\mathbb E_\pi\bigl(f\mid\mathcal F\bigr)^2\Bigr]
=\mathbb E_\pi\Bigl[\operatorname{Var}_\pi\bigl(f\mid A_{-(i,R)}\bigr)\Bigr].
\]
The same identity follows from the second expression in \eqref{eq:dirichlet-form}. Let $A_t\sim\pi$ and let the meeting $(i,R)$ occur. Then $A_{t+1}$ keeps $A_{t,-(i,R)}$ and redraws the block from $\pi(\cdot\mid A_{t,-(i,R)})$, which is also the conditional law of the block of $A_t$. Given $A_{t,-(i,R)}$, $f(A_t)$ and $f(A_{t+1})$ are therefore independent draws from the same conditional distribution, and half the expected squared difference of two such draws is their conditional variance. The Dirichlet form of a meeting is thus the variance of the statistic that remains once the entries outside the block are known, averaged over the stationary distribution of those entries. It is large when the block carries much of the statistic's variation given the rest of the profile.

Second, let $R\subseteq R'$, so that $\mathcal F'=\mathcal F_{i,R'}$ fixes fewer entries than $\mathcal F=\mathcal F_{i,R}$, and $\mathcal F'\subseteq\mathcal F$. The law of total variance under $\pi(\cdot\mid A_{-(i,R')})$ splits the conditional variance given the coarser information into two parts: the average conditional variance given the finer information, and the variance of the finer conditional mean:
\[
\operatorname{Var}_\pi\bigl(f\mid\mathcal F'\bigr)
=\mathbb E_\pi\Bigl[\operatorname{Var}_\pi\bigl(f\mid\mathcal F\bigr)\Bigm|\mathcal F'\Bigr]
+\operatorname{Var}_\pi\Bigl(\mathbb E_\pi\bigl[f\mid\mathcal F\bigr]\Bigm|\mathcal F'\Bigr).
\]
Taking expectations under $\pi$ and using the first step gives the exact gain:
\[
\mathcal E_{H_{i,R'}}(f)-\mathcal E_{H_{i,R}}(f)
=\mathbb E_\pi\!\left[
\operatorname{Var}_\pi\!\left(
\mathbb E_\pi[f\mid\mathcal F]\mid\mathcal F'
\right)\right]\geq0.
\]
The gain vanishes exactly when $\mathbb E_\pi[f\mid\mathcal F]$ is $\mathcal F'$-measurable: its value is already determined by the entries outside the larger meeting. Merely ignoring the added coordinates in $f$ itself is insufficient, because memberships can be correlated. Both sides use the same $\pi$, which is why the payoffs, $\tau$ and $\overline P$ are held fixed. No independence across memberships is needed.

Third, let $\gamma$ be the assumed coupling: a joint distribution of meetings $\bigl((i,R),(i,R')\bigr)$ whose first component has law $\rho$, whose second has law $\rho'$, and with $R\subseteq R'$ with probability one. For every statistic $f$, linearity of the Dirichlet form in the operator, the marginals of $\gamma$, and the second step for each coupled pair give
\[
\mathcal E_{K_\rho}(f)=\sum_{(i,R)}\rho(i,R)\,\mathcal E_{H_{i,R}}(f)
=\mathbb E_\gamma\bigl[\mathcal E_{H_{i,R}}(f)\bigr]
\leq\mathbb E_\gamma\bigl[\mathcal E_{H_{i,R'}}(f)\bigr]
=\mathcal E_{K_{\rho'}}(f).
\]
Dividing by $\operatorname{Var}_\pi(f)>0$ for nonconstant $f$ and minimizing over $f$ in \eqref{eq:gap-variational} gives $\operatorname{gap}(K_{\rho'})\geq\operatorname{gap}(K_\rho)$. The inequality is weak. If a minimizing statistic for $K_\rho$ satisfies the zero-gain condition above for $\gamma$-almost every coupled pair of meetings, its Rayleigh quotient is unchanged. The gap ordering then implies $\operatorname{gap}(K_{\rho'})=\operatorname{gap}(K_\rho)$.

Finally, suppose that under both laws the individual is uniform and the size is drawn independently of the individual, from distributions $w$ and $w'$ with $w'$ first-order stochastically dominating $w$. The set is then uniform among sets of that size. Draw $i$ uniformly, a uniform random number $U$, and a uniform random ordering of the $\overline P$ slots, all independently. Set $\ell=F_w^{-1}(U)$ and $\ell'=F_{w'}^{-1}(U)$, where $F^{-1}$ denotes the quantile function. Let $R$ and $R'$ be the first $\ell$ and the first $\ell'$ slots in the ordering. Each of $\ell$ and $\ell'$ has its intended distribution, and the first $\ell$ entries of a uniform ordering form a uniformly drawn set of size $\ell$. Hence $(i,R)$ has law $\rho$ and $(i,R')$ has law $\rho'$. Stochastic dominance gives $F_{w'}\leq F_w$, so $\ell\leq\ell'$ and $R\subseteq R'$. Under the uniform size rule, $w_{\ell,k}=1/m_k$ on $\{1,\ldots,m_k\}$ with $m_k=\min\{k,\overline P\}$. Since $m_{k+1}\geq m_k$, the uniform distribution on the larger range dominates, and the gap is nondecreasing in $k$.
\end{proof}

Example~\ref{ex:two-slot-switching} shows where larger meetings help.

\begin{example}[Switching projects]
\label{ex:two-slot-switching}
Take one individual, $n=1$, and two slots, $\overline P=2$, labeled $r$ and $s$. Let the potential reward holding exactly one of the two projects: $Q=0$ if $a_{i,r}+a_{i,s}=1$, and $Q=-c$ with $c>0$ otherwise. Any function on $\mathcal A$ is the exact potential of the game in which every payoff equals it, so the example satisfies the assumptions of the theorem.

Write $\epsilon:=e^{-c/\tau}$. Under $\pi$, the profiles $(a_{i,r},a_{i,s})=(1,0)$ and $(0,1)$ each have probability $1/(2+2\epsilon)$, and the profiles $(0,0)$ and $(1,1)$ each have probability $\epsilon/(2+2\epsilon)$. Take the statistic $f=a_{i,r}$, which records the project on which $i$ works. Each of its values has probability one half, so $\operatorname{Var}_\pi(f)=1/4$.

A meeting on $r$ redraws $a_{i,r}$ given $a_{i,s}$, and $a_{i,r}$ takes the value opposite to $a_{i,s}$ with probability $1/(1+\epsilon)$. Its conditional variance is therefore $\epsilon/(1+\epsilon)^2$, whatever $a_{i,s}$ is. A meeting on $s$ does not move $f$. With one-slot meetings on $r$ and $s$, each with probability one half,
\[
\frac{\mathcal E_{K_\rho}(f)}{\operatorname{Var}_\pi(f)}=\frac{2\epsilon}{(1+\epsilon)^2},
\qquad\text{so}\qquad
\operatorname{gap}(K_\rho)\leq2e^{-c/\tau}.
\]
The gap vanishes as the cost of the intermediate profiles grows relative to $\tau$: to move from $r$ to $s$ one slot at a time, $i$ must pass through $(0,0)$ or $(1,1)$.

A meeting on $\{r,s\}$ redraws the pair jointly. Nothing relevant is held fixed, so its Dirichlet form for $f$ is $\operatorname{Var}_\pi(f)=1/4$, and the move from $(1,0)$ to $(0,1)$ takes one step.
\end{example}

In the estimated model, the costs of the intermediate profiles come from project congestion when holding both projects and from the lost value of the project when holding neither.

\subsection{Identification of Payoffs and Participation Constraints}
\label{app:sorting-consent-identification}

This subsection proves the two results in Section~\ref{sec:identification-formal}: local identification within the maintained specification and separation of the participation penalty from broader additive payoffs. The local score condition addresses the first; a contrast across four configurations addresses the second. Section~\ref{app:identification-verification} reports the corresponding checks at the fitted parameters. The practical margin and effective potential are
\begin{equation}
\label{eq:identification-adjusted-slack}
e_{j,r,t}(A)=s_{j,r,t}(A)
+\min\{b_{jt}-\Delta C_P(p_{jt})+\log\bar P,\,0\},
\end{equation}
and
\begin{equation}
\label{eq:identification-practical-law}
F_t(A;\eta)=Q_t(A;\theta)-\kappa\bar{\mathcal V}_t(A;\theta),
\qquad
\bar\pi_t(A;\eta)=\frac{e^{F_t(A;\eta)}}{\sum_{D\in\mathcal A_t}e^{F_t(D;\eta)}}.
\end{equation}
These are the margin and law used in Section~\ref{sec:structural}, with $\bar\pi_t=\pi_t^\kappa$ and $\eta=(\theta,\kappa)$.

\begin{proof}[Proof of Proposition~\ref{prop:sorting-consent-local-identification}]
On the stated parameter region, $\bar{\mathcal V}_t(A;\theta)$ is constant in $\theta$ for every state. Consequently the gradient of the effective potential $F_t$ is $D_t$, and differentiating the finite normalizing sum gives
\[
\nabla_\eta\log\bar\pi_t(A;\eta_0)
=D_t(A)-\mathbb E_{\bar\pi_t}\{D_t(A)\}.
\]
Thus \eqref{eq:sorting-consent-information} is the sum of the conditional score covariance matrices. For any vector $u$,
\[
u^\top\mathcal Iu
=\sum_t\operatorname{Var}_{\bar\pi_t}\!\bigl(u^\top D_t(A)\bigr).
\]
Every state has positive probability. Hence this expression is zero precisely when $u^\top D_t(A)$ is a state-independent constant for each year. Positive definiteness rules out any nonzero infinitesimal parameter change that leaves every conditional probability unchanged. Equivalently, the Jacobian of the vector of conditional probabilities has full column rank. It therefore has a nonsingular square minor; the inverse function theorem applied to those probabilities gives local injectivity. This proves local identification.

Conditional on the other payoff parameters, the two score-generating statistics for $\gamma=\omega^{\mathrm{gap}}$ and $\kappa$ are $G_t$ and $-\bar{\mathcal V}_t$. Their information matrix has diagonal entries $\sum_t\operatorname{Var}(G_t)$ and $\sum_t\operatorname{Var}(\bar{\mathcal V}_t)$ and off-diagonal entry $-\sum_t\operatorname{Cov}(G_t,\bar{\mathcal V}_t)$. It is positive definite exactly when there is no nonzero common linear combination of these statistics that is constant within every year. If nuisance parameters are also unknown, projecting these centered columns onto the nuisance-score span gives the usual residual covariance, or Schur complement when the nuisance information is nonsingular. This residual covariance must be positive definite to identify the two coefficients separately as part of the regular local model.
\end{proof}

The exponential-family argument clarifies the role of linear independence. For fixed statistics, write $F_t(A)=\alpha^\top S_t(A)-\kappa v_t(A)$, where $S_t$ collects the payoff statistics and $v_t$ is the fixed violation count. Equality of two positive normalized laws requires
\[
(\alpha-\alpha')^\top S_t(A)-(\kappa-\kappa')v_t(A)=c_t
\qquad\text{for every }A,t.
\]
If the augmented statistics admit no nonzero linear combination that is constant within every year, then $\alpha=\alpha'$ and $\kappa=\kappa'$. This is the minimal-exponential-family condition \citep{WainwrightJordan2008}. It identifies canonical coefficients for a fixed statistic collection. In the empirical model, the nonlinear map from structural parameters to the potential and changes in violation classifications require the local score criterion of Proposition~\ref{prop:sorting-consent-local-identification}; the exponential form alone does not establish global structural identification.

Hold other payoff parameters, including the individual effects, fixed within the same violation-pattern region. Write $Q_t=Q_t^0+\gamma G_t$, where $Q_t^0$ omits the gap term. For a same-year contrast $(A_h,B_h)$, let $L_h=\log\{\bar\pi_t(A_h)/\bar\pi_t(B_h)\}$ and let $\Delta_h$ denote the difference between those states. Equation~\eqref{eq:sorting-consent-log-ratio} gives
\[
L_h-\Delta_h Q_t^0
=\gamma\Delta_h G_t-\kappa\Delta_h\bar{\mathcal V}_t.
\]
Two contrasts therefore identify the two coefficients whenever their rows $(\Delta_hG_t,-\Delta_h\bar{\mathcal V}_t)$ are linearly independent. In particular, if $\Delta_1\bar{\mathcal V}_t=0$, $\Delta_1G_t\ne0$, and $\Delta_2\bar{\mathcal V}_t\ne0$, then
\[
\gamma=\frac{L_1-\Delta_1Q_t^0}{\Delta_1G_t},
\qquad
\kappa=\frac{\Delta_2Q_t^0+\gamma\Delta_2G_t-L_2}
{\Delta_2\bar{\mathcal V}_t}.
\]
These are local constructions: the candidate values must lie in the region whose violation indicators were used. Different parameter regions require comparison of the full effective potentials. Indeed, equality of the normalized positive probabilities is equivalent to $F_t(A;\eta)-F_t(A;\eta')=c_t$ for every state, since the difference of log normalizing constants is independent of the state.

Proposition~\ref{prop:participation-nonabsorption} separates participation penalties from the additive payoff class defined in \eqref{eq:additive-payoff-class}. Its proof uses the four-configuration contrast in \eqref{eq:participation-four-profile-contrast}. Fix a year and suppress its index.

\begin{proof}[Proof of Proposition~\ref{prop:participation-nonabsorption}]
For any $\widetilde Q$ in the stated class, individual counts agree between $A_0$ and $B_0$ and between $A_2$ and $B_2$, so portfolio terms cancel in each difference. The weighted networks agree and are unchanged by solo projects, so network terms also cancel. Adding the same solo projects to the two configurations leaves the difference in project-composition values unchanged. Consequently $\mathcal R\widetilde Q=0$.

The contrast coefficients sum to zero, so the log normalizing constant cancels from $\mathcal R\log\pi$. At the maintained parameters,
\[
\mathcal R\log\pi
=\mathcal RQ-\kappa_0\mathcal R\bar{\mathcal V}
=-\kappa_0\mathcal R\bar{\mathcal V}\ne0.
\]
An unconstrained logit law generated by any $\widetilde Q$ in the additive class instead satisfies $\mathcal R\log\widetilde\pi=\mathcal R\widetilde Q=0$. The laws cannot coincide.
\end{proof}

This exclusion allows the alternative payoff functions to change arbitrarily within the displayed class. It does not require their parameters to remain near $\theta_0$, or their implied participation classifications to remain fixed, because the alternative has no participation penalty. The condition concerns the failure statistic at the original parameters. Joint identification of models with different positive penalties and payoff parameters remains a separate question addressed locally by Proposition~\ref{prop:sorting-consent-local-identification}.

\begin{example}[Portfolio congestion and team participation]
\label{ex:portfolio-team-consent}
Take six economists in the same section, $\bar P=400$ labeled slots, and a per-person cap of at least four projects. Let
\[
x=(0,0,0,1,1,2),\qquad \beta_i=x_i-\log400,
\]
\[
C_P(p)=\frac12(p-2)_+^2,\qquad
V_{\mathrm{size}}(n)=3\,\mathbf1\{n\geq2\},\qquad
w_{ij}=-(\beta_i-\beta_j)^2.
\]
Set section, tenure, past-collaboration, and degree terms to zero, and take $C_G(1)=0$. Consider the following memberships in the first four slots; the remaining slots are empty:
\begin{center}
\begin{tabular}{@{}ccc@{}}
\hline\hline
Project & $A_0$ & $B_0$\\
\hline
1 & $\{1,2,3\}$ & $\{1,2,4\}$\\
2 & $\{1,4,5\}$ & $\{1,3,5\}$\\
3 & $\{2,4,6\}$ & $\{2,3,6\}$\\
4 & $\{3,5,6\}$ & $\{4,5,6\}$\\
\hline\hline
\end{tabular}
\end{center}
Every economist has two projects, and every pair has exactly the same number of shared projects in the two configurations. For a member $j$ of a triple $r$, the solo-replacement surplus is
\[
s_{j,r}=3-\sum_{\ell\in N_r\setminus\{j\}}(x_j-x_\ell)^2.
\]
Initially $\Delta C_P(2)=0$ and $\beta_j+\log400=x_j\geq0$, so the empirical margin equals $s_{j,r}$. Evaluating these surpluses gives
\[
\bar{\mathcal V}(A_0)=4,\qquad \bar{\mathcal V}(B_0)=3.
\]
Form $A_2$ and $B_2$ by adding solo projects for economist 1 in slots 5 and 6. Her project count rises to four and $\Delta C_P(4)=3/2$. Her surpluses in the two original teams are $(3,1)$ in $A_0$ and $(2,2)$ in $B_0$. Since $\beta_1+\log400=0$, her adjusted exit margins after the additions are these surpluses minus $3/2$. One additional membership therefore fails in $A_2$, while both memberships still pass in $B_2$. Other members' checks are unchanged, and the new singleton memberships are not checked. Hence
\[
\bar{\mathcal V}(A_2)=5,\qquad
\bar{\mathcal V}(B_2)=3,\qquad
\mathcal R\bar{\mathcal V}=1.
\]
The stationary law consequently satisfies
\begin{equation}
\label{eq:participation-four-profile-odds}
\log\frac{\pi(A_2)\pi(B_0)}{\pi(B_2)\pi(A_0)}=-\kappa.
\end{equation}
Every unconstrained model in \eqref{eq:additive-payoff-class} instead assigns this log probability contrast the value zero.
\end{example}

The example shows how grouping the same bilateral relationships into different teams can make portfolio changes affect participation checks differently. A payoff premium for cross-section pairs with no prior collaboration is included in $\Psi(G^W)$ and satisfies the same zero-contrast restriction. Such a term can resemble the participation penalty on a subset of dyads, but cannot generally reproduce the contrast in \eqref{eq:participation-four-profile-odds}. The displayed checks have strict signs, so the example persists in a neighborhood of its parameters. It is an illustration of structural nonredundancy, not verification at the estimated parameters. Informative configurations can also have very small probability, leaving limited precision in an observed sample.

An unrestricted alternative game can reproduce the weights by defining $\widetilde U_i(A)=U_i(A)-\kappa\bar{\mathcal V}(A;\theta)$, holding the original failure statistic fixed. These payoffs have effective potential $Q-\kappa\bar{\mathcal V}$. Under Proposition~\ref{prop:participation-nonabsorption}'s condition, that potential lies outside \eqref{eq:additive-payoff-class}: it introduces dependence between portfolios and particular teams beyond those additive components. The stationary law therefore separates the penalty from the admitted payoff class, while its interpretation as stochastic consent uses the maintained behavioral construction.

A separate payoff overlap concerns product and gap sorting. For weighted degree $d_i^W=\sum_{j\ne i}g_{ij}$, define $T_t=\sum_{i<j}g_{ij}(\beta_i-\bar\beta)(\beta_j-\bar\beta)$. Direct expansion yields
\[
G_t=\sum_i\beta_i^2d_i^W-2\sum_{i<j}g_{ij}\beta_i\beta_j,
\qquad
T_t=\sum_{i<j}g_{ij}\beta_i\beta_j
-\bar\beta\sum_i\beta_id_i^W
+\bar\beta^2\sum_{i<j}g_{ij}.
\]
When the individual effects and every $d_i^W$ are fixed, $G_t+2T_t$ is constant. Within a fixed violation-pattern region, such rewiring contrasts identify only $\omega^{\mathrm{product}}-2\omega^{\mathrm{gap}}$ from the payoff component. Separating the two coefficients needs additional weighted-degree variation or a coefficient restriction; threshold crossings outside this region may provide further global information. Joint regular identification with consent additionally needs the rank condition in the proposition.

If no candidate state violates a check, $\kappa$ drops out. If all individual effects are equal, the gap statistic is zero. If a sorting statistic and the violation count are affinely dependent over all relevant states, their coefficients cannot be separated within that parameter region. Threshold crossings can supply additional information, but invalidate the smooth local score calculation; the proposition does not assert global identification across different violation patterns.

The individual effects are latent and must be learned jointly from activity and matching. Correlation between coauthors' fitted effects alone is therefore not independent evidence identifying sorting: changing the sorting parameter can change those effects too. Short histories, little variation near exit cutoffs, unobserved research opportunities, and persistence in unobserved pair affinity can weaken separation. A hierarchy improves precision but does not replace these identification requirements.

Under depth weighting, use $\bar{\mathcal V}_{\ell,t}(A;\theta)=\sum_{r:n_r>1}\sum_{j\in N_r}[-e_{j,r,t}(A)]_+$ in place of the count. Probability ratios still apply, but the payoff score becomes $\nabla_\theta Q_t-\kappa_\ell\nabla_\theta\bar{\mathcal V}_{\ell,t}$, since violation depths change with payoff parameters even away from a cutoff. Count and depth identification should therefore be assessed separately. Neither stationary law identifies the meeting bound $k$ or meeting probabilities; those require transition information. The project payoff normalization in Remark~\ref{rem:practical-slot-exit} is a maintained economic convention; the slot representation alone does not identify it.

\subsubsection{Verifying the Identification Conditions}
\label{app:identification-verification}

The preceding results require two distinct checks at the fitted parameters. Proposition~\ref{prop:participation-nonabsorption} asks whether the participation penalty can be distinguished from the broader additive payoff class; Proposition~\ref{prop:sorting-consent-local-identification} asks whether all free payoff and participation parameters are locally identified jointly. Both checks use the full fitted payoff specification, including the individual effects, while holding each year's population, characteristics, workload, collaboration history, slot universe, and participation cap fixed.

\medskip
\noindent\emph{Proposition~\ref{prop:participation-nonabsorption}: separation from additive payoffs.}
The required check is $\hat\kappa>0$ and $\mathcal R\bar{\mathcal V}(\cdot;\hat\theta)\neq0$ for the four-configuration contrast in \eqref{eq:participation-four-profile-contrast}. Choose actual economists in a given year and construct $A_0,B_0$ with identical individual project counts and weighted collaboration networks; obtain $A_2,B_2$ by adding two fresh solo projects for the same economist to each. All four configurations must respect the slot and per-person caps, but need not pass the participation checks, since the finite-penalty model gives them positive probability. Evaluate the failure counts at the fitted payoffs and adjusted exit benchmark, counting each failed nonsingleton membership once. One nonzero contrast in one year suffices to exclude every unconstrained potential in \eqref{eq:additive-payoff-class} from reproducing the fitted conditional distributions; no normalizing constants are needed.

Each construction starts from an observed annual network, draws six economists with room for the added memberships, and places the four triples of Example~\ref{ex:portfolio-team-consent} in empty slots under its two arrangements. Project counts and weighted networks agree within each pair, and $|\mathcal RQ|\leq1.2\times10^{-12}$, confirming the cancellation implied by the additive specification. At the posterior mean, 816 of 39{,}000 constructions (3{,}000 per year) have nonzero violation contrasts, with at least one in every year. In a 2013 example, the two solo projects lower one of the first economist's adjusted exit margins from $0.007$ in $B_0$ to $-0.138$ in $B_2$, while the failure count is unchanged between $A_0$ and $A_2$, giving $\mathcal R\bar{\mathcal V}=-1$. With 300 constructions per year at each joint state, every state has nonzero contrasts in at least two years; 62\% have them in all thirteen. Since $\kappa>0.17$ at all 500 states, Proposition~\ref{prop:participation-nonabsorption}'s condition holds at each of them and at the posterior mean.

\medskip
\noindent\emph{Proposition~\ref{prop:sorting-consent-local-identification}: joint local identification.}
The rank check concerns
\[
\mathcal I(\hat\eta)
=\sum_t\operatorname{Cov}_{\bar\pi_t(\cdot;\hat\eta)}\!\bigl(D_t(A)\bigr),
\qquad \hat\eta=(\hat\theta,\hat\kappa),
\]
with $D_t$ as in \eqref{eq:sorting-consent-information}. All 470 columns are included: the 20 free common payoff parameters, the 449 individual effects, and $-\bar{\mathcal V}_t$. Positive definiteness requires both full rank of the nuisance payoff scores and variation in the violation count after projection onto them. Configurations are centered within each year before covariances are summed, always at fixed parameters; pooling different posterior states would mix configuration variation with parameter uncertainty.

This calculation covers the posterior mean and four joint states, numbered 1, 127, 253, and 377, rather than all 500 states. At each parameter vector, chains start from the observed networks, run for $10^7$ updates, and record 1{,}000 configurations per chain, $2\times10^6$ updates apart. Four chains per year yield 52{,}000 configurations at the posterior mean; two per year yield 26{,}000 at each joint state. Common payoff scores are central differences of the potential with step $10^{-4}\max\{1,|\theta_k|\}$. Individual-effect scores are analytic; spot checks against central differences agree to within $3\times10^{-9}$.

The estimated matrix has full rank at all five parameter vectors. After scaling to a unit diagonal, its smallest eigenvalue ranges from $5.5\times10^{-3}$ to $8.7\times10^{-3}$; independent halves of the chains reproduce it to within 4\%. The weakest direction combines $v_{\mathrm{size},2}$ and $\rho_{\mathrm{size}}$, which govern the same team-size profile. Between 38\% and 44\% of the violation count's variance remains after projection onto the 469 nuisance columns. After removing the other 468 columns, the correlation between $G_t$ and $-\bar{\mathcal V}_t$ is only $-0.07$ to $-0.02$, and 30--38\% of the variance of $G_t$ remains. Thus the numerical rank check separates the participation penalty and sorting from the remaining payoff parameters.

The simulated configurations also supply a finite-configuration version of the proposition's local argument. Full rank of the within-year covariance sum means that their within-year score differences span all 470 parameter directions. No nonsingleton membership has a zero margin in these configurations, so their failure counts remain fixed in a common neighborhood of each checked parameter vector. Select 470 within-year log probability ratios with independent score differences. Their Jacobian is nonsingular by \eqref{eq:sorting-consent-log-ratio}, and the inverse function theorem gives local identification from these ratios. This argument uses the simulated configurations as a numerical rank witness and does not require the chains to have reached stationarity; stationarity matters for interpreting the eigenvalues and variance shares as information under the fitted law. It also avoids claiming that simulation verifies unchanged violation indicators over the entire state space.

The neighborhoods can be small: the minimum absolute margins are $4\times10^{-7}$ to $4\times10^{-6}$. Perturbing each of the 20 common payoff parameters by the finite-difference step changes $\bar{\mathcal V}_t$ in 0.03--0.09\% of evaluations. This diagnostic measures changes in the aggregate count, not every membership indicator, and does not cover perturbations of the individual effects. The results support local identification at the five checked vectors; they neither establish global identification nor verify rank throughout the posterior.

\begin{table}[!htbp]
\centering
\caption{Identification Checks at the Estimated Parameters}
\label{tab:identification-checks}
\footnotesize
{\setlength{\tabcolsep}{4pt}\renewcommand{\arraystretch}{1.2}\begin{tabular}{@{}l c cc ccccccc@{}}
\hline\hline
 & & \multicolumn{2}{c}{Prop.~\ref{prop:participation-nonabsorption}} & \multicolumn{7}{c}{Prop.~\ref{prop:sorting-consent-local-identification}} \\
\cline{3-4}\cline{5-11}
 & & & $\mathcal R\bar{\mathcal V}\neq0$ & Min.\ eig. & & \multicolumn{2}{c}{Retained} & Resid. & Min.\ $|e|$ & Changes \\
\cline{7-8}
Parameters & $\kappa$ & Years & (\%) & ($10^{-3}$) & Cond. & $\bar{\mathcal V}$ & $G$ & corr. & ($10^{-6}$) & (\%) \\
\hline
Posterior mean & 0.34 & 13 & 2.1 & 8.7 & 802 & 0.40 & 0.38 & $-$0.02 & 3.1 & 0.03 \\
Joint state 1 & 0.26 & 13 & 1.8 & 6.4 & 1{,}101 & 0.38 & 0.36 & $-$0.06 & 1.4 & 0.03 \\
Joint state 127 & 0.40 & 13 & 2.3 & 6.2 & 1{,}179 & 0.44 & 0.34 & $-$0.06 & 3.7 & 0.06 \\
Joint state 253 & 0.32 & 10 & 0.9 & 5.5 & 1{,}308 & 0.43 & 0.30 & $-$0.07 & 0.4 & 0.04 \\
Joint state 377 & 0.35 & 13 & 1.9 & 7.6 & 978 & 0.41 & 0.31 & $-$0.04 & 2.6 & 0.09 \\
\hline
All 500 joint states & 0.18--0.48 & 2--13 & 0.1--6.3 & & & & & & & \\
\hline\hline
\end{tabular}}
\par\smallskip
\begin{minipage}{0.97\textwidth}\floatnotesfont
Notes. Posterior mean: means of the 21 free common parameters and 449 individual effects. Proposition~\ref{prop:participation-nonabsorption}: years with at least one nonzero violation contrast, and percentage of constructions with such contrasts. The last row gives ranges over all 500 states for this check only. Proposition~\ref{prop:sorting-consent-local-identification}: results at the posterior mean and the four listed states. Min.\ eig.\ and Cond.: smallest eigenvalue and condition number of the estimated covariance matrix scaled to a unit diagonal. Retained $\bar{\mathcal V}$: share of $\sum_t\operatorname{Var}(\bar{\mathcal V}_t)$ remaining after projection onto the other 469 columns. Retained $G$ and Resid.\ corr.: variance share of $G_t$ and correlation of $G_t$ with $-\bar{\mathcal V}_t$ after projection onto the other 468 columns. Min.\ $|e|$: smallest absolute nonsingleton membership margin. Changes: percentage of evaluations at $\theta_k\pm10^{-4}\max\{1,|\theta_k|\}$, for the 20 common payoff parameters only, in which the aggregate violation count changes.
\end{minipage}
\end{table}

\clearpage

\section{Additional Analysis and Discussion}
\label{app:additional-analysis}

This appendix develops the comparisons and calculations used in the main analysis. It first relates project stability to stability of the induced collaboration network, then derives individual payoff margins and illustrates the role of consent.

\subsection{Relationship to Equilibria with Consent in Social Networks}
\label{sec:network-correspondence}

This section compares a project equilibrium with the equilibrium of its induced collaboration network. We adapt the setting of \citet{Badev2021} by dropping endogenous behavior and triangle effects, retaining heterogeneous pair values, degree congestion, and consent to links. The comparison is conditional: network stability describes the choice of collaborators, while project equilibrium also requires an optimal number of projects. We use $K$ for the number of project memberships under consideration and $q$ for the number of individuals in a network consideration set.

\subsubsection{Network Stability and Induced Payoffs}

\begin{definition}[Network stability, adapted from \citet{Badev2021}]
\label{def:net-stability}
Consider a simple undirected network $G=(g_{ij})$ on $N$, with payoffs $U_i^G(G)$. For $2\leq q\leq n$, let $I_q\subseteq N$ contain $i$ and have $q$ members. Write $\mathcal H_i(G;I_q)$ for the networks that agree with $G$ except possibly on links $ij$ with $j\in I_q\setminus\{i\}$. The network is \emph{$q$-stable} if, for every $i$ and $I_q$, the current network solves
\begin{equation}
\label{eq:net-stability}
\begin{aligned}
\max_{H\in\mathcal H_i(G;I_q)}\quad &U_i^G(H)\\
\text{subject to}\quad &h_{ij}=1\ \Longrightarrow\ 
\Delta_{g_{ij}}U_j^G(H)\geq0,
\qquad j\in I_q\setminus\{i\}.
\end{aligned}
\end{equation}
Here $\Delta_{g_{ij}}U_j^G(H)=U_j^G(H^{+ij})-U_j^G(H^{-ij})$, where $H^{+ij}$ and $H^{-ij}$ set the undirected link $ij$ to one and zero, respectively, holding all other links fixed.
\end{definition}

Consent is evaluated at the candidate network for retained or added links within the consideration set. Links outside that set are held fixed and are not reassessed. Thus $q$ individuals permit reconsidering $q-1$ incident links. The case $q=2$ gives pairwise stability, while $q=n$ permits reconsidering all incident links; see Proposition~2 of \citet{Badev2021}.

To obtain a network representation of project payoffs, restrict every project to one or two participants and allow each pair to share at most one project. These restrictions apply to both profiles and deviations. Then $g_{ij}\in\{0,1\}$ and $G^W=G^U=G$. Project creation, solo projects, and changes in project counts remain available as in Definition~\ref{def:equilibrium}. We retain the normalizations $V(X_i,\{X_i\})=0$ and $C_G(0)=0$.

\begin{definition}[Induced network game]
\label{def:induced-network}
In this restricted project environment, define the net value of a relationship by
\begin{equation}
\label{eq:induced-pair-value}
\theta_{ij}:=V(X_i,\{X_i,X_j\})+w(X_i,X_j)-\Delta C_G(0).
\end{equation}
The induced network game has payoffs
\begin{equation}
\label{eq:induced-network-payoff}
U_i^G(G)=\sum_{j\ne i}g_{ij}\theta_{ij}
-\psi_D\left[\frac{d_i(d_i+1)}2+\sum_{j\ne i}g_{ij}d_j\right].
\end{equation}
Its link increment also equals the mixed membership difference
\[
\Delta_{g_{ij}}U_i^G
=\Delta_{a_{i,r}}\Delta_{a_{j,r}}U_i
=U_i(A^{11})-U_i(A^{10})-U_i(A^{01})+U_i(A^{00}),
\]
where $r$ has no participants other than $i,j$, $A^{xy}$ sets their memberships in $r$ to $x,y$, all other memberships are fixed, and the pair shares no other project.
\end{definition}

Each dyadic project contributes one relationship and each solo project contributes only its own-project return. Since $\Delta C_G(0)=C_G(1)$, direct substitution into \eqref{eq:payoffs-project} gives
\begin{equation}
\label{eq:project-network-decomposition}
U_i(A)=B_i(p_i)+U_i^G(G(A)),
\qquad B_i(p):=\beta_i p-C_P(p).
\end{equation}
The mixed difference in Definition~\ref{def:induced-network} follows because adding $j$ to $r$ changes $i$'s payoff only when $i$ participates there; it converts her solo project into the dyadic project without changing her project count.

Assumption~\ref{ass:payoffs}(ii)--(iii) makes $\theta_{ij}=\theta_{ji}$: the two members have the same dyadic team value $\Phi(\{i,j\})$, and $w$ is symmetric. For a present link, with degrees evaluated with the link present,
\begin{equation}
\label{eq:network-link-margin}
\Delta_{g_{ij}}U_i^G(G)
=\Delta_{g_{ij}}U_j^G(G)
=\theta_{ij}-\psi_D(d_i+d_j).
\end{equation}
Replacing a dyadic membership by a solo project preserves the member's project count and removes precisely this link. Hence, for $N_r=\{i,j\}$, the individual slack from Section~\ref{sec:participation-slacks} is
\begin{equation}
\label{eq:dyadic-participation-slack}
\sigma_{i,r}(A)
=\theta_{ij}-\psi_D(d_i+d_j)-\lambda_i(p_i),
\qquad \lambda_i(p_i)=[\Delta C_P(p_i)-\beta_i]_+.
\end{equation}
Thus project participation agrees with network consent whenever the solo-project return is nonnegative. Otherwise, participation requires the link benefit to cover the additional shortfall $\lambda_i(p_i)$.

\subsubsection{Conditional Equivalence}

\begin{proposition}[Conditional equivalence with network stability]
\label{prop:team_implies_network_stability}
Let $n\geq2$, suppose Assumption~\ref{ass:payoffs}(ii)--(iii) holds, $C_P(0)=C_G(0)=0$, $\Delta C_P$ is nondecreasing, and $\psi_D\geq0$. Impose the dyadic and no-repetition restrictions above on profiles and deviations. Fix $K\geq2$ and a profile $(P,A)$ in which every individual has at least one solo project. This last requirement applies to the profile, not to its deviations.

Then $(P,A)$ is $K$-team stable under Definition~\ref{def:equilibrium}, within the restricted project environment, if and only if both of the following hold:
\begin{enumerate}[label=(\roman*)]
\item each individual's project count is optimal for her own-project payoff,
\begin{equation}
\label{eq:network-project-count-optimality}
\Delta C_P(p_i)\leq\beta_i\leq\Delta C_P(p_i+1),
\qquad i\in N;
\end{equation}
\item the induced network $G$ is $n$-stable for the payoffs \eqref{eq:induced-network-payoff}.
\end{enumerate}
Condition (i) is equivalent to $p_i\in\arg\max_{p\in\mathbb N_0}B_i(p)$. Whenever these conditions hold, $G$ is $q$-stable for every $2\leq q\leq n$. Moreover, no individual can gain from any unilateral project deviation satisfying the dyadic and no-repetition restrictions, even without the participation constraints.
\end{proposition}

\begin{proof}
\emph{Project-count optimality.} Suppose $(P,A)$ is $K$-team stable. Individual $i$ can delete one of her solo projects, with payoff gain $\Delta C_P(p_i)-\beta_i$. This leaves the network unchanged, weakly increases her slacks in remaining teams because $\Delta C_P$ is nondecreasing, and leaves other individuals' slacks unchanged. The deletion is feasible, so $\Delta C_P(p_i)\leq\beta_i$.

Adding a new solo project gives payoff gain $\beta_i-\Delta C_P(p_i+1)$. If positive, the solo-project return is positive both before and after the addition. Thus $\lambda_i$ is zero at both profiles, all existing team slacks are unchanged, and the addition is feasible. Equilibrium therefore requires $\beta_i\leq\Delta C_P(p_i+1)$. These one-membership deviations prove \eqref{eq:network-project-count-optimality}. Since $B_i(p)-B_i(p-1)=\beta_i-\Delta C_P(p)$ is nonincreasing, the two inequalities are exactly the conditions for $p_i$ to maximize $B_i$ globally. In particular, $\lambda_i(p_i)=0$ for every individual.

\emph{Network optimality from project stability.} Current individual slacks are nonnegative by Remark~\ref{rem:slack-equilibrium}. Equations~\eqref{eq:network-link-margin}--\eqref{eq:dyadic-participation-slack}, together with $\lambda_i(p_i)=0$, imply consent on every current link. Fix $i$ and hold links between other individuals fixed. Define
\[
d_j^{-i}:=\sum_{\ell\ne i,j}g_{j\ell},
\qquad t_{ij}:=\theta_{ij}-\psi_D(d_j^{-i}+1).
\]
For a chosen set of neighbors $S$, $i$'s network payoff is
\[
U_i^G(S)=\sum_{j\in S}t_{ij}
-\frac{\psi_D}{2}|S|(|S|+1).
\]
Write $S^*=\{j:g_{ij}=1\}$ and $d=|S^*|$. Current-link consent gives
\begin{equation}
\label{eq:team_network_existing}
t_{i\ell}\geq\psi_D d,
\qquad \ell\in S^*.
\end{equation}

If a missing link $ij$ satisfied $t_{ij}>\psi_D(d+1)$, individual $i$ could leave one of her solo projects and join a solo project of $j$. This changes two memberships and preserves every individual's project count. The payoff gain is $t_{ij}-\psi_D(d+1)>0$. By symmetry and \eqref{eq:dyadic-participation-slack}, this is also the slack of each member of the new dyadic project: both retain $\lambda_i(p_i)=\lambda_j(p_j)=0$. The deviation is feasible and profitable, contradicting $K\geq2$ stability. Hence
\begin{equation}
\label{eq:team_network_missing}
t_{ij}\leq\psi_D(d+1),
\qquad j\notin S^*,\quad j\ne i.
\end{equation}

If instead $j\notin S^*$ and $\ell\in S^*$ satisfied $t_{ij}>t_{i\ell}$, individual $i$ could leave her project with $\ell$ and join a solo project of $j$. This two-membership replacement preserves her project count and degree. Its payoff gain is $t_{ij}-t_{i\ell}>0$, and both members' slacks in the new dyadic project equal
\[
t_{ij}-\psi_D d>t_{i\ell}-\psi_D d\geq0.
\]
The old project becomes a singleton. As in Definition~\ref{def:equilibrium}, the participation checks apply to reconsidered projects in which $i$ participates at the candidate profile; projects outside that set supply no additional veto. Thus this deviation is feasible and profitable, again a contradiction. We obtain
\begin{equation}
\label{eq:team_network_ranking}
t_{ij}\leq t_{i\ell},
\qquad j\notin S^*,\quad \ell\in S^*.
\end{equation}
Both constructions change only two project memberships and are available for every $K\geq2$.

Order the $n-1$ values as $t_{i,(1)}\geq\cdots\geq t_{i,(n-1)}$ and let
\[
F_i(s):=\sum_{h=1}^{s}t_{i,(h)}
-\frac{\psi_D}{2}s(s+1),\qquad 0\leq s\leq n-1.
\]
Equation~\eqref{eq:team_network_ranking} makes $S^*$ a set of the $d$ largest values, allowing ties, so $U_i^G(S^*)=F_i(d)$. The increments
\[
F_i(s)-F_i(s-1)=t_{i,(s)}-\psi_D s
\]
are nonincreasing because $\psi_D\geq0$. Equations~\eqref{eq:team_network_existing} and \eqref{eq:team_network_missing} give a nonnegative last increment into $d$ when $d>0$ and a nonpositive first increment beyond $d$ when $d<n-1$. Hence $F_i(d)\geq F_i(s)$ for every $s$, including the boundary cases. The current neighbor set therefore maximizes $U_i^G$ over all incident-link choices, even without consent. Since current links satisfy consent, $G$ is $q$-stable for every $q$, establishing (ii).

\emph{Reverse implication.} Suppose (i)--(ii) hold. Fix $i$ and let $\widehat H$ maximize $U_i^G$ over all incident-link configurations, initially ignoring consent. Every link $ij$ present in $\widehat H$ has $\Delta_{g_{ij}}U_i^G(\widehat H)\geq0$, because deleting it cannot improve the maximizer's payoff. Equation~\eqref{eq:network-link-margin} makes this equal to $j$'s link increment. Thus $\widehat H$ satisfies all consent constraints in the $n$-player problem. Since $G$ solves that problem, it also maximizes $U_i^G$ over unrestricted incident-link choices. This argument applies to every individual and gives nonnegative link increments on current links.

Condition (i) gives $\lambda_i(p_i)=0$, so every current dyadic membership has nonnegative individual slack. Any unilateral project change respecting the dyadic and no-repetition restrictions changes only the mover's project count and incident links. By \eqref{eq:project-network-decomposition}, its payoff change is the sum of a change in $B_i$, which cannot be positive by (i), and a change in $U_i^G$, which cannot be positive by the preceding argument. Thus no such deviation is profitable, even before imposing participation constraints. The current profile is feasible and optimal in every project consideration set, proving $K$-team stability. The same unrestricted network optimality and current-link consent give $q$-stability for every $q$.
\end{proof}

Solo projects have two roles in this correspondence. They provide an existing project into which another individual can enter and allow the mover to give up a solo project while adding a relationship, keeping her workload fixed. Retaining a solo project in equilibrium also implies $\lambda_i(p_i)=0$, so the individual participation slack becomes precisely the network link-consent margin. The assumption is needed only at the profile being compared: a deviation may use an individual's last solo project. Without solo opportunities, the link implication can fail. For example, with two individuals, $\beta_i=0$, $C_P(p)=p^2$, dyadic team value $3$, and other payoff terms zero, the empty project profile is stable because any unilateral new project is solo and yields $-1$. Yet a dyadic project gives each individual payoff $2$, and the induced network has $\theta_{12}=3$, so the empty network is not pairwise stable.

Project-count optimality supplies the reverse implication. A network does not record solo projects and cannot rule out excessive or insufficient project activity. Condition~\eqref{eq:network-project-count-optimality} makes the own-project block optimal while network stability makes the collaboration block optimal; project loads are endogenous throughout. In particular, a given $n$-stable graph admits a representation covered by the proposition only if one can select optimal counts with $p_i\geq d_i+1$ for every individual. One then creates one project per edge and $p_i-d_i$ solo projects for individual $i$. The proposition does not guarantee such a representation for arbitrary own-project returns.

The comparison also clarifies the consideration bounds. Two project memberships suffice to replace one collaborator with another, an adjustment involving two incident links and three individuals in the network formulation. Consequently, $K=2$ project stability need not coincide with $q=2$ network stability. The stronger conclusion of stability for every $q$ uses the symmetric link values and convex degree cost in \eqref{eq:induced-network-payoff}: additions and replacements already identify a globally optimal neighbor set. It is a property of this specialization of \citet{Badev2021}, not a claim about the model with endogenous behavior and triangle effects restored.

\begin{samepage}
Finally, the dyadic and no-repetition restrictions make the collaboration network sufficient for the team and collaboration blocks of payoff. For three individuals, compare one triple project plus a solo project for each member with the three dyadic projects. Both profiles have $p_i=2$ and $g_{ij}=1$ for every pair, but give individual $i$ team benefits $V(X_i,\{X_1,X_2,X_3\})$ and $\sum_{j\ne i}V(X_i,\{X_i,X_j\})$, respectively, which need not agree. Repeated collaborations create a further distinction between $G^W$ and $G^U$. The general project model therefore retains information and participation restrictions beyond those summarized by a simple network.
\par
\end{samepage}

\subsection{Payoff Margins and a Constrained Equilibrium}
\label{app:setup-details}

The payoff blocks govern starting a project, joining existing collaborators, and forming new relationships. The calculations below derive these margins and then illustrate why an equilibrium with consent need not maximize the unconstrained potential. The participation setup is shown in Figure~\ref{fig:intro-setup}.

\paragraph{Team values with the potential structure.} Two examples of Assumption~\ref{ass:payoffs}(ii), combined in Section~\ref{sec:structural}, are a value of team size, $V(X_i,\{X_j\}_{j\in S})=V_{\mathrm{size}}(|S|)$ with $V_{\mathrm{size}}(1)=0$ and $\Phi(S)=\sum_{k=1}^{|S|}V_{\mathrm{size}}(k)$, and a value of team structure that depends on the number $m_{S,c}=\sum_{j\in S}\1\{c(j)=c\}$ of members in $i$'s group $c(i)$ (in the application, the section), $V(X_i,\{X_j\}_{j\in S})=V_{\mathrm{section}}(m_{S,c(i)})$ with $V_{\mathrm{section}}(1)=0$ and $\Phi(S)=\sum_c\sum_{k=1}^{m_{S,c}}V_{\mathrm{section}}(k)$. A value that depends on the share of same-group members, or a curvature specific to each member, does not in general have this form. A value common to every project, solo ones included, would add the same amount to every $\beta_i$, which is why solo projects are normalized to have no team value.

The payoff blocks enter three decisions: starting a solo project, joining a team of existing collaborators, and joining a team that includes new collaborators. Throughout, $N_r$ and $n_r$ describe the team after entry, including $i$.

First, starting a solo project changes only the own-project block. The gain relative to not undertaking the project is the baseline return net of the additional project cost,%
\footnote{Let $p_i$ be the total number of projects including the contemplated addition, i.e. the margin is $\beta_i p_i - C_P(p_i) - \left( \beta_i (p_i-1) - C_P(p_i-1)\right)$.}
\begin{equation}
\label{eq:alone}
\Delta_{a_{i,r}}U_i^{\mathrm{alone}}
=
\beta_i - \Delta C_P(p_i).
\end{equation}
Under Assumption~\ref{ass:payoffs}(i), the marginal cost is nondecreasing in $p_i$, so this net gain is nonincreasing.

Second, joining a team of existing collaborators adds team and relationship benefits to the solo-project gain. No new relationships are formed, so degrees and network congestion remain unchanged. Each colleague contributes the value $w(X_i,X_j)$ net of the cost of one more common project, $\Delta C_G(g_{ij})$, where $g_{ij}\geq1$ counts shared projects before entry:
\begin{align}
\label{eq:team-vs-alone}
\Delta_{a_{i,r}}U_i^{\mathrm{team}}
={}&
\Delta_{a_{i,r}}U_i^{\mathrm{alone}}
+ V(X_i,\{X_j\}_{j\in N_r})
+ \sum_{\substack{j\in N_r\\j\neq i}}
\Bigl[w(X_i,X_j) - \Delta C_G\bigl(g_{ij}\bigr)\Bigr].
\end{align}
The team is preferable whenever its team benefit and net relationship benefits sum to a positive amount, even if working with a particular member yields a negative pair value.

Third, joining a team with new collaborators also changes the network congestion block. Each new colleague incurs the first-project cost $\Delta C_G(0)$, and the new relationships raise both $i$'s degree and the degrees of those colleagues. After entry, $n_r^{(i)}\leq n_r-1\leq d_i$: the new colleagues are a subset of the other team members, who in turn are a subset of all of $i$'s collaborators. The resulting gain is
\begin{align} \notag
\Delta_{a_{i,r}}U_i^{\mathrm{new}}
={}&
\Delta_{a_{i,r}}U_i^{\mathrm{team}}
-\sum_{j\in N_r^{(i)}}\Delta C_G(0)
-\psi_D
\left[
\frac{1}{2}n_r^{(i)}
\left(2d_i-n_r^{(i)}+1\right)
+
\sum_{j\in N_r^{(i)}}d_j
\right]\\
\label{eq:new-vs-existing}
={}&
\Delta_{a_{i,r}}U_i^{\mathrm{team}}
-\Delta C_G(0)\,  n_r^{(i)}
-\psi_D \left(d_i n_r^{(i)} -
\frac{n_r^{(i)}\left(n_r^{(i)}-1 \right)}{2}
+ \sum_{j\in N_r^{(i)}}d_j \right),
\end{align}
where $\Delta_{a_{i,r}}U_i^{\mathrm{team}}$ is \eqref{eq:team-vs-alone} with the cost $\Delta C_G(g_{ij})$ counted for the existing collaborators only, and the degrees are those after $i$ joins. If $\Delta C_G(0)\geq0$ and $\psi_D\geq0$, the costs of establishing new relationships imply $\Delta_{a_{i,r}}U_i^{\mathrm{new}}\leq \Delta_{a_{i,r}}U_i^{\mathrm{team}}$, since $d_i\geq n_r^{(i)}$; these costs increase in $n_r^{(i)}$, $d_i$, and $d_j$ for all $j\in N_r^{(i)}$. Substituting \eqref{eq:team-vs-alone} into \eqref{eq:new-vs-existing}, with $g_{ij}=0$ for $j\in N_r^{(i)}$, gives \eqref{eq:new-vs-alone}.

The following example shows a $k$-team stable equilibrium that is not $k$-team optimal in the absence of the stability constraint, and compares it with the global maximizer of the potential.

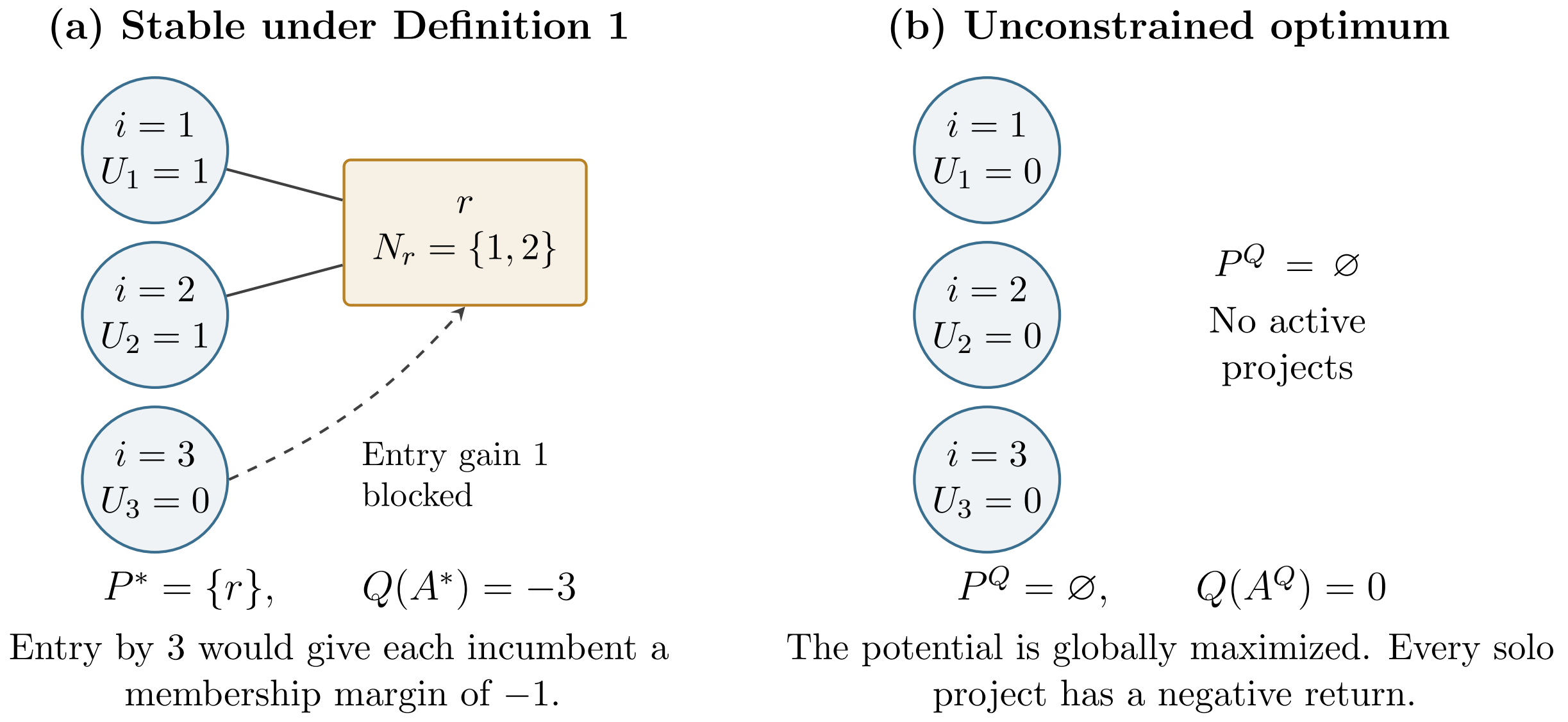
\begin{figure}[!htbp]
\centering
\caption{A Constrained Equilibrium and a Global Potential Maximizer}
\label{fig:stable-versus-potential}
\begin{minipage}[t]{0.48\textwidth}
\centering
\textbf{(a) Stable under Definition~\ref{def:equilibrium}}\\[6pt]
\begin{tikzpicture}[
    person/.style={individual node},
    project/.style={project node},
    every node/.style={font=\small}
]
\path[use as bounding box] (-0.9,-2.5) rectangle (4.7,2.5);
\node[person] (c1) at (0,1.7) {$i=1$\\$U_1=1$};
\node[person] (c2) at (0,0) {$i=2$\\$U_2=1$};
\node[person] (c3) at (0,-1.7) {$i=3$\\$U_3=0$};
\node[project] (cr) at (3.2,0.85) {$r$\\$N_r=\{1,2\}$};
\draw[diagram edge] (c1) -- (cr);
\draw[diagram edge] (c2) -- (cr);
\draw[diagram edge,dashed,->,>=stealth] (c3.east) to[bend right=12]
    node[pos=0.42,below right=3pt,align=left,font=\footnotesize] {Entry gain $1$\\blocked} (cr.south);
\end{tikzpicture}\\[-2pt]
$P^\ast=\{r\},\qquad Q(A^\ast)=-3$\\[3pt]
\parbox[t]{\linewidth}{\centering\small Entry by 3 would give each incumbent a membership margin of $-1$.}
\end{minipage}\hfill
\begin{minipage}[t]{0.48\textwidth}
\centering
\textbf{(b) Unconstrained optimum}\\[6pt]
\begin{tikzpicture}[
    person/.style={individual node},
    every node/.style={font=\small}
]
\path[use as bounding box] (-0.9,-2.5) rectangle (4.7,2.5);
\node[person] (q1) at (0,1.7) {$i=1$\\$U_1=0$};
\node[person] (q2) at (0,0) {$i=2$\\$U_2=0$};
\node[person] (q3) at (0,-1.7) {$i=3$\\$U_3=0$};
\node[align=center,text width=26mm] at (3.1,0) {$P^Q=\varnothing$\\[3pt]No active\\projects};
\end{tikzpicture}\\[-2pt]
$P^Q=\varnothing,\qquad Q(A^Q)=0$\\[3pt]
\parbox[t]{\linewidth}{\centering\small The potential is globally maximized. Every solo project has a negative return.}
\end{minipage}
\par\medskip
{\RaggedRight\floatnotesfont Notes. The two panels use the same parameters from Example~\ref{ex:stable-not-unconstrained}. Blue circles represent individuals and the amber rectangle represents an active project. Solid edges are memberships; the dashed arrow is individual 3's profitable but infeasible entry. The network in panel (a) is stable under Definition~\ref{def:equilibrium} but violates unconstrained optimality. The network in panel (b) globally maximizes $Q$ and therefore satisfies unconstrained optimality.\par}
\end{figure}

\begin{example}[Stability without unconstrained optimality]
\label{ex:stable-not-unconstrained}
Consider three individuals with $\beta_1=\beta_2=-3$ and $\beta_3=-1$. Retain only own-project returns, quadratic project costs, and geometric team benefits:
\begin{align*}
U_i(A)&=\beta_i p_i-p_i^2+\sum_{r\in P_i}V_{\mathrm{size}}(n_r),\\
V_{\mathrm{size}}(n)&=[n\geq2]\,5\left(\frac35\right)^{n-2}.
\end{align*}
This is a special case of the estimation model, with $\psi_P=2$, $p_0=0$, $v_{\mathrm{size},2}=5$, $\rho_{\mathrm{size}}=3/5$, and the other payoff terms zero. In the notation of Section~\ref{sec:intro-payoffs}, $V(X_i,\{X_j\}_{j\in N_r})=V_{\mathrm{size}}(n_r)$.

Let $P^\ast=\{r\}$ and $N_r=\{1,2\}$, so individual 3 undertakes no project. Then
\[
A^\ast=\begin{pmatrix}1\\1\\0\end{pmatrix},
\qquad (U_1(A^\ast),U_2(A^\ast),U_3(A^\ast))=(1,1,0).
\]
Let $s$ denote the available new project. Holding the other individuals' memberships fixed, the payoffs from every unilateral choice are:
\begin{center}
{\small\upshape\renewcommand{\arraystretch}{1.2}
\begin{tabular}{@{}lcc@{}}
\hline
\textbf{Individual's choice} & \textbf{Individuals 1 and 2} & \textbf{Individual 3}\\
\hline
Neither project & $0$ & $0$\\
Existing project $r$ only & $1$ & $1^{\dagger}$\\
New solo project $s$ only & $-4$ & $-2$\\
Both projects & $-5$ & $-3$\\
\hline
\end{tabular}\\[3pt]
$\dagger$ Profitable entry that violates the stability constraint.}
\end{center}
The existing team is stable: each incumbent's membership margin is $1$, and every solo-project option has a negative return, so the right-hand side of \eqref{stability_constraint} is zero. The table shows that neither incumbent has a profitable deviation. Individual 3 would gain $1$ by joining $r$, but her entry lowers team value from $5$ to $3$, giving each incumbent the membership margin
\[
\Delta_{a_{j,r}}U_j=-3-1+3=-1<0,\qquad j=1,2.
\]
This entry is infeasible because both incumbents would prefer to exit. Individual 3's remaining choices yield no gain. Since the table covers all choices over $P^\ast\cup\{s\}$, $A^\ast$ is a $k$-team stable equilibrium for every $k\geq2$, but $A^\ast$ is not $k$-team optimal in the absence of stability constraints.

For comparison, the same model has a global potential maximizer with no active projects, denoted $A^Q$. Let $W_{\mathrm{size}}(n)=\sum_{m=1}^nV_{\mathrm{size}}(m)$, so $W_{\mathrm{size}}(1)=0$, $W_{\mathrm{size}}(2)=5$, and $W_{\mathrm{size}}(3)=8$. Its potential satisfies
\begin{align*}
Q(A)&=\sum_i\bigl(\beta_i p_i-p_i^2\bigr)+\sum_{r\in P}W_{\mathrm{size}}(n_r)\\
&\leq\sum_{r\in P}\left[\sum_{i\in N_r}(\beta_i-1)+W_{\mathrm{size}}(n_r)\right]
\leq-|P|,
\end{align*}
using $p_i^2\geq p_i$. The bracket equals $-4$, $-4$, or $-2$ for a solo project; $-3$, $-1$, or $-1$ for a pair; and $-2$ for the three-person team. Thus every nonempty network has negative potential, whereas $Q(A^Q)=0>Q(A^\ast)=-3$. At $A^Q$, every unilateral project is solo and unprofitable, so unconstrained optimality holds. Figure~\ref{fig:stable-versus-potential} compares the two equilibria.
\end{example}

\FloatBarrier

For $k\geq2$, unconstrained optimality rules out both dropping a membership and replacing it with a new solo project. Hence every member satisfies the stability constraint, and restricting the individual's choice set preserves optimality, as Lemma~\ref{lem:optimality_implies_stability} states. The converse fails in Example~\ref{ex:stable-not-unconstrained}: the stability constraint excludes a deviation that benefits its mover but leaves the incumbents unwilling to remain in the team.

\FloatBarrier

\clearpage

\end{document}